\documentclass{aa}  

\usepackage{graphicx}
\usepackage{txfonts}
\usepackage{scalefnt}
\usepackage{multirow}

\usepackage[allcolors=blue]{hyperref}
\usepackage[normalem]{ulem}

\begin{document} 

   \title{Photo-chemo-dynamical analysis and the origin of the bulge globular cluster Palomar~6  \thanks{Observations collected at the European Southern Observatory, Paranal, Chile (ESO), under programmes 0103.D-0828A (PI: M. Valentini); based on observations with the NASA/ESA Hubble Space Telescope, obtained at the Space Telescope Science Institute, which is operated by AURA, Inc. under NASA contract NAS 5-26555 associated with programme GO-14074.}}

   \author{
          S. O. Souza\inst{1}
          \and
          M. Valentini\inst{2}
         \and
         B. Barbuy\inst{1}
          \and
           A. Pérez-Villegas\inst{3}
          \and
          C. Chiappini\inst{2}
          \and
          S. Ortolani\inst{4,5,6}
          \and 
          D. Nardiello\inst{7,5}
          \and
          B. Dias\inst{8}
           \and
          F. Anders\inst{9}
          \and
          E. Bica\inst{10}
          }
         \offprints{S. O. Souza}

   \institute{
              Universidade de S\~ao Paulo, IAG, Rua do Mat\~ao 1226, Cidade Universit\'aria, S\~ao Paulo 05508-900, Brazil\\
              \email{stefano.souza@usp.br}
            \and
             Leibniz-Institut für Astrophysik Potsdam (AIP), An der Sternwarte 16, Potsdam, 14482, Germany
             \and
             Instituto de Astronom\'ia, Universidad Nacional Aut\'onoma de M\'exico, A. P. 106, C.P. 22800, Ensenada, B. C., M\'exico
             \and
              Universit\`a di Padova, Dipartimento di Astronomia, Vicolo dell'Osservatorio 2, I-35122 Padova, Italy
              \and 
             INAF-Osservatorio Astronomico di Padova, Vicolo dell'Osservatorio 5, I-35122 Padova, Italy 
             \and
             Centro di Ateneo di Studi e Attivit\`a Spaziali "Giuseppe Colombo" - CISAS. Via Venezia 15, 35131 Padova, Italy
             \and
              Aix Marseille Univ, CNRS, CNES, LAM, F-13013, Marseille F-13013, France
              \and
              Instituto de Alta Investigación, Universidad de Tarapacá, Casilla 7D, Arica, Chile
              \and
              Institut de Ci\`encies del Cosmos, Universitat de Barcelona (IEEC-UB), Carrer Mart\'{\i} i Franqu\`es 1, 08028 Barcelona, Spain
             \and
              Universidade Federal do Rio Grande do Sul, Departamento de Astronomia,CP 15051, Porto Alegre 91501-970, Brazil
}  

\date{Received 13 July 2021; accepted 25 August 2021}
 \abstract
   {Palomar 6 (Pal~6) is a moderately metal-poor globular cluster projected towards the Galactic bulge. 
   A full analysis of the cluster can give hints on the early chemical enrichment of the Galaxy and a plausible origin of the cluster.}
   {The aim of this study is threefold: a detailed analysis of high-resolution spectroscopic data obtained with the UVES spectrograph at the Very Large Telescope {(VLT)} at ESO, the derivation of the age and distance of Pal~6 from {\it Hubble Space Telescope} {(HST)} photometric data, and an orbital analysis to determine the probable origin of the cluster.}
   {High-resolution spectra of six red giant stars in the direction of Palomar 6 were obtained at the $8$m VLT UT2-Kueyen telescope equipped with the UVES spectrograph in FLAMES$+$UVES configuration. Spectroscopic parameters were derived through excitation and ionisation equilibrium of \ion{Fe}{I} and \ion{Fe}{II} lines, and the abundances were obtained from spectrum synthesis. From HST photometric data, the age and distance were derived through a statistical isochrone fitting. Finally, a dynamical analysis was carried out for the cluster assuming two different Galactic potentials.}
   {Four stars that are members of Pal~6 were identified in the sample, which gives a mean radial velocity of $174.3\pm1.6$ km\,s$^{-1}$ and a mean metallicity of [Fe/H]$\,=-1.10\pm0.09$ for the cluster. We found an enhancement of $\alpha$-elements (O, Mg, Si, and Ca) $0.29<\,$[X/Fe]$\,<0.38$ and the iron-peak element Ti of [Ti/Fe]$\,\sim+0.3$. The odd-Z elements (Na and Al) show a mild enhancement of [X/Fe]$\,\sim +0.25$. The abundances of both first- (Y and Zr) and second-peak (Ba and La) heavy elements are relatively high, with $+0.4<\,$[X/Fe]$\,<+0.60$ and $+0.4<\,$[X/Fe]$\,<+0.5$, respectively. The r-element Eu is also relatively high with [Eu/Fe]$\,\sim +0.6$. One member star presents enhancements in N and Al, with [Al/Fe]$\,>+0.30$, this being evidence of a second stellar population, further confirmed with the NaON-Al (anti)correlations. For the first time, we derived the age of Pal~6, which resulted to be $12.4\pm0.9$ Gyr. We also found a low extinction coefficient $R_V=2.6$ for the Pal~6 projection, which is compatible with the latest results for the highly extincted bulge populations. The derived extinction law results in a distance of $7.67\pm0.19$ kpc from the Sun with an A$_{\rm V} = 4.21\pm0.05$. The chemical and photometric analyses combined with the orbital-dynamical analyses point out that Pal~6 belongs to the bulge component probably formed in the main-bulge progenitor.} 
   {The present analysis indicates that the globular cluster Pal~6 is located in the bulge volume and that it was probably formed in the bulge in the early stages of the Milky Way formation, sharing the chemical properties with the family of intermediate metallicity very old clusters M~62, NGC~6522, NGC~6558, and HP~1. }
\titlerunning{The Globular Cluster Palomar 6}
\authorrunning{Souza et al.}
\maketitle
%

\section{Introduction}

The stellar populations in the Galactic bulge can provide information on its complex formation processes \citep[e.g.][]{barbuy18a,queiroz20a,queiroz20b,rojas-arriagada20}. The system of globular clusters (GCs) is an important tracer for the study of the formation and evolution of the Galaxy since they retain the chemo-dynamical signatures of the first stages of the Milky Way formation. 

It is expected that the oldest stars of the Galaxy have metallicities of [Fe/H]$\sim -3$ and are mostly found in the Galactic halo. However, the oldest stars might instead reside mainly in the Galactic bulge \citep[e.g.][]{Tumlinson10} with a 
higher metallicity of [Fe/H]$>-1.5$ due to the fast early chemical enrichment in the inner Galaxy
\citep{Chiappini11,wise12,barbuy18a}. Analyses of Galactic bulge GCs have demonstrated that the metallicity distribution of their members peaks at [Fe/H]$\sim -1.0$ \citep[][and references therein]{bica16}  and that some of these GCs are older than 12.5 Gyr \citep[][]{miglio16,barbuy16,barbuy18a,kerber19,ortolani19,oliveira20}.


Palomar~6 (Pal~6) is a GC projected towards the Galactic bulge ($l = 2.10^\circ$ and $b = 1.78^\circ$), 
located in a highly-extincted region with $A_V > 4.3$ \citep[][2010 edition]{harris96}\footnote{http://physwww.mcmaster.ca/~harris/mwgc.dat}. Despite being a very interesting cluster, information about Pal~6 is conflicting, preventing further analysis, in particular concerning its distance, and consequently in terms of the Galactic component to which Pal~6 should belong.
Pal~6 has been considered to belong to the Galactic bulge due to its current position with respect to the Galactic centre \citep{ortolani95,bica16}. \cite{lee04} suggested that, based on its chemical and kinematic determinations, Pal~6 should belong to an internal component related to a contribution of the halo (inner). \cite{perezvillegas2020} discussed the case of Pal~6 using a distance of d$_{\odot} = 5.8$ kpc \citep{Baumgardt19}, and from their dynamical orbital analysis, they classified the cluster as a thick disc member. This result is opposite to that of \cite{ortolani95}, which found a distance of d$_{\odot}=8.9$ kpc, placing Pal~6 in the Galactic bulge. Recently, \cite{massari19} presented a classification of clusters in terms of their plausible progenitors, indicating whether a cluster originates in a well-defined component of the Galaxy or if it came from one of the merger processes that occurred in the history of the Galaxy, besides other possibilities. They indicated Pal~6 as an unassociated low-energy cluster, again based on the distance estimated by \citet{Baumgardt19}. 

The controversy on  with Galactic component Pal~6 is part of is also due to an uncertain metallicity. The first metallicity estimations of Pal~6 by \citet{Malkan81} from a reddening-free index resulted in [Fe/H] $\sim -1.30$. \citet{ortolani95}, from the $V$ versus $V-I$ colour-magnitude diagram (CMD) based on data observed at the ESO NTT-EMMI, found [Fe/H] $\sim -0.40$ by the slope of the red giant branch (RGB)  and the presence of a red horizontal branch (RHB). \cite{lee02} obtained [Fe/H]$\,=-1.22\pm0.18$ by analysing the slope of the RGB on the near-infrared (NIR) CMD with NICMOS3 JHK bands. Spectroscopic analysis from high-resolution NIR spectra of three RGB stars by the same authors resulted in [Fe/H] =  $-1.08\pm0.06$. In \cite{lee04}, a metallicity of [Fe/H]$=-1.0\pm0.1$ was confirmed from a high-resolution spectroscopic analysis of five probable member stars observed with the CSHELL spectrograph at the NASA Infrared Telescope Facility. 

As part of the present work, we carried out a detailed analysis of Pal~6 from high-resolution spectra obtained with the FLAMES-UVES spectrograph at the ESO Very Large Telescope (VLT).  We also provide the first age derivation of Pal~6 and a consistent distance determination based on photometric data from the {\it Hubble Space Telescope} (HST). Furthermore, to connect the spectroscopic and photometric analyses, we perform  orbital calculations determining the most probable Galactic component to which Pal~6 belongs. Finally, we indicate the probable progenitor for Pal~6. 

This work is organised as follows. The spectroscopic and photometric data are described in Section \ref{sec_data} along with the membership analysis of the observed stars. Section \ref{sec:photo_param} gives  the derivation of photometric stellar parameters as a first guess for the spectroscopic analysis. The final spectroscopic stellar parameters  and  abundance derivation are presented in Section \ref{sec:chemical_analysis}. The photometric analysis and derivation of the fundamental parameters age and distance are described in Section \ref{sec:Age_Distance}. The orbital analysis and discussion on the origin of Pal~6 are presented in Section \ref{sec:discussion}. Finally, our conclusions are drawn in Section \ref{sec:conclusions}.

 
\section{Data}\label{sec_data}

In this section, we describe the spectroscopic and photometric data, and the proper motion analysis.

\subsection{Spectroscopy}\label{sec:data_spec}

The UVES spectra were obtained using the FLAMES-UVES setup centred at 580 nm in the ESO programme 0103.D-0828 (A) (PI: M. Valentini). The ESO programme was coordinated with programme GO11126 (PI: M. Valentini) for Campaign 11 of the K2 satellite \citep[K2 is the re-purposed \textit{Kepler} mission;][]{Howell2014}: the goal was to obtain asteroseismology for the giants in the proposed GCs. K2 observed four giants in Pal 6, but their UVES spectra were not collected due to clouds and strong winds that affected ESO observations. UVES spectra have a coverage ranging from 480 nm to 680 nm. Six giant stars of Pal 6 were observed, and the log of observations is given in Table 1. The $JHK_S$-combined image of Pal 6 is shown in Figure 1 and was obtained from the Vista Variables in the Via Lactea VVV survey \citep{saito12}.

 \begin{table}
    \caption{Log of the spectroscopic  FLAMES-UVES observations of programme 0103.D-0828 (A), carried out in 2019. The quoted seeing and air mass are the mean values along the exposures. In the last column, we give the corresponding GIRAFFE setup in which additional stars were observed.}
    \small
    \centering
    \begin{tabular}{lccccc}
    \noalign{\smallskip}
    \hline
    \noalign{\smallskip}
    \hline
    \noalign{\smallskip}
    Date & UT & exp    & Air mass  & Seeing & GIRAFFE \\
         &    &   ( s )  &          & ($''$) &  \\
    \noalign{\smallskip}
    \hline
    \noalign{\smallskip}
    \multicolumn{6}{c}{ Programme 0103.D-0828 (A) } \\
    \hline
    \noalign{\smallskip}
    2019-06-24 & 23:44:40 & 2400 & 1.810 & 0.87$''$ & H13-1 \\
    2019-06-25 & 01:30:48 & 2400 & 1.190 & 0.85$''$ & H13-3 \\
    2019-06-25 & 02:31:52 & 2700 & 1.057 & 0.95$''$ & H14-1 \\
    2019-06-25 & 03:20:48 & 2700 & 1.012 & 0.91$''$ & H14-2 \\
    2019-06-25 & 05:35:43 & 2700 & 1.097 & 0.93$''$ & H14-3 \\
    2019-06-25 & 06:24:35 & 2700 & 1.223 & 0.93$''$ & H14-4 \\
    \noalign{\smallskip}
    \hline
    \noalign{\smallskip}
    \hline
    \end{tabular}
    \label{logobs}
\end{table}

\begin{figure}
    \centering
    \includegraphics[width=0.75\columnwidth]{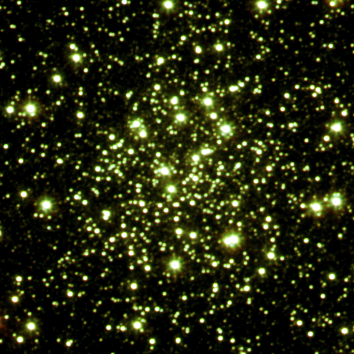}
    \caption{$JHK_s$-combined colour image from the VVV survey for Pal~6. The image has a size of $2\times2$ arcmin$^2$. North is at $45^{\circ}$anticlockwise.}
    \label{fig:image_2mass}
\end{figure}

The data were reduced using the ESO-Reflex software with UVES-Fibre pipeline \citep{ballester00,modigliani04}. After reduction, we are left with six spectra for each star. The corresponding spectra of each star were corrected by the radial velocity. To compute the radial velocities and the barycentric corrections, we used the python library \texttt{PyAstronomy} cross-correlating the spectra with  the Arcturus spectrum \citep{hinkle00}. 

The values of heliocentric radial velocity of each spectrum and their mean are presented in Table \ref{tab:velocities}. From these values, we calculate a mean heliocentric radial velocity for Pal~6 of $174.3\pm1.6$ km s$^{-1}$, excluding the stars \texttt{ID730} and \texttt{ID030} for which the radial velocities are very discrepant compared with the other stars. Our mean radial velocity determination is in good agreement with the recent value of $176.3\pm1.5$ km s$^{-1}$ given by \citet{Baumgardt19}.  Finally, each spectrum is normalised and combined through the median flux to obtain the final stellar spectrum.

\begin{table}
\caption{Radial velocity obtained for each extracted spectra and the average value for each star.}
\scalefont{0.85}
\centering
\begin{tabular}{lcclcc}
\noalign{\smallskip}
\hline
\noalign{\smallskip}
\hline
\noalign{\smallskip}
Target     &  V$_{r}^{hel}$   & $\sigma_{V_{r}}$ & Target  & V$_{r}^{hel}$    & $\sigma_{V_{r}}$ \\
           &  km s$^{-1}$     &  km s$^{-1}$     &         &  km s$^{-1}$     &  km s$^{-1}$     \\
\noalign{\smallskip}
\hline
\noalign{\smallskip}
730\_1     &   $-87.83$    &   $6.12$   &  243\_1      &  $+172.80$  &   $6.06$   \\
730\_2     &   $-87.79$    &   $6.10$   &  243\_2      &  $+172.34$  &   $6.38$   \\
730\_3     &   $-87.61$    &   $5.84$   &  243\_3      &  $+172.50$  &   $6.18$   \\
730\_4     &   $-86.73$    &   $6.05$   &  243\_4      &  $+172.85$  &   $5.87$   \\
730\_5     &   $-86.56$    &   $5.87$   &  243\_5      &  $+172.84$  &   $6.00$   \\
730\_6     &   $-87.43$    &   $6.02$   &  243\_6      &  $+173.02$  &   $6.54$   \\
\hline
\noalign{\smallskip}
{ 730}  & ${-87.33}$ & ${2.65}$   &  {243}   & ${172.73}$ & ${2.61}$ \\
\hline
\noalign{\smallskip}
030\_1     &   $-62.94$    &   $6.34$   &  785\_1      &  $+175.41$  &   $8.79$   \\
030\_2     &   $+14.63$    &   $9.36$   &  785\_2      &  $+175.72$  &   $7.77$   \\
030\_3     &   $-32.31$    &   $9.82$   &  785\_3      &  $+174.17$  &   $5.56$   \\
030\_4     &   $-56.20$    &   $6.81$   &  785\_4      &  $+174.01$  &   $6.58$   \\
030\_5     &   $+14.36$    &   $8.82$   &  785\_5      &  $+175.50$  &   $7.20$   \\
030\_6     &   $+14.86$    &   $12.0$   &  785\_6      &  $+174.69$  &   $7.80$   \\
\hline
\noalign{\smallskip}
{030}  & ${-12.36}$ & ${17.44}$  &  {785}    & ${174.99}$ & ${3.25}$ \\
\hline
\noalign{\smallskip}
145\_1     &   $+179.81$   &   $7.07$   &  401\_1      &  $+170.16$  &   $6.12$   \\
145\_2     &   $+178.59$   &   $7.39$   &  401\_2      &  $+170.82$  &   $7.13$   \\
145\_3     &   $+178.38$   &   $6.27$   &  401\_3      &  $+169.62$  &   $6.82$   \\
145\_4     &   $+178.33$   &   $5.67$   &  401\_4      &  $+168.72$  &   $6.18$   \\
145\_5     &   $+179.65$   &   $6.51$   &  401\_5      &  $+170.87$  &   $6.06$   \\
145\_6     &   $+179.23$   &   $7.27$   &  401\_6      &  $+171.00$  &   $6.34$   \\
\hline
\noalign{\smallskip}
{145}  & ${179.02}$ & ${2.98}$   &  {401}    & ${170.21}$ & ${2.96}$ \\
\noalign{\smallskip}
\hline
\noalign{\smallskip}
\hline
\end{tabular}
\label{tab:velocities}
\end{table}

\subsection{Photometry}\label{sec:cmd}
\begin{figure*}
    \centering
    \includegraphics[width=0.95\textwidth]{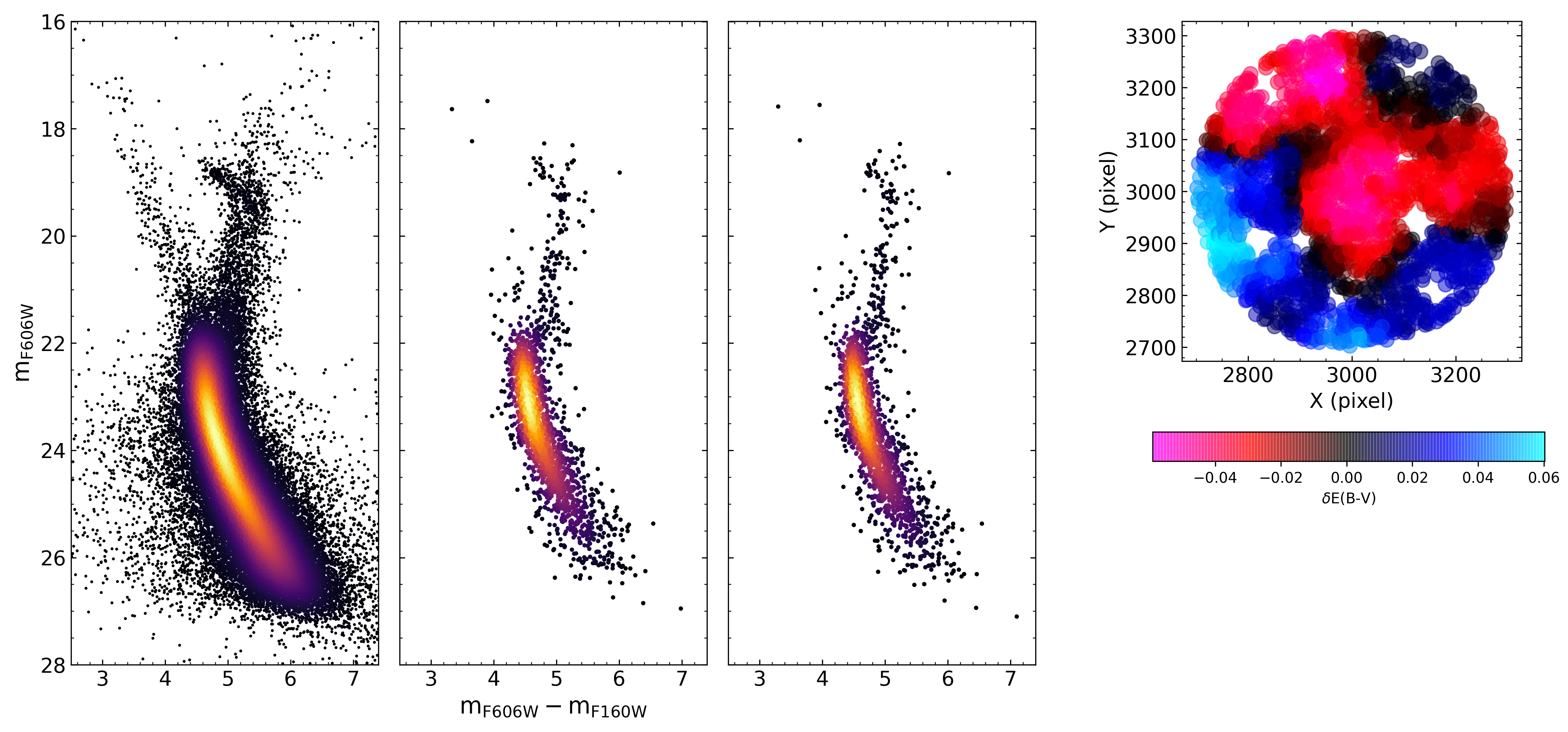}
    \caption{ Procedure to obtain the photometry of Pal~6. First panel: HST photometry from \citet{cohen18}. Second panel: Stars selected by the quality method within a radius of $\sim 0.66$ arcmin from the cluster centre. Third panel: Differential reddening corrected CMD. Last panel: Differential reddening map.} 
    \label{fig:phot_proc}
\end{figure*}

For the photometric analysis, we used the HST data collected during the GO-14074 \citep[PI: Cohen,][]{cohen18} in {F110W and F160W (WFC3-IR), and in F606W (ACS-WFC)} (first panel of Figure \ref{fig:phot_proc}). Data were reduced using the pipeline described in \cite{nardiello18}. We also followed their recipe (based on the use of the quality-of-fit and photometric error parameters) to select well-measured stars and reject poor photometric measurements. Additionally, we selected the stars within a radius of $300$ pixels from the cluster centre that is equivalent to a core radius of $\sim 0.66$ arcmin \citep[][2010 edition]{harris96}. The cleaned CMD is shown in the second panel of Figure \ref{fig:phot_proc}, which contains the final selected stars.

Another important effect in the photometric data is the differential reddening. Mainly for the clusters with a high reddening value, differential reddening increases the spread on the CMD. This is the case of Pal~6, which has an extinction of  $A_V > 4$.  We perform a reddening correction with a method similar to that described in \cite{milone12}. The third panel of Figure \ref{fig:phot_proc} presents the final CMD after the reddening correction is applied, and the map of differential reddening is on the last panel of Figure \ref{fig:phot_proc}. The contamination by field stars, combined with the high extincted region, results in low values of differential reddening. However, we obtained clearer main-sequence (MS) turn-off (TO) and sub-giant-branch (SGB) structures for the Pal~6 CMD.

\subsection{Membership selection}\label{sec:memb}
\begin{figure}
   \centering
   \includegraphics[width=\columnwidth]{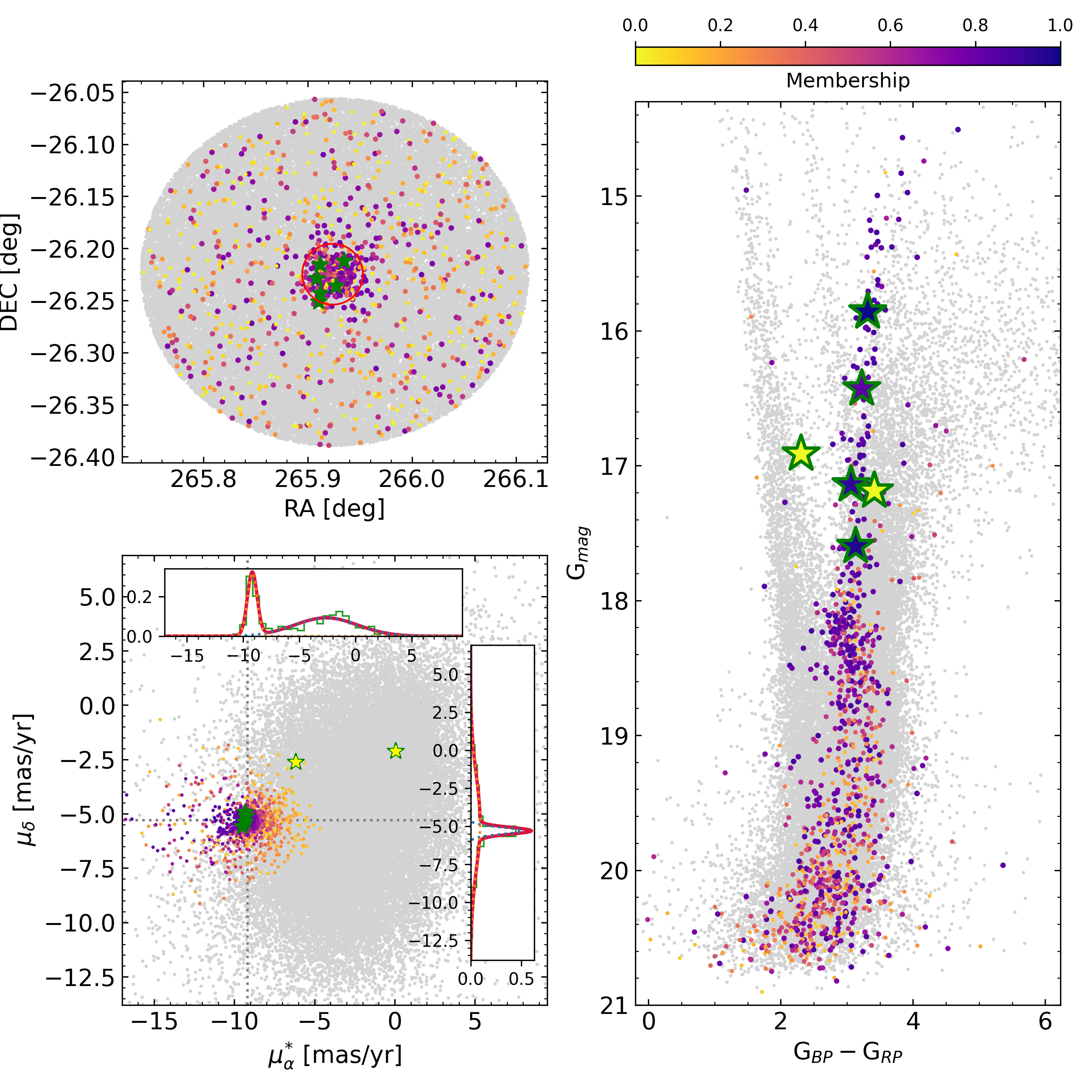}
   \caption{Proper motion analysis to obtain the cluster members. Top left panel: Sky distribution of stars within 10 arcmin of the cluster centre. Bottom left panel: Vector point diagram with the cluster (coloured dots) and field (grey dots) stars; the green star symbols are the observed stars with FLAMES-UVES, and the insert plots show the density distributions found using GMMs. Right panel: Gaia EDR3 $G$ versus $G_{BP}-G_{RP}$ CMD; the green star symbols are the observed stars. From the bottom left and right panels,  we can identify that two observed stars have zero membership probability (yellow star symbols).}
   \label{fig:gaia}
\end{figure}

We performed a membership analysis to find out which stars observed spectroscopically are members of Pal~6. We selected the Gaia Early Data Release 3 \citep[EDR3;][]{gaiaedr3} stars within $10'$ of the cluster centre (top left panel of Figure \ref{fig:gaia}). For the proper-motion distribution presented in  the bottom left panel of Figure \ref{fig:gaia}, we applied the Gaussian mixture models \citep[GMMs;][]{Pedregosa11} clustering method to separate the cluster members from the field stars. The derived mean proper motion for Pal~6 is $<\mu_{\alpha}^{*}> = -9.19\pm0.06$ mas~yr$^{-1}$ and $<\mu_{\delta}> = -5.30\pm0.05$ mas~yr$^{-1}$, in excellent agreement with the new values computed by \cite{vasiliev21}. 

The membership probabilities are computed taking into account both cluster and field distributions, which are derived using GMMs \citep[see][for the mathematical description of the membership distribution]{bellini09}. Once we obtained the membership probability, we cross-matched our sample stars with the Gaia data (Table \ref{tab:gaia}), indicated as green stars in Figure \ref{fig:gaia}. We found that two stars of our sample have zero membership probability (non-members) and four stars have probabilities above $80\%$. The non-member stars are the same stars with discrepant radial velocities (ID730 and ID030).

\begin{table}
\caption{Gaia EDR3 information about the observed stars; the last column shows the membership probabilities.}
\small
\tabcolsep 0.15cm
\begin{tabular}{lccccc}
\noalign{\smallskip}
\hline
\noalign{\smallskip}
\hline
\noalign{\smallskip}
  ID  & $^\dag \mu_{\alpha}^{*}$   & $\mu_{\delta}$  &   $G$  & $G_{RP}$ & $\mathcal{P}_{\rm memb}$     \\
      &  (mas/yr) &  (mas/yr) &  (mag) & (mag)    & (\%)     \\
\noalign{\smallskip}
\hline
\noalign{\smallskip}
  730 & $-6.18\pm0.10$ & $-2.61\pm0.06$ & $17.187$ & $15.873$ & $0  $ \\
  243 & $-9.32\pm0.07$ & $-5.37\pm0.04$ & $15.859$ & $14.493$ & $100$ \\
  30  & $+0.04\pm0.10$ & $-2.11\pm0.06$ & $16.913$ & $15.820$ & $0  $ \\
  785 & $-9.26\pm0.14$ & $-5.12\pm0.08$ & $17.598$ & $15.997$ & $97 $ \\
  145 & $-9.49\pm0.12$ & $-5.58\pm0.07$ & $17.141$ & $15.788$ & $93 $ \\
  401 & $-9.33\pm0.08$ & $-4.92\pm0.05$ & $16.430$ & $15.056$ & $83 $ \\
\noalign{\vskip 0.2cm}
\noalign{\hrule}
\noalign{\smallskip} \hline
\end{tabular}  
\label{tab:gaia}
\\
$^\dag \mu_{\alpha}^{*} = \mu_{\alpha} \cos \delta. $
\end{table}

\section{Atmospheric stellar paramaters} \label{sec:photo_param}

The photometric effective temperature ($T_{\rm eff}$) and surface gravity ($\log g$) are derived from  the $VIJHK_S$ magnitudes given in Table \ref{starmag}. For comparison purposes, we also obtained the effective temperature from the Transiting Exoplanet Survey Satellite (TESS) input catalogue \citep[TIC;][]{stassun18} for five of our six observed stars. We collected the 2MASS $J$, $H$, and $K_S$ magnitudes from \citet{skrutskie06} and the VVV survey \citep[][]{saito12}. Finally, according to \cite{alonso99}, the colour $V-I$ is the best colour index to derive the effective temperature of giant stars. To obtain the $V-I$ colour for our sample, we employed the photometric systems' relationships $G - V = f(G_{BP} - G_{RP})$ and $G - I = f(G_{BP} - G_{RP})$ from Gaia EDR3 \citep{riello2021}.

\begin{table*}
\caption[1]{Identifications, coordinates, and magnitudes. $JHK_{s}$ are given from both 2MASS and VVV surveys. }
\small
\tabcolsep 0.15cm
\begin{tabular}{lcccccccccccc}
\noalign{\smallskip}
\hline
\noalign{\smallskip}
\hline
\noalign{\smallskip}
  ID          &     ID              &   RA        &       DEC      &   K$_P$ &    $V$ &   $V-I$ &  $J$   &  $H$   &  $K_S$   & $J$      &   $H$   &   $K$ \\
  & 2MASS &   ( hh:mm:ss )  & (dd:mm:ss)  &       &     &       & \multicolumn{3}{c}{ 2MASS }   & \multicolumn{3}{c}{ VVV }  \\
\noalign{\smallskip}
\hline
\noalign{\smallskip}
730$^\dag$    &  $17433876-2612551$ & $17:43:38.75$ & $-26:12:55.2$ & $16.05$ & $18.29$ & $3.22$ & $13.43$ & $12.23$ & $11.83$ & $13.22$ & $12.21$ & $11.82$  \cr
243           &  $17434250-2614101$ & $17:43:42.51$ & $-26:14:10.2$ & $14.75$ & $16.91$ & $3.12$ & $11.86$ & $10.66$ & $10.25$ & $11.95$ & $11.32$ & $10.73$  \cr
30$^\dag$     &  $17433862-2615013$ & $17:43:38.47$ & $-26:15:04.8$ & $15.82$ & $17.62$ & $2.19$ & $13.58$ & $12.46$ & $11.93$ & $14.02$ & $13.53$ & $13.26$  \cr
785           &  $17434440-2612418$ & $17:43:44.38$ & $-26:12:42.5$ & $16.44$ & $18.56$ & $2.94$ & $13.35$ & $12.36$ & $11.87$ & $13.65$ & $12.77$ & $12.37$  \cr
145           &  $17433889-2614359$ & $17:43:38.86$ & $-26:14:34.7$ & $15.98$ & $18.07$ & $2.87$ & $12.76$ & $11.61$ & $11.51$ & $13.28$ & $12.27$ & $11.98$  \cr  
401           &  $17433806-2613426$ & $17:43:38.05$ & $-26:13:42.7$ & $15.28$ & $17.43$ & $3.03$ & $12.55$ & $11.46$ & $11.08$ & $12.52$ & $11.45$ & $11.23$  \cr

\noalign{\vskip 0.2cm}
\noalign{\hrule}
\noalign{\smallskip} \hline

\end{tabular}                                             
\\ $^\dag$ Stars classified as non-members based on proper-motion and radial velocities.                
\label{starmag}                                                                                      
\end{table*}                                                                                                                    

\subsection{Effective temperatures}

Effective temperatures $T_{\rm eff}$ were derived from $V-I$, $V-K$, and $J-K$ using the colour-temperature calibrations from \citet{casagrande10}. The VVV $JHK$ colours were transformed into the 2MASS $JHK_S$ system using relations given by \citet{soto13}. For Pal~6, the distance modulus of ($m-$M)$_0 = 13.87$, extinction A$_V=4.53$, and metallicity [Fe/H]$=-0.91$ were used \citep[][2010 edition]{harris96} to perform the reddening correction of the colours. Table \ref{tabteff} lists the derived photometric effective temperatures. The $<T_{\rm eff}>$ is the mean effective temperature considering only values below 5000 K.

\begin{table*}
\caption{ Atmospheric parameters derived from photometry using calibrations by \cite{casagrande10} for $V-I$, $V-K$, $J-K,$ and spectroscopic analysis of
Fe lines. }
\label{tabteff}
\small
\resizebox{\linewidth}{!}{%
\tabcolsep 0.15cm
\begin{tabular}{lccccccccccccccccc}
\noalign{\smallskip}
\hline
\noalign{\smallskip}
\hline
\noalign{\smallskip}
& & \multicolumn{9}{c}{Photometric parameters} & & \multicolumn{6}{c}{Spectroscopic parameters}\cr
\cline{3-11}  \cline{13-18}    \cr
         &  &          &  \multicolumn{2}{c}{ 2MASS }   &   \multicolumn{2}{c}{ VVV } &  & & & & & & \cr
{\rm ID} & T$_{\rm TESS}$  & T$_{(V-I)}$ & T$_{(V-K)}$ & T$_{(J-K)}$ & T$_{(V-K)}$ & T$_{(J-K)}$ & $<$T$_{\rm eff}>$ & ${\rm BC_V}$ & ${\rm M_{bol}}$ & $\log g$ & & T$_{\rm eff}$ & $\log g$ & [\ion{Fe}{I}/H] & [\ion{Fe}{II}/H] & [Fe/H] &  $v_t$ \cr
      &    (K)      &   (K)       &   (K)       &   (K)       &   (K)       &   (K)          &    (K)           &                &                 &          & &  (K)      &          &       &          &              & (km s$^{-1}$)    \cr
\noalign{\smallskip}
\hline
\noalign{\smallskip}

$730$ & $3973$  & $4267$ & $4764$ & $4240$ & $4742$ & $4534$ & $4535$ & $-0.721$ & $-0.89$ & $1.67$ & &    $4857$ & $1.40$ & $-1.09$ & $-1.10$ & $-1.10$ & $2.5$  \cr
$243$ & $4323$  & $4385$ & $4592$ & $4212$ & $5024$ & $5304$ & $4385$ & $-0.623$ & $-2.18$ & $1.09$ & &    $4350$ & $0.80$ & $-0.93$ & $-0.91$ & $-0.92$ & $1.0$  \cr
$ 30$ & $5058$  & $7610$ & $5632$ & $4103$ & $8780$ & $8095$ & $4103$ & $-0.093$ & $-0.94$ & $1.48$ & &    $4800$ & $1.50$ & $-1.65$ & $-1.61$ & $-1.63$ & $2.3$  \cr
$785$ &  --     & $4659$ & $4568$ & $4601$ & $5013$ & $4985$ & $4630$ & $-0.446$ & $-0.35$ & $1.92$ & &    $4860$ & $2.40$ & $-1.21$ & $-1.20$ & $-1.21$ & $2.0$  \cr
$145$ & $4865$  & $4790$ & $4676$ & $5455$ & $5120$ & $4871$ & $4790$ & $-0.380$ & $-0.77$ & $1.81$ & &    $4800$ & $1.90$ & $-1.31$ & $-1.26$ & $-1.28$ & $2.5$  \cr
$401$ & $4387$  & $4511$ & $4866$ & $4634$ & $5002$ & $4888$ & $4750$ & $-0.534$ & $-1.57$ & $1.48$ & &    $4500$ & $1.50$ & $-1.00$ & $-0.99$ & $-1.00$ & $1.0$  \cr

\noalign{\vskip 0.2cm}
\noalign{\hrule}
\noalign{\smallskip} \hline

\end{tabular}}
\end{table*}

\subsection{Surface gravities}

To derive the photometric surface gravities $\log g$, we used the ratio $\log(g_* / g_\odot)$ where $\log g_\odot = 4.44$:
\begin{eqnarray}
\log g_*=4.44+4\log \frac{T_{\rm eff}*}{T_{\odot}}+0.4(M_{\rm bol}-M_{\rm bol\odot})+\log \frac{M_*}{M_{\odot}}.
\end{eqnarray}

We adopted the values of <$T_{\rm eff}$> from Table \ref{tabteff}, $M_*=0.85 M_{\odot}$, and $M_{\rm bol \odot} = 4.75$. The derived values of the photometric $T_{\rm eff}$ and $\log g$ are given in the left columns of Table \ref{tabteff}.

\section{Abundance analysis}\label{sec:chemical_analysis}

We carried out a detailed abundance analysis by means of ionisation and excitation equilibrium to derive stellar parameters, and line-by-line spectrum synthesis  for the derivation of abundance ratios.

\subsection{Spectroscopic stellar parameters}

To determine the final stellar parameters T$_{\rm eff}$, log~g, metallicity [Fe/H], and microturbulence velocity v$_{\rm t}$ of Pal~6, we measured the equivalent width (EW) for a list of \ion{Fe}{I} and \ion{Fe}{II} lines using DAOSPEC \citep{stetson08}. With the purpose of evaluating the impact of blending lines, we remeasured some lines with IRAF, mainly for \ion{Fe}{II}. In the line list of Table \ref{tab:fe}, we also give the adopted oscillator strengths ($\log gf$) for \ion{Fe}{I} lines obtained from VALD3 and NIST databases \citep{piskunov95, martin02} and for \ion{Fe}{II} lines from \citet{melendez09}.

Using the MARCS grid of atmospheric models \citep{gustafsson08}, we extracted the 1D photospheric models for our sample. These CN-mild models consider [$\alpha$/Fe]$=+0.20$ for [Fe/H]$=-0.50,$ while [$\alpha$/Fe]$=+0.40$ for [Fe/H]$\leq -1.00$. For the solar Fe abundance, we adopted $\epsilon(\rm Fe) = 7.50$ \citep{grevesse98}.

\begin{figure}
    \centering
    \includegraphics[width=\columnwidth]{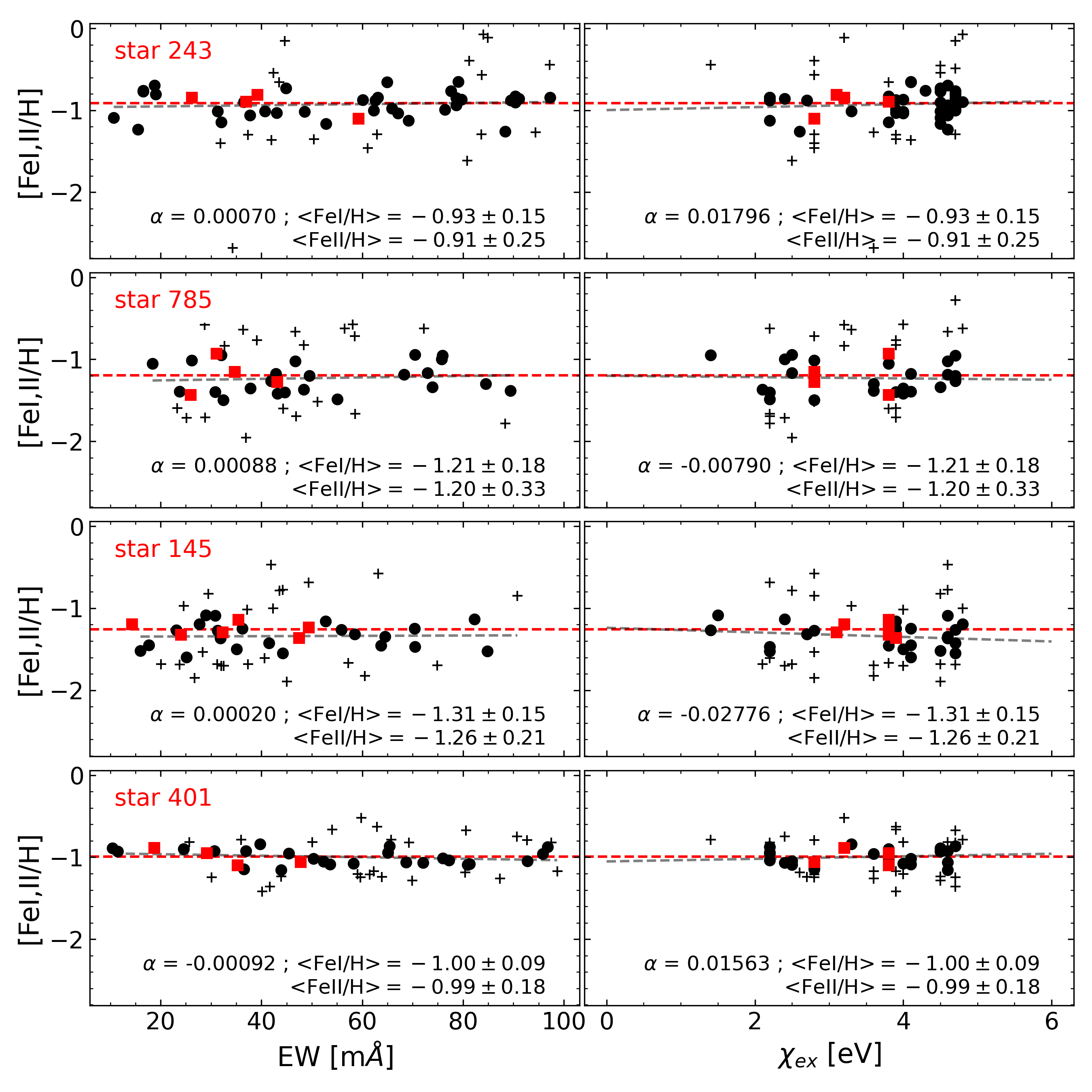}
    \caption{Excitation and ionisation equilibria of \ion{Fe}{I} and \ion{Fe}{II} lines for the four member stars. The black dots are the values considered to compute the metallicity of \ion{Fe}{I} lines after a sigma-clipping of $1-\sigma$. The crosses are the omitted values. The red squares are the values of \ion{Fe}{II} lines. The $\alpha$ values show the slope of the trends of \ion{Fe}{I} lines.}
    \label{fig:excioneq}
\end{figure}

Adopting the mean photometric <T$_{\rm eff}$> and $\log g$ calculated in Section \ref{sec:photo_param} as initial guesses, we derived the spectroscopic parameters. Through an iterative method, we obtained the excitation and ionisation equilibrium. The excitation equilibrium means a constant distribution of \ion{Fe}{I} versus $\chi_{\rm exc}$ and is obtained iterating the value of $T_{\rm eff}$. The similar values of [\ion{Fe}{I}/H] and [\ion{Fe}{II}/H] indicate that the ionisation equilibrium is reached, obtained by iterating in $\log g$. Finally, the microturbulence velocity $v_t$ is obtained by imposing a constant distribution of \ion{Fe}{I} abundance versus EW. Figure \ref{fig:excioneq} shows the excitation and ionisation equilibrium for the four member stars.

The derived spectroscopic parameters T$_{\rm eff}$, $\log g$, [\ion{Fe}{I}/H], [\ion{Fe}{II}/H], [Fe/H], and $v_t$ are presented in the right columns of Table \ref{tabteff}. Our metallicity determination, based on the four member stars, is [Fe/H]$= - 1.10\pm0.09$ dex. This metallicty is in excellent agreement with the spectroscopic determinations of \citet{lee02} and \citet{lee04},  which are [Fe/H] = $-1.08\pm0.06$ and $-1.0\pm0.1$, respectively.

\subsection{Spectrum synthesis}
\begin{figure}
    \centering
    \includegraphics[width=\columnwidth]{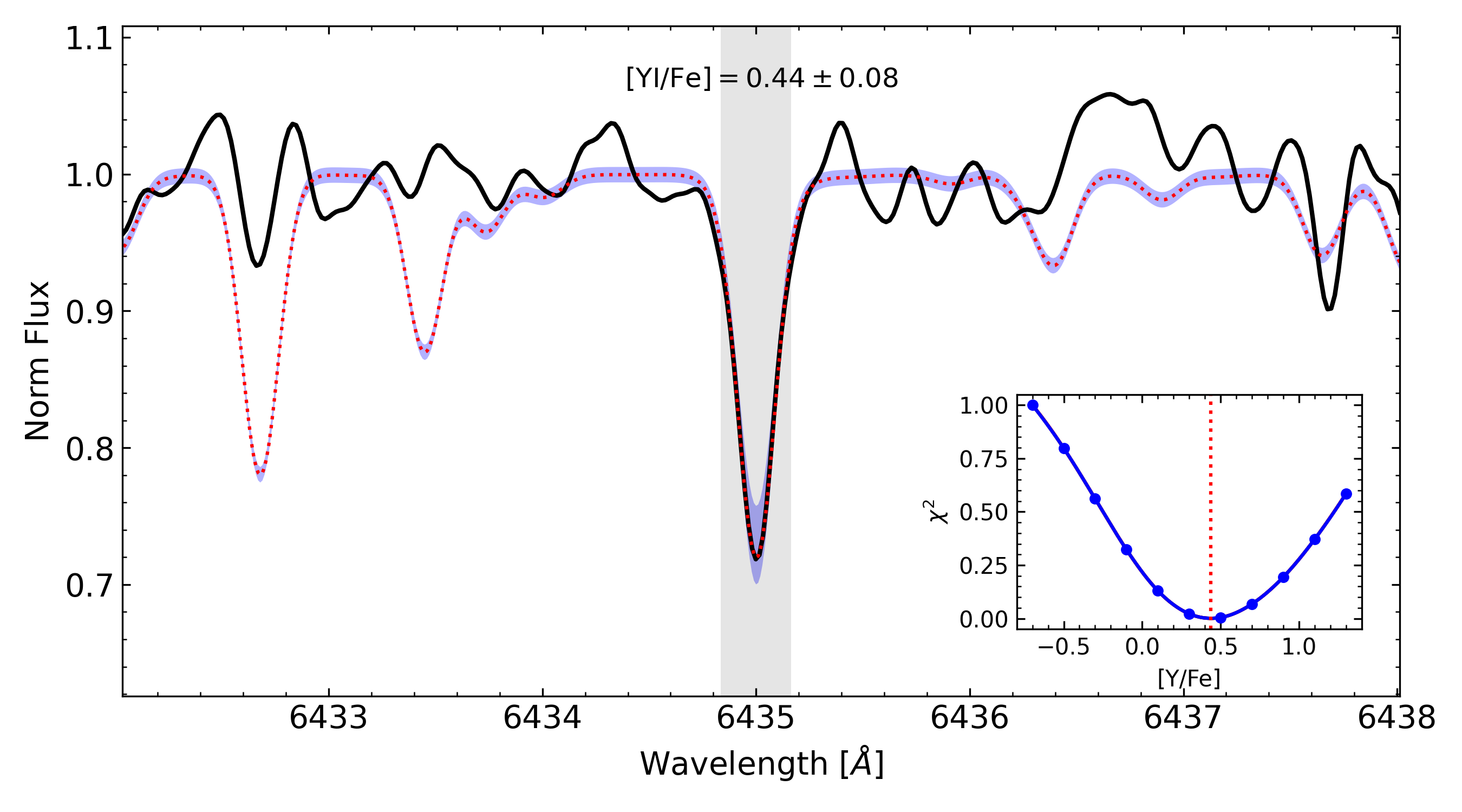}
    \caption{Fit  to \ion{Y}{I} $6435.004$ {\rm \AA} line for star 243. The red dotted lines are the synthetic spectra, the blue strip represents the $1\sigma$ region, and the observed spectrum is the black solid line. The values of $\chi^2$ are in the insert plot.}
    \label{fig:example_fit}
\end{figure}

We derived the abundance ratios for the elements C, N, O, Na, Mg, Al, Si, Ca, Ti, Y, Zr, Ba, La, and Eu. For the spectrum synthesis, we employed the PFANT code described in \citet{barbuy18c}. The code is an update of the Meudon code by M. Spite, and it adopts the local thermodynamic equilibrium (LTE). The basic atomic line list is from VALD3 \citep{Ryabchikova15}. To obtain the best abundance value, we performed a chi-square minimisation algorithm that fits different values to a region of the spectrum. When needed, a variation on the level of the continuum was taken into account. Figure \ref{fig:example_fit} shows an example of the result obtained with this algorithm for the line of \ion{Y}{I} $6435.004$ {\rm \AA} of star 243. The blue shaded region represents the best-fit spectrum within $1-\sigma$, while the grey vertical stripe shows the fit region. The solar abundances A(X) were taken from \cite{grevesse15}.

The CNO abundances are listed in Table \ref{tab:cno}, as detailed below.
For the odd-Z, $\alpha$, and heavy elements, we used the line list from \cite{barbuy16}. In Table \ref{tab:lines_ratios}, we give the line-by-line abundance ratios of the odd-Z elements Na and Al; the $\alpha$-elements Mg, Si, Ca, and Ti; neutron-capture dominant s-elements Y, Zr, La, and Ba; and the r-element Eu. We did not measure Sr lines because they are faint in the observed spectra. The mean values for each star, as well as the cluster mean (considering only the mean of the member stars), are given in Table \ref{tab:mean_abunds}.

\subsection{CNO abundances}
\begin{figure}
    \centering
    \includegraphics[width=\columnwidth]{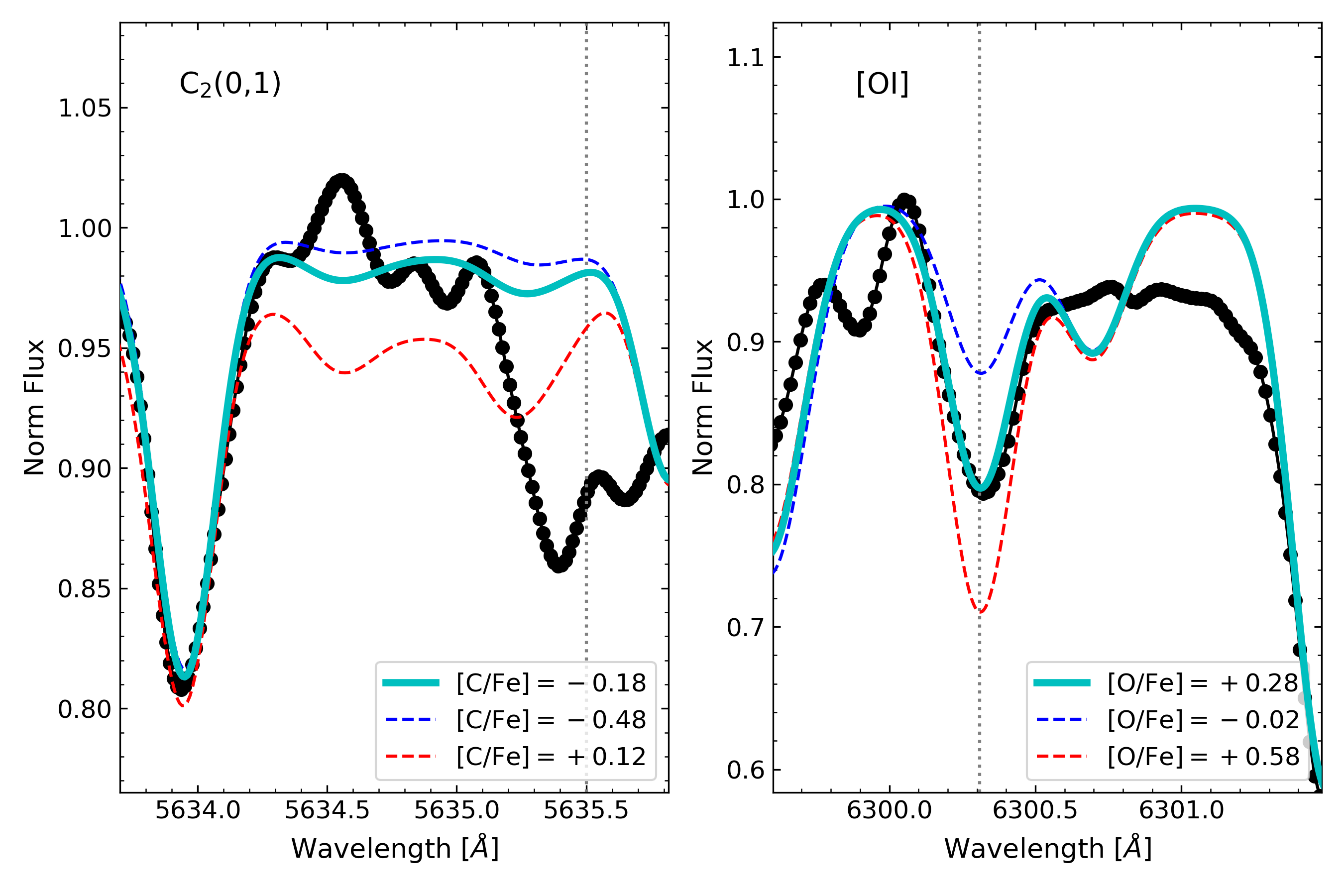}
    \caption{Example for star 243 line fit of the bandhead C$_2$(0,1) (left) and [OI] (right). The solid cyan line is the best-fit abundance ratio, while the dotted lines consider [X/Fe]$=$[X/Fe]$_{\rm best} \pm 0.20$ (red, plus - blue, minus). }
    \label{fig:CO_specs}
\end{figure}

\begin{figure}
    \centering
    \includegraphics[width=\columnwidth]{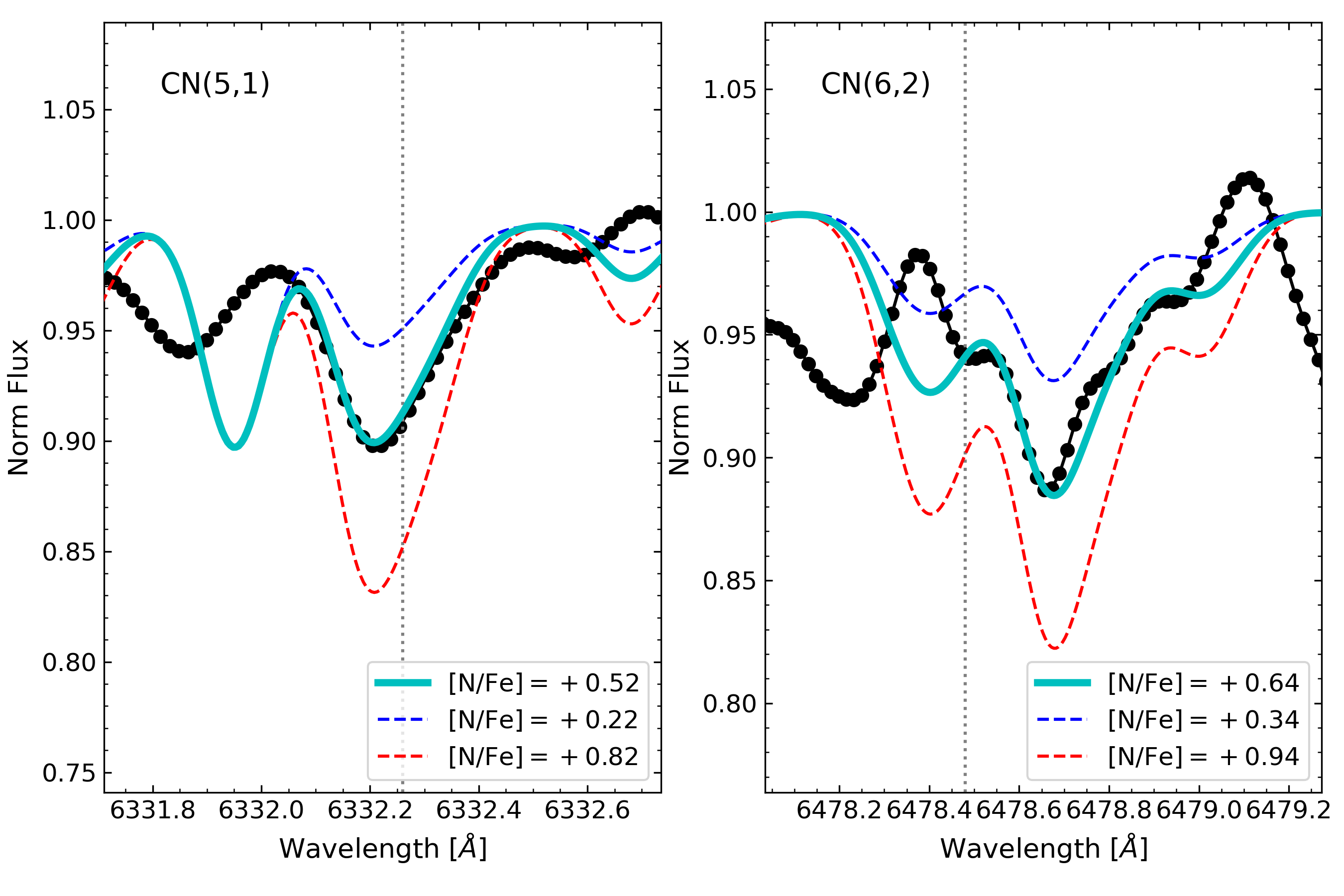}
    \caption{Same as Figure \ref{fig:CO_specs}, but for N from CN(5,1) (left) and CN(6,2) (right).}
    \label{fig:N_specs}
\end{figure}

  To measure the CNO abundances we performed an iterative fitting
  of C, N, and O abundances. For the C abundance, we use the extended C$_2$(1,0) Swan molecular bandhead at $5635.3$\AA. We considered the average fit of the region (left panel, Figure \ref{fig:CO_specs}) and assumed the abundances as upper limits. For the oxygen (Figure \ref{fig:CO_specs}) forbidden line at [OI] $6300.31$ {\rm \AA}, a selection among the original spectra where telluric lines did not contaminate the line was needed, since most of the observations were contaminated, showing that these spectra seem to have been observed at too high air masses. A few spectra could be retrieved showing a clean [OI] $6300.31${\rm \AA} line, and the oxygen abundance could be derived. The nitrogen abundance is derived from the CN(5,1) at 6332.2 {\rm \AA} and CN(6,2) at 6478.48 {\rm \AA} of the $A^2 \Pi X^2 \Sigma$ system  bandheads (Figure \ref{fig:N_specs}). The derived abundances are listed in Table \ref{tab:cno}. 
 
\begin{table}
\caption{Carbon, nitrogen, and oxygen abundances [X/Fe] from C$_2$, CN bandheads, and [OI], respectively.}
\centering
\tabcolsep 0.15cm
\begin{tabular}{lcccc}
\noalign{\smallskip}
\hline
\noalign{\smallskip}
\hline
\noalign{\smallskip}
     &     [C/Fe]   & \multicolumn{2}{c}{[N/Fe]}  & [O/Fe] \\
Star &     C$_2$    &    CN(5,1)   &    CN(6,2)   &     ${\rm [OI]}$   \\
     & $5635.50$ {\rm \AA} & $6332.26$ {\rm \AA} & $6478.60\AA$ {\rm \AA} & $6300.31$ {\rm \AA} \\  
 \noalign{\smallskip}
\hline
\noalign{\smallskip}
730  &    $\leq+0.04$  &    ---       &  $+0.98$     &  $+0.37$     \\
243  &    $\leq-0.18$  &  $+0.52$     &  $+0.64$     &  $+0.28$     \\
030  &    $\leq+0.00$  &  $+0.82$     &  $+0.77$     &  $+0.16$     \\
785  &    $\leq+0.10$  &  $+0.34$     &   ---        &  $+0.38$     \\
145  &    $\leq+0.05$  &  $+0.62$     &   ---        &  $+0.42$     \\
401  &    $\leq-0.12$  &  $+0.90$     &  $+0.74$     &  $+0.45$     \\
\noalign{\vskip 0.2cm}
\noalign{\hrule}
\noalign{\smallskip} \hline
\end{tabular}  
\label{tab:cno}
\end{table}

\subsection{Odd-Z elements}

We derived the sodium abundances using three NaI lines, one of which is located in the blue arm at $5682.633 \AA$. The blue-arm spectrum has a S/N lower than the red-arm one. Due to the lower S/N values in all stars, these lines show a higher noise. For this reason, the abundance ratios were essentially derived from the lines located in the red arm, $6154.23$ {\rm \AA} and $6160.753$ {\rm \AA}.

\begin{figure}
    \centering
    \includegraphics[width=\columnwidth]{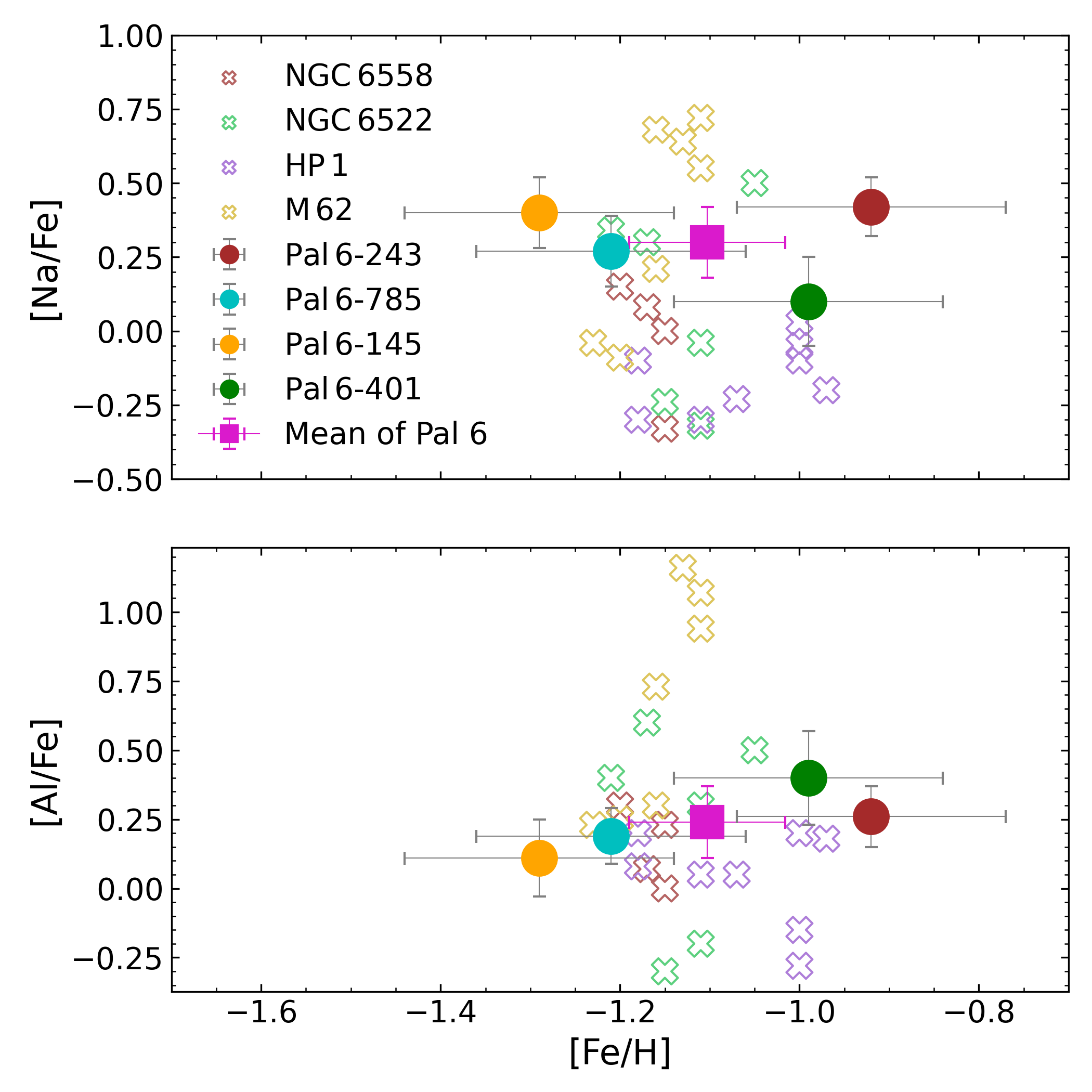}
    \caption{Odd-Z elements, Na (top panel) and Al (bottom panel), abundances as a function of metallicity [Fe/H]. Symbols: Crosses
    correspond to M~62 \citep[gold;][]{yong14}, NGC~6558 \citep[red;][]{barbuy18b}, NGC~6552 (green; Barbuy et al., 2021), and HP~1 \citep[purple;][]{barbuy16}. The pink square shows the mean value of Pal~6. The filled dots are the mean abundances, together with the error bars.}
    \label{fig:na_al}
\end{figure}

The aluminium abundances were derived from lines at $6696.185${\rm \AA} and $6698.673${\rm \AA}. In Figure \ref{fig:na_al}, we show the Na and Al abundances compared with literature abundances of four other bulge GCs with similar [Fe/H]: M62 \citep[gold;][]{yong14}, NGC~6558 \citep[red;][]{barbuy18b}, NGC~6522 (green; Barbuy et al. 2021), and HP~1 \citep[purple;][]{barbuy16}. For Pal~6, only the member stars are plotted. In general, the abundances are consistent with the other GCs within the uncertainties. The mean value of Pal~6 (pink square) is in good agreement with the other GCs except for M62.

\subsection{$\alpha$-elements}
The fast early enrichment of the proto-cluster gas by supernovae type II (SNII) can be seen through the abundances of $\alpha$-elements O, Mg, Ca, and Si, together with Eu produced through the rapid neutron capture process. We obtained a consistent enrichment for all $\alpha$-elements with a mean value of [$\alpha$/Fe]$=+0.35$ and a dispersion of $0.06$.  

\begin{figure*}
    \centering
    \includegraphics[width=0.8\textwidth]{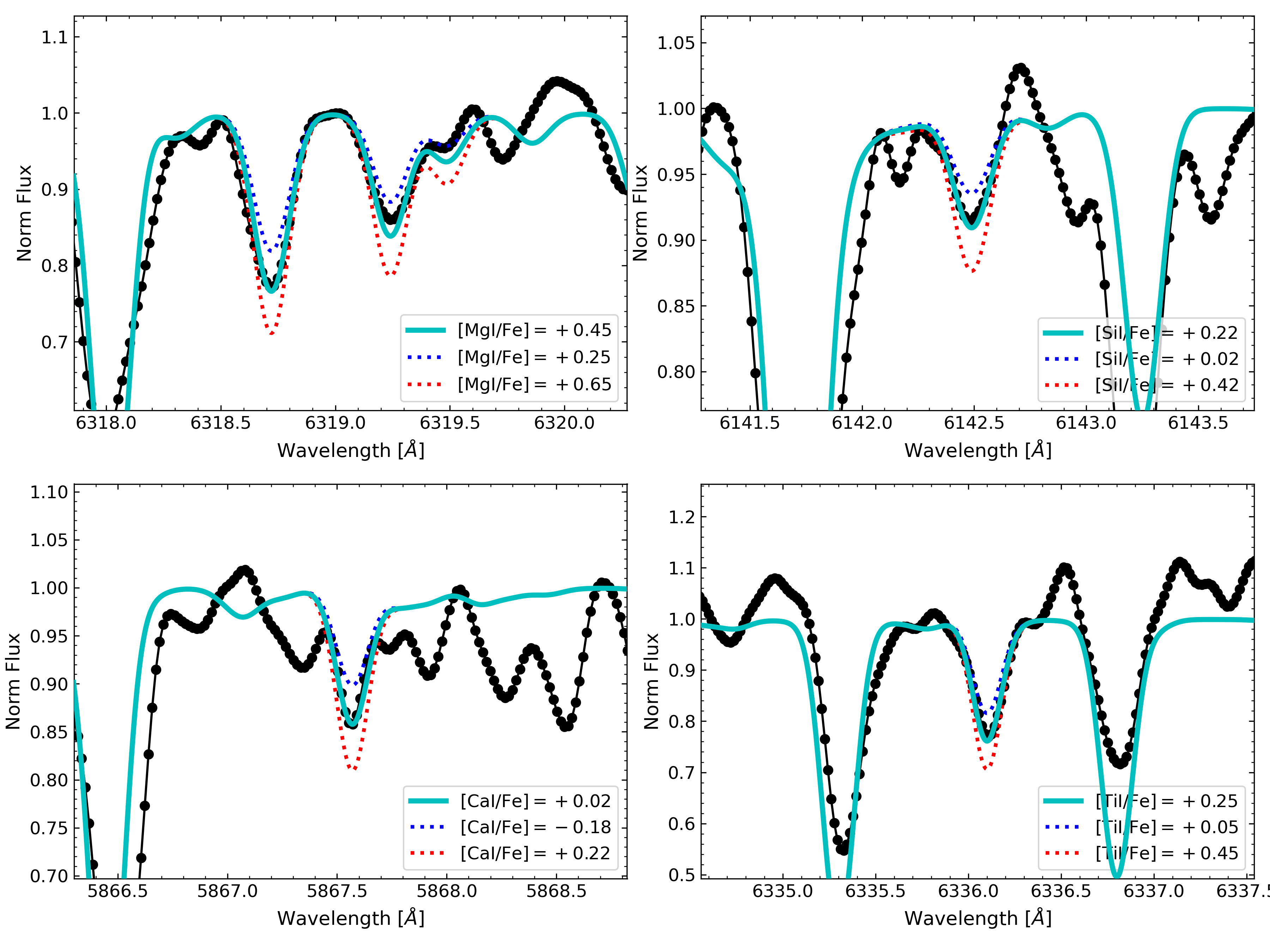}
    \caption{Same as Figure \ref{fig:CO_specs}, but for $\alpha-$elements Mg (top left) Si (top right), Ca (bottom left), and Ti (bottom right). }
    \label{fig:alpha-specs}
\end{figure*}

 Figure \ref{fig:alpha-specs} shows the line profile fitting of the \ion{Mg}{I} $6318.720$ \AA, \ion{Si}{I}  $6142.494$ \AA, \ion{Ca}{I}  $5867.562$ \AA, and \ion{Ti}{I}  $6336.113$ {\rm \AA} for the member star 243. The best fit is represented by the cyan line. We also show the lines considering a variation of $0.20$ dex plus (red) and minus (blue) with regard to the best abundance.

\begin{figure}
    \centering
    \includegraphics[width=\columnwidth]{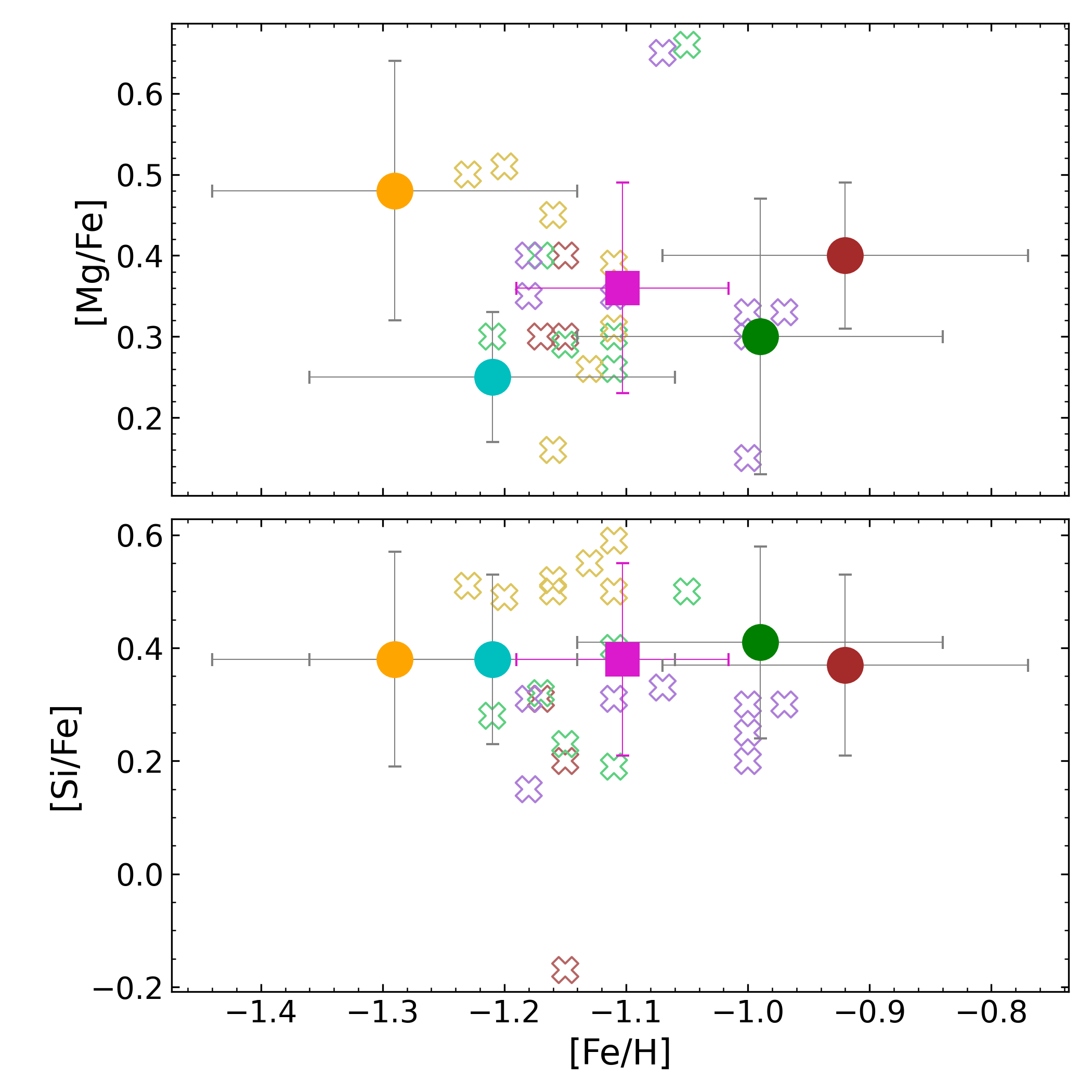}
    \caption{$\alpha-$elements Mg (top panel) and Si (bottom panel) as functions of metallicity [Fe/H]. The colour code is the same in Figure \ref{fig:na_al}.  }
    \label{fig:mg_si}
\end{figure}

\begin{figure}
    \centering
    \includegraphics[width=\columnwidth]{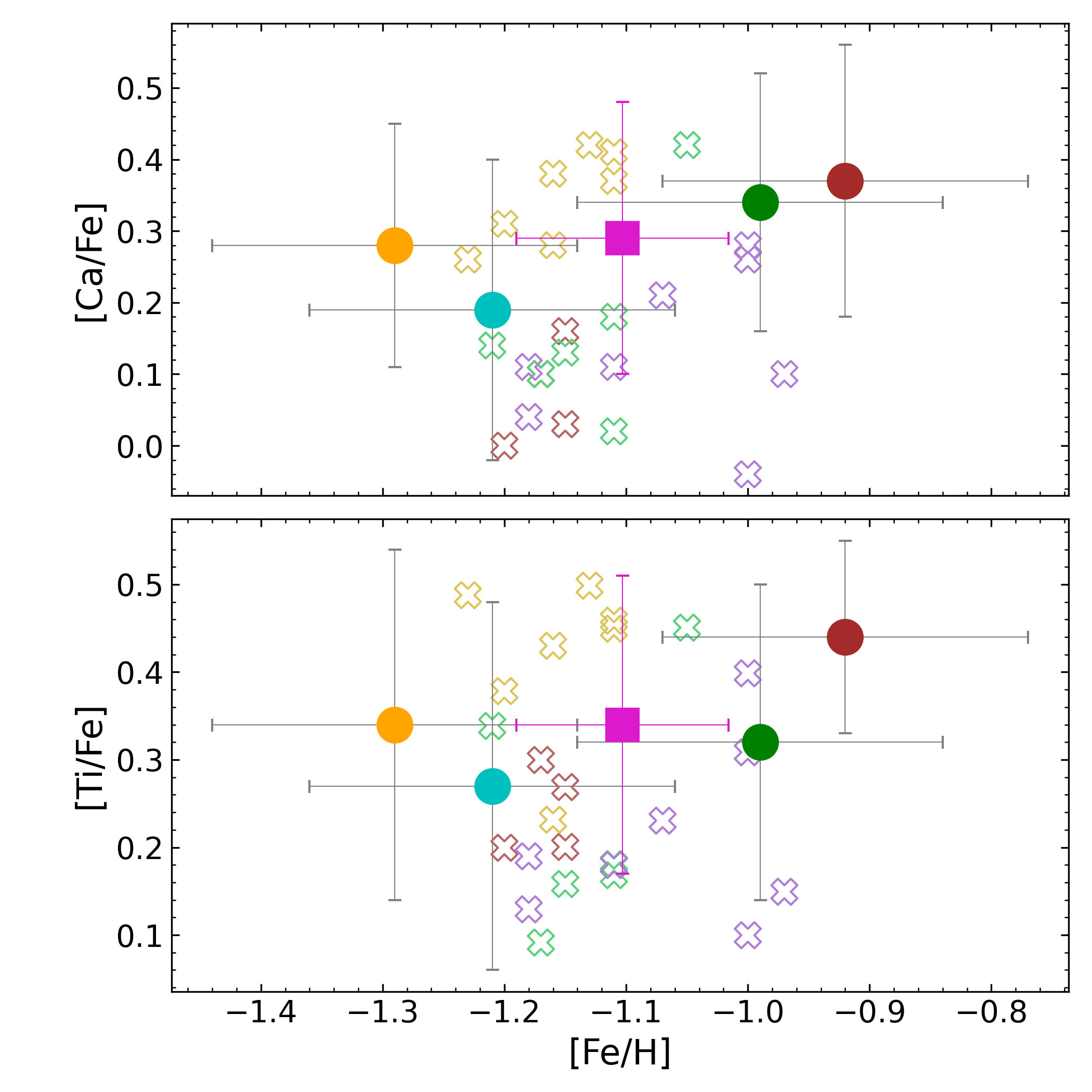}
    \caption{$\alpha-$elements Ca (top panel) and Ti (bottom panel) as functions of metallicity [Fe/H]. The colour code is the same in Figure \ref{fig:na_al}. }
    \label{fig:ca_ti}
\end{figure}

We compare the literature abundances of $\alpha$-elements for the same four GCs NGC~6522, NGC~6558, HP~1, and M~62 in Figure \ref{fig:mg_si} (Mg in the top panel and Si in the bottom panel) and Figure \ref{fig:ca_ti} (Ca in the top panel and Ti in the bottom panel). These GCs show $\alpha$ enrichment and abundances between $\sim 0.0$ and $\sim +0.60$, which means an average value of $\sim 0.35$. For all elements, the abundances are uniform as functions of [Fe/H].

\subsection{Heavy elements}
  
\begin{figure*}
    \centering
    \includegraphics[width=0.8\textwidth]{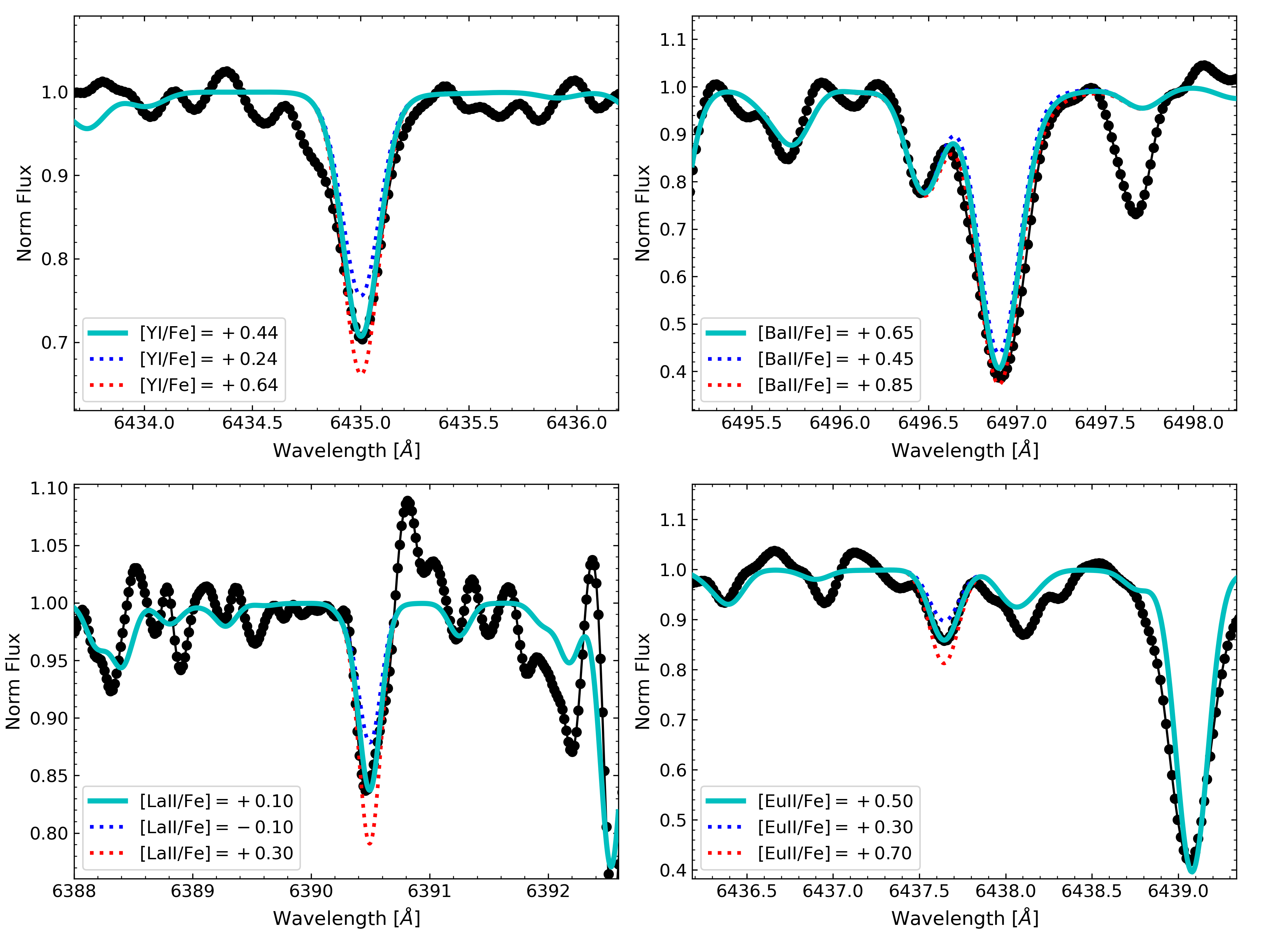}
    \caption{Same as Figure \ref{fig:CO_specs}, but for heavy elements Y (top left), Ba (top right), La (bottom left), and Eu (bottom right).}
    \label{fig:heavy-specs}
\end{figure*}

\begin{figure}
    \centering
    \includegraphics[width=\columnwidth]{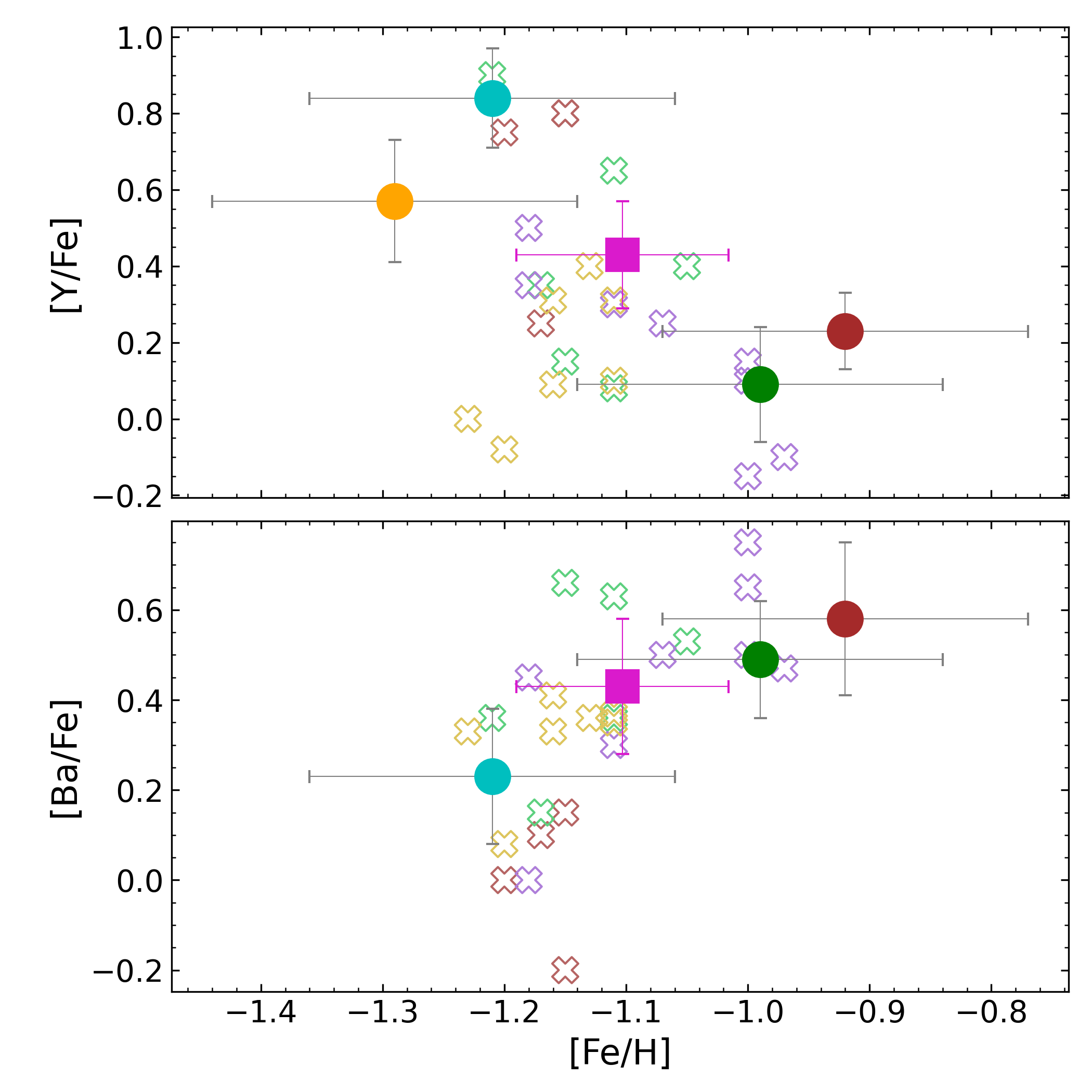}
    \caption{Heavy-elements Y (top panel) and Ba (bottom panel) as a functions of metallicity [Fe/H]. The colour code is the same as in Figure \ref{fig:na_al}.  }
    \label{fig:y_ba}
\end{figure}

 We derived the abundances of the heavy neutron-capture elements Y, Zr, Ba, La, and Eu. The Eu abundance is essentially the reference for the r-process.  We measured the \ion{Y}{I} $6435.004\AA$ and the \ion{Y}{II} $6613.73\AA$ lines.
 For the final [Y/Fe] values, we assumed that the ionised species of Y contributes $99\%$ to the abundance.
 Figure \ref{fig:heavy-specs} shows the line profile fitting of the \ion{Y}{I} $6435.004$ \AA, \ion{Ba}{II} $6496.897$ {\rm \AA}, \ion{La}{II} $6390.477$ \AA, and \ion{Eu}{II} $6437.640$ {\rm \AA} for the member star 243. The [Y/Fe] is systematically enhanced for Pal~6 and follows the same pattern observed for the bulge GCs with the same metallicity (top panel of Figure \ref{fig:y_ba}).

 The barium abundance was measured considering only the \ion{Ba}{II} $5853.675\AA$ and $6496.897\AA$ lines. In the bottom panel of Figure \ref{fig:y_ba}, we show the barium abundances as a function of [Fe/H] compared with the other three bulge GCs. It is possible to observe an opposite pattern compared with [Y/Fe]. The [Ba/Fe] has an enhanced abundance value.

 For zirconium, we fit four \ion{Zr}{I} lines: $6127.47\AA$, $6134.58\AA$, $6140.53\AA$, and $6143.25\AA$. We neglected the strong lines of \ion{Zr}{I} located in the blue arm.

\begin{figure}
    \centering
    \includegraphics[width=\columnwidth]{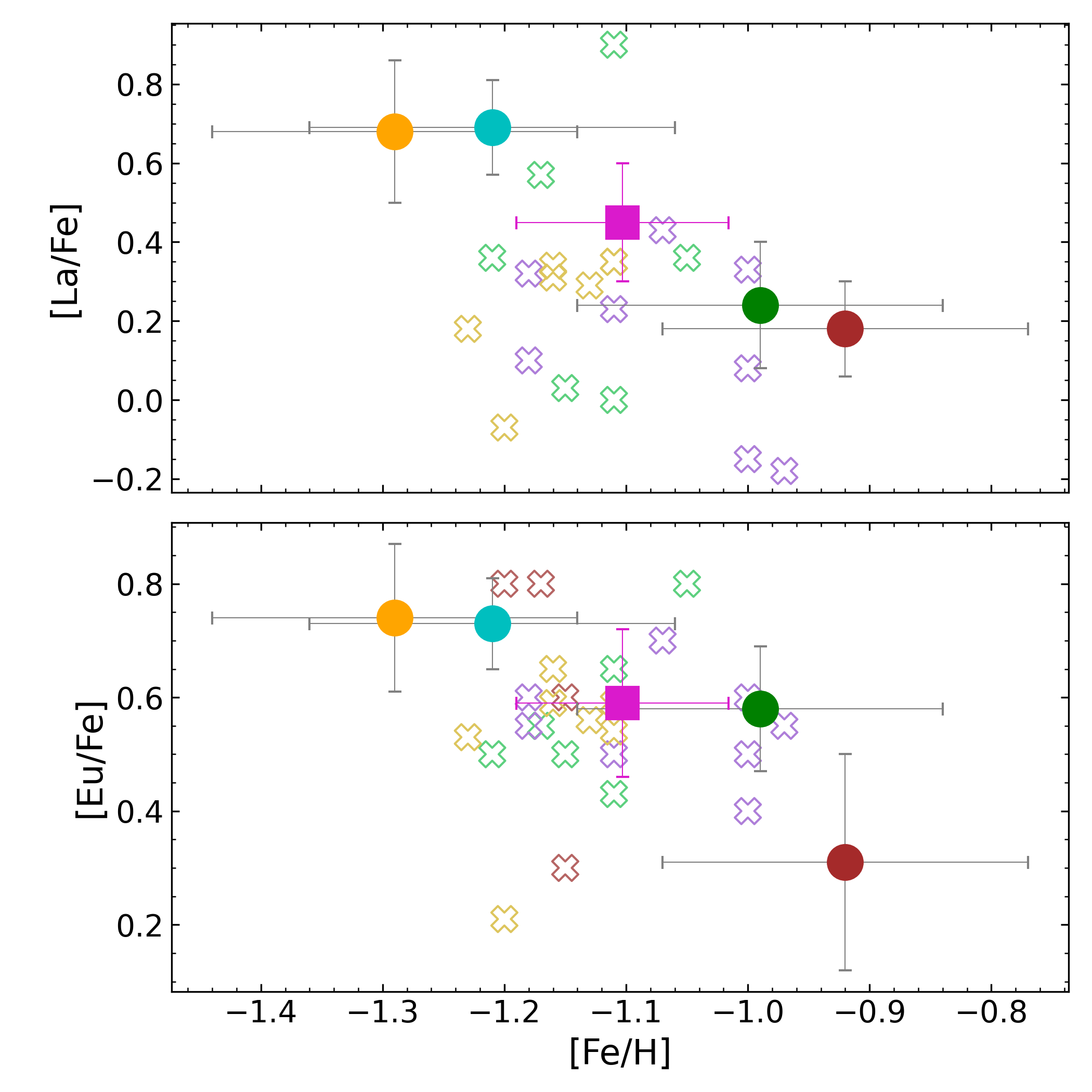}
    \caption{Heavy-elements La (top panel) and Eu (bottom panel) as functions of metallicity [Fe/H]. The colour code is the same as in Figure \ref{fig:na_al}.  }
    \label{fig:la_eu}
\end{figure}

 The lanthanum abundances are based on five \ion{La}{II} lines, which are located at $6172.72 \AA$, $6262.287 \AA$, $6296.079 \AA$, $6320.376 \AA$, and $6390.477 \AA$. In Figure \ref{fig:la_eu}, we show the comparison of La abundances with the bulge GCs (top panel). The abundances are in good agreement with the values of the reference GCs. Finally, for the europium abundances, we adopted the lines of \ion{Eu}{II} $6437.6\AA$ and $6645.1\AA$. The literature comparison of europium is shown in Figure \ref{fig:la_eu} (bottom panel).

\begin{table*}
\centering
\caption{Abundances in the six UVES sample stars. The mean abundance for the clusters is computed using only the four member stars.}            
\label{tab:mean_abunds}      
\small
\centering
\tabcolsep 0.15cm
\begin{tabular}{l|cccccc|c}
\noalign{\smallskip}
\hline
\noalign{\smallskip}
\hline
\noalign{\smallskip}
[$X$/Fe] &     star 730        &         star 243        &         star 030        &         star 785        &         star 145        &         star 401        &       $<$Pal~6$>$   \\
\noalign{\smallskip}
\hline
\noalign{\smallskip}
    C & $+0.04\pm0.15$  & $-0.18\pm0.15$  & $+0.00\pm0.15$  & $+0.10\pm0.15$  & $+0.05\pm0.15$  & $-0.12\pm0.15$  & $-0.04\pm0.15$ \\ 
    N & $+0.98\pm0.15$  & $+0.58\pm0.16$  & $+0.79\pm0.12$  & $+0.34\pm0.15$  & $+0.62\pm0.15$  & $+0.82\pm0.08$  & $+0.59\pm0.14$ \\ 
    O & $+0.37\pm0.15$  & $+0.28\pm0.15$  & $+0.16\pm0.15$  & $+0.38\pm0.15$  & $+0.42\pm0.15$  & $+0.45\pm0.15$  & $+0.38\pm0.15$ \\ 
\hline    
   Na & $+0.29\pm0.22$  & $+0.42\pm0.10$  & $+0.17\pm0.26$  & $+0.27\pm0.12$  & $+0.40\pm0.12$  & $+0.10\pm0.15$  & $+0.30\pm0.12$ \\ 
   Al & $+0.44\pm0.12$  & $+0.26\pm0.11$  & $+0.49\pm0.10$  & $+0.19\pm0.10$  & $+0.11\pm0.14$  & $+0.40\pm0.17$  & $+0.24\pm0.13$ \\  
\hline  
   Mg & $+0.43\pm0.12$  & $+0.40\pm0.09$  & $+0.53\pm0.14$  & $+0.25\pm0.08$  & $+0.48\pm0.16$  & $+0.30\pm0.17$  & $+0.36\pm0.13$ \\ 
   Si & $+0.33\pm0.19$  & $+0.37\pm0.16$  & $+0.32\pm0.15$  & $+0.38\pm0.15$  & $+0.38\pm0.19$  & $+0.41\pm0.17$  & $+0.38\pm0.17$ \\ 
   Ca & $+0.30\pm0.26$  & $+0.37\pm0.19$  & $+0.11\pm0.35$  & $+0.19\pm0.21$  & $+0.28\pm0.17$  & $+0.34\pm0.18$  & $+0.29\pm0.19$ \\ 
   Ti & $+0.32\pm0.19$  & $+0.44\pm0.11$  & $+0.29\pm0.20$  & $+0.27\pm0.21$  & $+0.34\pm0.20$  & $+0.32\pm0.18$  & $+0.34\pm0.17$ \\ 
\hline  
    Y & $+0.33\pm0.16$  & $+0.23\pm0.10$  & $+0.49\pm0.32$  & $+0.84\pm0.13$  & $+0.57\pm0.16$  & $+0.09\pm0.15$  & $+0.43\pm0.20$ \\ 
   Zr & $+0.76\pm0.17$  & $+0.68\pm0.19$  & $+0.65\pm0.13$  & $+0.61\pm0.12$  & $+0.74\pm0.24$  & $+0.41\pm0.35$  & $+0.61\pm0.22$ \\ 
   Ba &      ---        & $+0.58\pm0.17$  &      ---        & $+0.23\pm0.15$  &      ---        & $+0.49\pm0.13$  & $+0.43\pm0.18$ \\ 
   La & $+0.46\pm0.15$  & $+0.24\pm0.13$  & $+0.57\pm0.28$  & $+0.69\pm0.12$  & $+0.68\pm0.18$  & $+0.24\pm0.16$  & $+0.46\pm0.15$ \\ 
   Eu & $+0.43\pm0.12$  & $+0.31\pm0.19$  & $+0.49\pm0.12$  & $+0.73\pm0.08$  & $+0.74\pm0.13$  & $+0.58\pm0.11$  & $+0.59\pm0.13$ \\ 
\noalign{\vskip 0.2cm}
\noalign{\hrule}
\noalign{\smallskip} \hline
\end{tabular}
\end{table*}

\subsection{Errors}

 Uncertainties in spectroscopic parameters are given in Table \ref{errors} for star 243. For each stellar parameter, we adopted the usual uncertainties as for similar samples \citep{barbuy14,barbuy16,barbuy18b}: $\pm$100 K in effective temperature, $\pm$0.2 on gravity, and $\pm$0.2 km~s$^{-1}$ on the microturbulence velocity. The sensitivities are computed by employing models with these modified parameters and recomputing lines of different elements considering changes of $\Delta$T${\rm eff}=+100$ K, $\Delta$log  g$ =+0.2$, $\Delta$v$_{\rm  t}$ $=  0.2$  km~s$^{-1}$. The given error is the difference between the new abundance and the adopted one. Uncertainties due to non-LTE effects are negligible for these stellar parameters as discussed in \cite{ernandes18}. The same error analysis and estimations can be applied to other stars in our sample. The abundance derivations from strong lines are, in general, avoided, since they are too sensitive to stellar parameters and spectral resolution, as can be seen for the sensitivity of \ion{the Ba}{II} lines in Table \ref{errors}. The La lines are, on the other hand, faint, and they are at least not affected by the same problem. 
 Finally, it is important to note that the main uncertainties in stellar parameters are due to uncertainties in the effective temperature, as can be seen in Table \ref{tabteff}. Other significantly important sources of error are the EWs, given the limited S/N of the spectra, which can be estimated by the formula from \cite{cayrel04}: $\sigma_{EW}$ = 1.5 $\sqrt{FWHM.\delta_{x}}/(S/N)$, where $\delta_{x}$ is the pixel size. 

\begin{table}
\centering
\caption{Sensitivity in abundances due to variation in atmospheric parameters
 for the star 243, considering uncertainties of $\Delta$T$_{\rm eff}$ = 100 K and
$\Delta$log g = 0.2, $\Delta$v$_{\rm t}$ = 0.2 km s$^{-1}$. Last column is the total error. The errors are to be added to reach the reported abundances. }
 
\label{errors}
\begin{tabular}{lcccc@{}c@{}}
\noalign{\smallskip}
\hline
\noalign{\smallskip}
\hline
\noalign{\smallskip}
\hbox{Element} & \hbox{$\Delta$T} & \hbox{$\Delta$log $g$} & 
\phantom{-}\hbox{$\Delta$v$_{t}$} & \phantom{-}\hbox{($\sum$x$^{2}$)$^{1/2}$}&
\\
\hbox{} & \hbox{100 K} & \hbox{0.2 dex} & \hbox{0.2 kms$^{-1}$} & & \\
\hbox{(1)} & \hbox{(2)} & \hbox{(3)} & \hbox{(4)} & \hbox{(5)}  &\\
\noalign{\smallskip}
\hline
\noalign{\smallskip}
\noalign{\hrule\vskip 0.1cm}
\hbox{[FeI/H]}    & $-0.05$ & $+0.03$ & $+0.06$ & $+0.08$ \\
\hbox{[FeII/H]}   & $+0.13$ & $-0.15$ & $+0.02$ & $+0.20$ \\
\hbox{[C/Fe]}     & $+0.02$ & $+0.02$ & $+0.00$ & $+0.03$ \\
\hbox{[N/Fe]}     & $+0.15$ & $+0.10$ & $+0.00$ & $+0.18$ \\
\hbox{[O/Fe]}     & $+0.00$ & $+0.05$ & $+0.00$ & $+0.05$ \\
\hbox{[NaI/Fe]}   & $+0.13$ & $+0.05$ & $+0.04$ & $+0.15$ \\
\hbox{[AlI/Fe]}   & $+0.10$ & $+0.03$ & $-0.01$ & $+0.10$ \\
\hbox{[MgI/Fe]}   & $+0.07$ & $+0.03$ & $+0.00$ & $+0.08$ \\
\hbox{[SiI/Fe]}   & $+0.02$ & $+0.12$ & $+0.08$ & $+0.14$ \\
\hbox{[CaI/Fe]}   & $+0.18$ & $+0.10$ & $-0.05$ & $+0.21$ \\
\hbox{[TiI/Fe]}   & $+0.25$ & $+0.09$ & $-0.04$ & $+0.27$ \\
\hbox{[TiII/Fe]}  & $-0.04$ & $+0.10$ & $-0.03$ & $+0.11$ \\
\hbox{[YI/Fe]}    & $+0.13$ & $+0.13$ & $-0.12$ & $+0.22$ \\
\hbox{[YII/Fe]}   & $+0.07$ & $+0.08$ & $-0.02$ & $+0.11$ \\
\hbox{[ZrI/Fe]}   & $+0.22$ & $+0.06$ & $-0.12$ & $+0.26$ \\
\hbox{[BaII/Fe]}  & $+0.05$ & $+0.12$ & $-0.16$ & $+0.21$ \\
\hbox{[LaII/Fe]}  & $+0.08$ & $+0.16$ & $+0.07$ & $+0.19$ \\
\hbox{[EuII/Fe]}  & $-0.01$ & $+0.10$ & $+0.00$ & $+0.10$ \\
\noalign{\smallskip} 
\hline 
\end{tabular}
 \end{table}


\subsection{Comparison with previous results}

The metallicity derived in this work is in very good agreement with the values derived by \cite{lee02} ([Fe/H]$\,=-1.08\pm0.06$) and \cite{lee04} ([Fe/H]$\,=-1.0\pm0.1$) from high-resolution spectroscopy. It is also in good agreement with the \cite{carretta09b} metallicity scale, where Pal~6 has [Fe/H]$\,=-1.06\pm0.09$. The metallicity scale of \cite{dias16} gives a value of [Fe/H]$\,=-0.85\pm0.11$ for Pal~6. For comparison purposes, we selected the stars of \cite{dias16} and calculated their membership probabilities. The stars Pal~6-9 and Pal~6-13 in their sample seem to be members of Pal~6 with metallicities [Fe/H]$\,=-0.76\pm0.18$ and [Fe/H]$\,=-1.14\pm0.28$, respectively. Therefore, we can suppose that the star Pal~6-13 is the most probable member of Pal~6. This shows the power of Gaia, which was not available until very recently, and membership should be verified in all samples preceding the Gaia data.

Recently, \cite{kunder21} analysed Pal~6 in the context of the data release 16 (DR16) of the Apache Point Observatory Galactic Evolution Experiment
(APOGEE) survey for five observed stars. We inspected the membership probabilities of their sample. With our analysis, all stars are members of the cluster. Their mean radial velocity of $174.5\pm1.5$ is in agreement with our derivation. Their mean metallicity given by the three stars with good \texttt{ASPCAPFLAG} is [Fe/H]$\,=-0.92\pm0.10$, which is compatible within $1-\sigma$ with our result. 

We also have abundances for C, N, O, Na, Mg, Si, and Ca elements from APOGEE DR16. The CNO abundances are [C/Fe]$\,=-0.05\pm0.04$, [N/Fe]$\,=+0.31\pm0.27$, and [O/Fe]$\,=+0.22\pm0.05$. These values agreed with our results considering our derived errors; the carbon abundance, which is in excellent agreement with our determination. The abundances of $\alpha$-elements [Mg/Fe]$\,=+0.34\pm0.03$, [Si/Fe]$\,=+0.22\pm0.07$, and [Ca/Fe]$\,=+0.20\pm0.03$, individually are following the results of Table \ref{tab:mean_abunds}. Additionally, the abundances of $\alpha$-elements give a value of [$\alpha$/Fe]$\,=+0.25\pm0.06$, which agrees with our UVES analysis. 
This value is also in agreement with \cite{coelho05}: [$\alpha$/Fe]$\,=+0.28\pm0.05$. Finally, only the two stars with \texttt{ASPCAPFLAG}$\,\neq0$ have [Na/Fe] values with a mean of [Na/Fe]$\,=+0.35\pm0.10$. However, it is expected that Na should show variations due to
the probable presence of first and second generation stars, as discussed below.

\subsection{Heavy element analysis}
The presence of heavy elements in old stars can be explained
through the r-process contribution to these elements, as first
suggested by \cite{truran81}. Otherwise, if an s-process contribution can be identified,
the early enhancement of heavy elements can be explained by the ignition of the s-process for the first generation of stars with high rotation, the fast-rotating massive stars \citep{Chiappini11,cescutti13,cescutti15,frischknecht16,choplin18}. The rotation transports the $^{12}$C from the internal layers to external ones to burn into $^{14}$N and $^{13}$C. The activation of the s-process occurs when the $^{14}$N is converted into $^{22}$Ne. Therefore, this mechanism does not predict carbon enhancements.  

 An alternative explanation is an s-process contribution within a binary system in which the main companion has gone through the asymptotic giant branch (AGB) phase \citep[][and references therein]{beers05,sneden08}. Due to the mass transfer from AGB, the second companion receives s-process yields \citep[see discussion in][]{barbuy21}.
 
\begin{figure}
    \centering
    \includegraphics[width=\columnwidth]{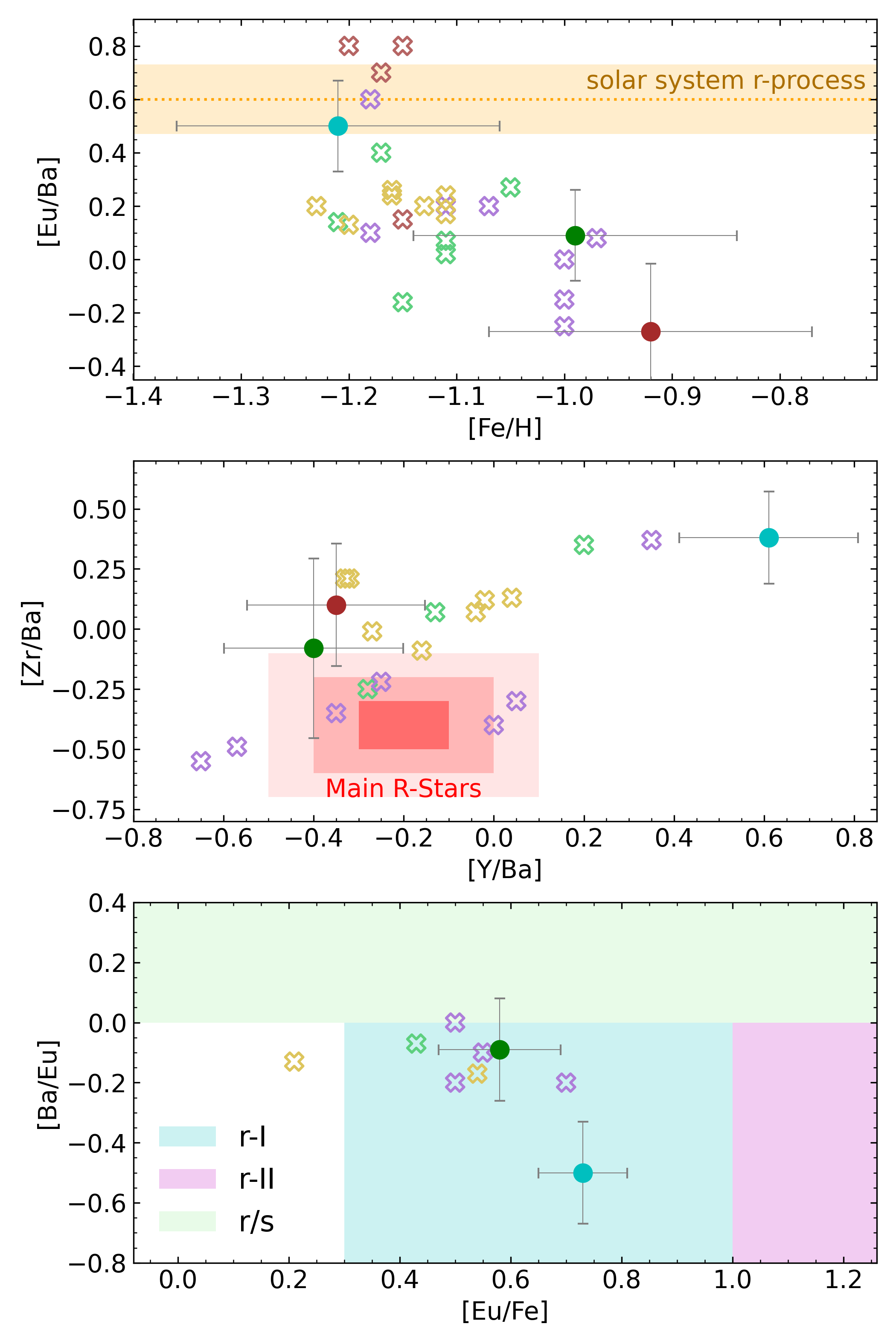}
    \caption{ Heavy-element enhancement diagnostic. \textit{Top panel:} [Eu/Ba] versus [Fe/H] diagram for the four reference GCs. The Solar System r-process region is highlighted by the orange strip for [Eu/Ba]$=+0.60\pm0.13$ (see text). \textit{Middle panel:} [Zr/Ba] versus [Y/Ba] diagram for the three reference GCs that have Zr, Y, and Ba abundance determinations. {The main r-process stars region is represented with the red-square ($3-\sigma$)}. \textit{Bottom panel:} [Ba/Eu] versus [Eu/Fe] diagram for the selected r-rich stars from the upper panel (see text). The light green region represents the regime of enhancement by both the r- and s-processes. The cyan and magenta regions show the domains of mainly r-process enhancement. The dot and cross colour code is the same as in Figure \ref{fig:na_al}.} 
    \label{fig:typeR}
\end{figure}

The top panel of Figure \ref{fig:typeR} highlights the region for the Solar System r-process abundance ratio of [Eu/Ba]$=+0.60\pm0.13$ \citep{simmerer04}, which would characterise r-II stars. Otherwise, r-I stars are defined to have 0.3 $\leq$ [Eu/Fe] $\leq$ +1.0 and [Ba/Eu] $<$ 0, and r/s stars to have 0.0 $<$ [Ba/Eu] $<$ +0.5 \citep{beers05}. These ratios are shown for the present sample of stars in  the bottom panel of Figure \ref{fig:typeR}.
 
 We also tentatively investigated the nature of heavy element enhancement through the diagnostic plots of Figure \ref{fig:typeR} using the [Zr/Ba] ratio.
 The use of [Zr/Ba]  as presented by \cite{siqueira-mello16} consisted of using [Y/Ba] and
 values from the six r-rich halo stars compiled in \cite{sneden08}, as representatives of the main r-process, which have a mean of [Y/Ba] $= -0.42\pm 0.12$. On the other hand, \cite{siqueira-mello16} gathered another six halo metal-poor stars showing enhancement of the first peak of heavy elements, which have [Y/Ba] $= +0.58\pm 0.18$ on the other extreme.
 The same is applied to Zr-to-Ba, with [Zr/Ba] $= -0.18 \pm 0.12$ and +0.95$\pm$0.15 in the two extremes.

In the middle panel of Figure \ref{fig:typeR}, we show the [Zr/Ba] versus [Y/Ba] diagram for Pal~6 and three other reference bulge GCs. For diagnostics, we highlighted the region of main r-process stars (red region) at [Y/Ba]$_r = -0.4\pm0.1$  \citep{sneden08} and [Zr/Ba]$_r = -0.2\pm0.1$ \citep{siqueira-mello16}. Only three of the six observed stars are plotted due the absence of Ba abundance. The member stars 785 and 401 are consistent with r-rich stars considering the errors. Besides that, the star 401 is located at the highest star density; consequently, it is compatible with the reference GCs.

The bottom panel of Figure \ref{fig:typeR} shows the further inspection of the r- and s-process to the r-rich stars selected by the [Eu/Ba] versus [Fe/H] and [Zr/Ba] versus [Y/Ba] diagrams. The two member stars (785 and 401) classified as r-rich are compatible with the definition of r-I, which is in agreement with that observed for the reference GCs.

\subsection{Are there two stellar populations?}

 According to \cite{martocchia18,martocchia19}, stellar clusters older than 2 Gyr show a presence of multiple stellar populations (MPs), evidenced by their chemical abundances. From a spectroscopic point of view, \cite{osborn71} observed anomalous variations in carbon molecules in one star of M5 and another of M10. Later, \cite{hartwick72} also discovered the anomaly in nitrogen. Currently, it is known that the phenomenon of MPs is also caused by star-by-star variations in light elements and helium mass fraction (Y) \citep{gratton04,carretta10,gratton12,milone18,meszaros20}. Specifically, the major variation is in N with a maximum enrichment of $\delta [{\rm N/Fe}] \sim 1.20$ dex \citep{milone18}. For that reason, many works have sought N-enhanced stars in field stars and GCs as evidence of second-generation stars \citep[e.g.][]{barbuy16,schiavon17b,dasilveira18,fernandez-trincado20,fernandez-trincado21}.
 
 It is important to determine whether the GC hosts MPs because this is related to the origin of the GC itself. For example, \cite{bellini17} analysed the complex Type II GC \citep{milone17} $\omega-$Cen (NGC~5139). They found that this cluster hosts at least five stellar populations, and that the populations can be split into 15 sub-populations. Their results show that $\omega-$Cen is much more complex than the majority of GCs. Additionally, \cite{massari19} associated $\omega-$Cen with the Gaia-Enceladus \citep{belokurov18,helmi18} progenitor. This could be pointing to a cluster that originated from a merger event. However, $\omega-$Cen seems to be more compatible with a core of a dwarf galaxy \citep{meszaros21}. For the `normal' \citep[Type I;][]{milone17} GCs, we could expect them to have an origin from main components of the Galaxy, as observed in \cite{massari19} with 62 GCs associated with a so-called main-progenitor.

\begin{figure}
    \centering
    \includegraphics[width=\columnwidth]{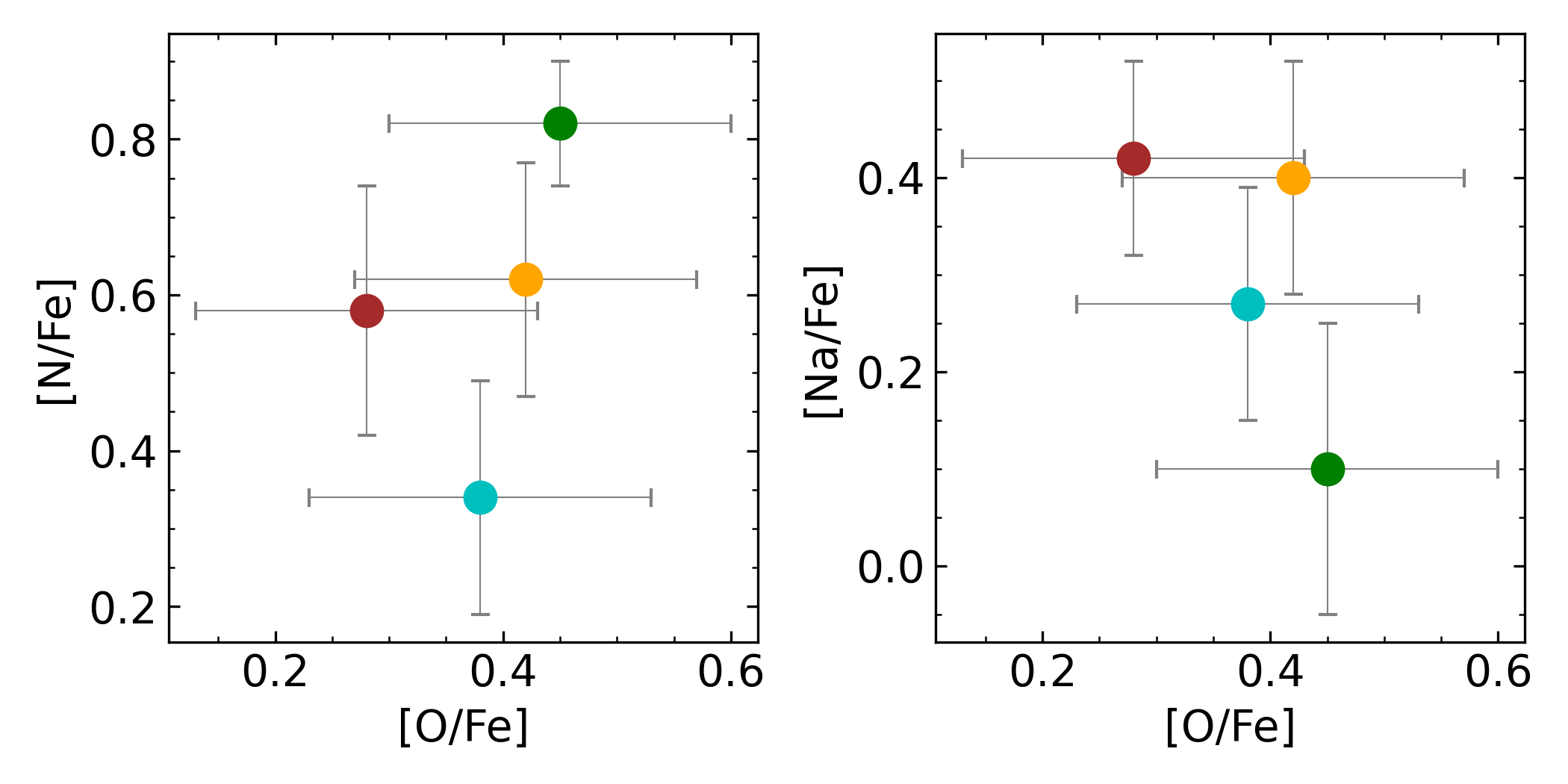}
    \caption{Anti-correlations N-O (left) and Na-O (right) for Pal~6 member stars.}
    \label{fig:cn_nao_corr}
\end{figure}

 The expected N-O anti-correlation \citep{carretta10,gratton04,gratton12} is given in left panel of Figure \ref{fig:cn_nao_corr}. We also found two N-rich non-member stars, which are possible field members, with [N/Fe]$>+0.70$. These could be stars that were Pal~6 or other cluster members trapped by the Galactic bulge \citep{schiavon17a}. Another indicator of MPs is the Na-O anti-correlation. \cite{carretta09} demonstrated that this anti-correlation is more likely to be seen in massive clusters. Since Pal~6 is a relatively low-mass cluster \citep[with an absolute magnitude of $M_V = -6.79$;][2010 edition]{harris96}, in Figure \ref{fig:cn_nao_corr} (right panel) we can observe a slight Na-O anti-correlation.

\begin{figure}
    \centering
    \includegraphics[width=\columnwidth]{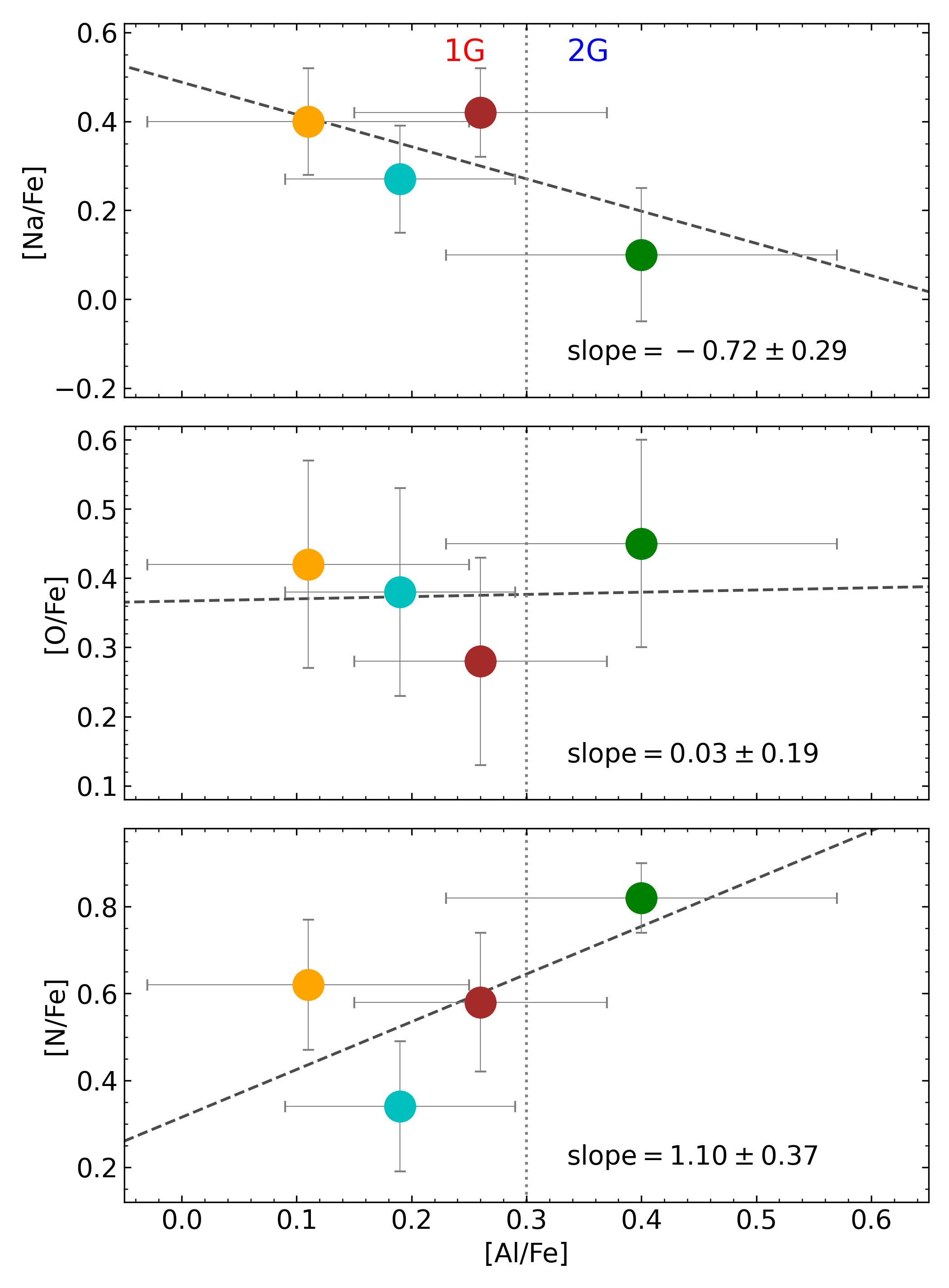}
    \caption{Al-NaON (anti)correlations. The dots are coloured according to the same colour code of Figure \ref{fig:cn_nao_corr}. The dotted grey line represents the generation split around ${\rm [Al/Fe]}=+0.30$ \citep{meszaros20}. The black dashed lines show the obtained linear regression.}
    \label{fig:mps_evi}
\end{figure}

 To verify if our N-enhanced star 401 is a probable second-generation member, we investigated the Al-NaON relations (Figure \ref{fig:mps_evi}). \cite{meszaros20} analysed stars observed with the APOGEE for $31$ GCs. They observed that at [Al/Fe]$=+0.30$, the stars are split reasonably well into two populations. We investigated these patterns and observed that our N-enhanced star has [Al/Fe]$>+0.30$, while the other three member stars have [Al/Fe]$<+0.30$. 
 Even though the phenomenon of MPs \citep{bastian18} is a characteristic of the majority of GCs \citep{piotto15}, \cite{lagioia19} presented the first evidence of a GC consistent with hosting a simple stellar population (Terzan 7). For that reason, the abundance pattern observed for Pal~6 is important in order to check if it hosts at least two stellar populations. 

\section{Age and distance} \label{sec:Age_Distance}

 Previous photometric studies did not attempt to derive the age of Pal~6,
 and there are controversies on its distance in the literature. These are largely due to the absence of observed standard candle stars in Pal~6, and  different values result from different methods. \citet{ortolani95} derived a distance of $\sim 8.9$ kpc from the horizontal branch (HB) magnitude method with an extinction of $A_V = 4.12$. \citet{lee02}, comparing the Pal~6 HB magnitude to the 47 Tuc one, obtained a distance of $\sim 7.2$ kpc with $A_V=4.1$ mag. 
\cite{harris96} gave a distance of $5.80$ kpc, which was adopted by 
 Baumgardt et al. (2019) and used in Massari et al. (2019) and P\'erez-Villegas et al. (2020).
 
With the final corrected CMD (Section \ref{sec:cmd}), we used the \texttt{SIRIUS} code \citep{souza20} to perform the statistical isochrone fitting to obtain the accurate probability distributions for the fundamental parameters of Pal~6. We employed isochrones from the MESA Isochrones \& Stellar Tracks database \citep[MIST; ][]{dotter16,choi16} with the metallicity [Fe/H] ranging from 0.0 to -2.0 dex in steps of 0.01 dex and ages from 10 Gyr to 15 Gyr with an interval of 0.1 Gyr; the reddening and distance modulus can vary freely. To obtain a consistent analysis, we used a Gaussian prior for the metallicity with information from the high-resolution spectroscopic determination by this work. 

We also obtained the temperature-dependent second-order extinction corrections $\Delta C_{\lambda}$ ($\Delta A_{\lambda}/A_V^{\rm reff}$) by comparing the MIST isochrones with A$_V = 0.00$ and $6.0$ for each value of T$_{\rm eff}$. The correction is given by the second-order polynomial function $\Delta C_\lambda = a_0\times (\log T_{\rm eff})^2 + a_1\times \log T_{\rm eff} + a_2$, and the A$_V^{\rm reff}=6.0$. As mentioned in \cite{oliveira20}, the second-order correction is obtained by interpolation considering the desired A$_V$. The coefficients $a_{0,1,2}$ are listed in Table \ref{tab:teffcoefs}.

\begin{table}
\centering
\caption{Coefficients for effective temperature second-order correction to different passbands. The coefficient orders are given by the following equation: $\Delta C_\lambda = a_0\times (\log T_{\rm eff})^2 + a_1\times \log T_{\rm eff} + a_2$. }     
\label{tab:teffcoefs}      
\centering
\tabcolsep 0.15cm
\begin{tabular}{lccc}
\noalign{\smallskip}
\hline
\noalign{\smallskip}
\hline
\noalign{\smallskip}
$\Delta C_{\lambda}$ & $a_0$  & $a_1$ & $a_2$ \\
\noalign{\smallskip}
\hline
\noalign{\smallskip}
 F606W    & $-0.325$ & $+2.555$ & $-5.041$ \\
 F110W    & $+0.056$ & $-0.365$ & $+0.571$ \\
 F160W    & $+0.012$ & $-0.078$ & $+0.127$ \\
 $V$*     & $-0.328$ & $+2.515$ & $-4.840$ \\ 
 $I$      & $-0.056$ & $+0.442$ & $-0.878$ \\ 
 $G$      & $-0.506$ & $+4.129$ & $-8.495$ \\
 $G_{BP}$ & $-0.191$ & $+1.723$ & $-3.847$ \\
 $G_{RP}$ & $-0.302$ & $+2.342$ & $-4.584$ \\
\noalign{\vskip 0.2cm}
\noalign{\hrule}
\noalign{\smallskip} \hline
\end{tabular}
\end{table} 

\begin{figure}
    \centering
    \includegraphics[scale=0.5]{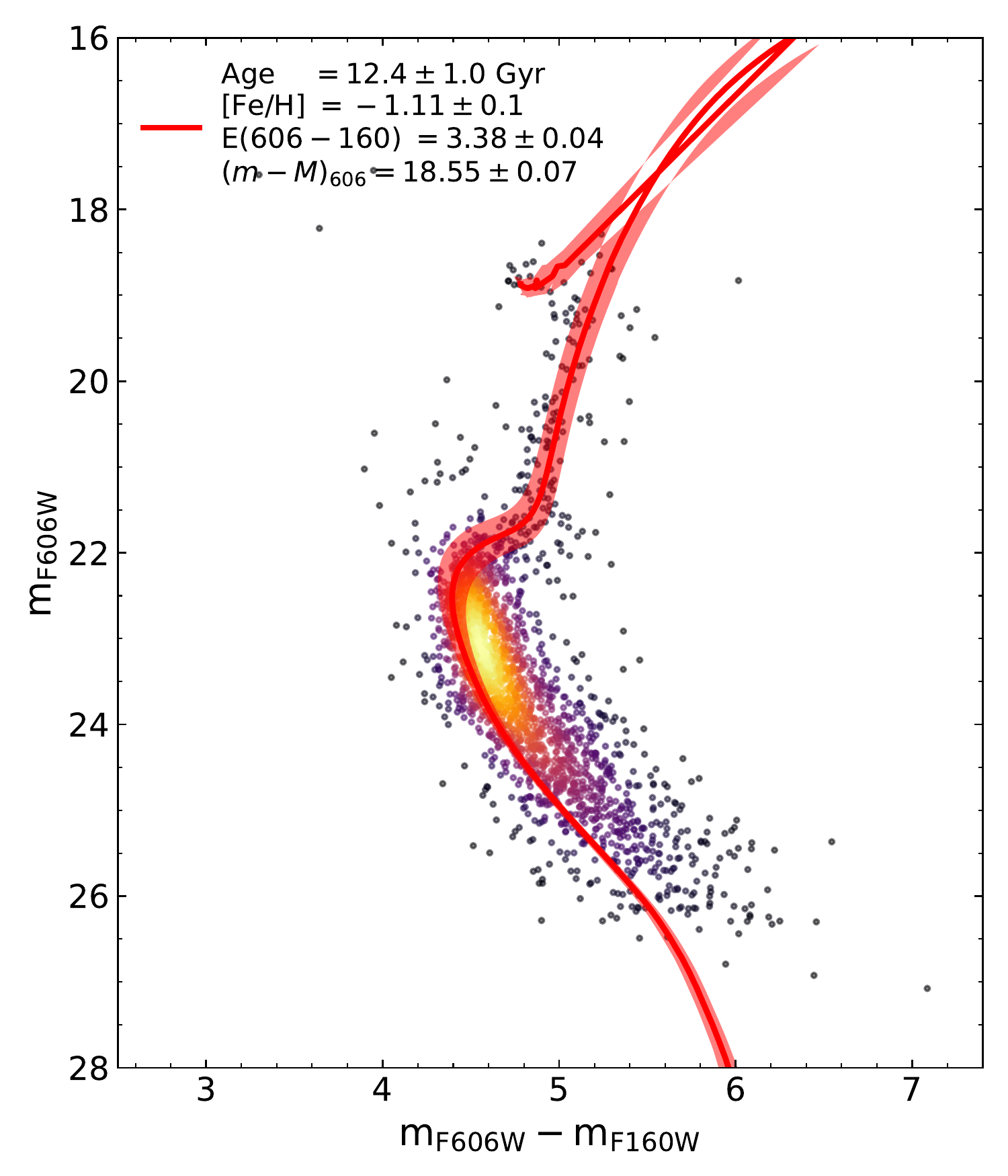}
    \caption{Best-fit from isochrone fitting (solid line) and results for $\pm1\sigma$ (red region). }
    \label{fig:cmdbestft}
\end{figure}

We adopted the $50^{\rm th}$ percentile as the best solution and $50^{\rm th}-16^{\rm th}$ and $84^{\rm th}-50^{\rm th}$ percentiles for the uncertainties. The red line in Figure \ref{fig:cmdbestft} represents the best fit, while the red strip shows the region of $1-\sigma$ solutions. We want to stress that the HB model fits well to the HB region in the CMD. Also, this technique allows us to obtain a better distance determination with low uncertainty. The best distance, reddening, and well-constrained metallicity values provided us with the first derivation of age for Pal~6 as $12.4\pm0.9$ Gyr, therefore it is among the oldest GCs in the Galaxy.

The reddening E($606-160$)$=3.38\pm0.04$ and distance modulus (m$-$M)$_{606} = 18.55\pm0.07$ obtained from isochrone fitting can be converted in E(B$-$V) and (m$-$M)$_0$ using the following relations:
\begin{eqnarray}
     E(606-160) = R_V\times  E(B-V)\times ( C_{606,R_V} - C_{160,R_V} ), \\
     (m-M)_{606} = (m-M)_0 + R_{V}\times C_{606,R_V} \times E(B-V)
,\end{eqnarray}
where $C_{\lambda,R_V}$ is the ratio of temperature- and gravity-dependent coefficients at the $\lambda$ for an extinction law with $R_V$ \citep[][and references therein]{pallanca21}. Since the extinction is given by $A_\lambda = R_{V}\times C_{\lambda,R_V} \times E(B-V)$, the reddening is inversely proportional to $R_V$. Therefore, the assumption of $R_V$ will affect the fundamental parameters of the cluster. 

For the isochrone fitting,  it is common to use the extinction law setting $R_V = 3.1$, which is the case for all the fundamental parameters calculated for Pal~6 in the literature. With this extinction law, our determination is $A_V=4.56\pm0.06$ ($E(B-V) = 1.47\pm0.02$), which is compatible with the reddening used to derive the photometric temperatures. However, \cite{nataf16} argue that a lower value of $R_V$ is more compatible with the Galactic bulge population, where it could reach down to $R_V = 2.5$ at least for (absolute) Galactic latitudes between 2 and 7 degrees and 
$-10^{\circ}<l<10^{\circ}$. \cite{vasiliev21} compared the distance from literature to the Gaia EDR3 parallaxes. They found a discrepancy between the photometric distances and the inverse of parallaxes for the bulge GCs, precisely those with high reddening values (E(B$-$V)>1.0). Also, \cite{pallanca21} show that the $R_V = 3.1$ needs different values of reddening and distance moduli to fit the CMD well with different colours in the case of the bulge GC Liller 1. They demonstrated that to fit the three CMDs simultaneously with a unique set of reddening and distance values, it is necessary to adopt an extinction law with $R_V=2.5$. They also conclude that the variation in the extinction law results in variations in the reddening and distance modulus determinations (consequently in the distance).

\begin{figure}
    \includegraphics[width=0.95\columnwidth]{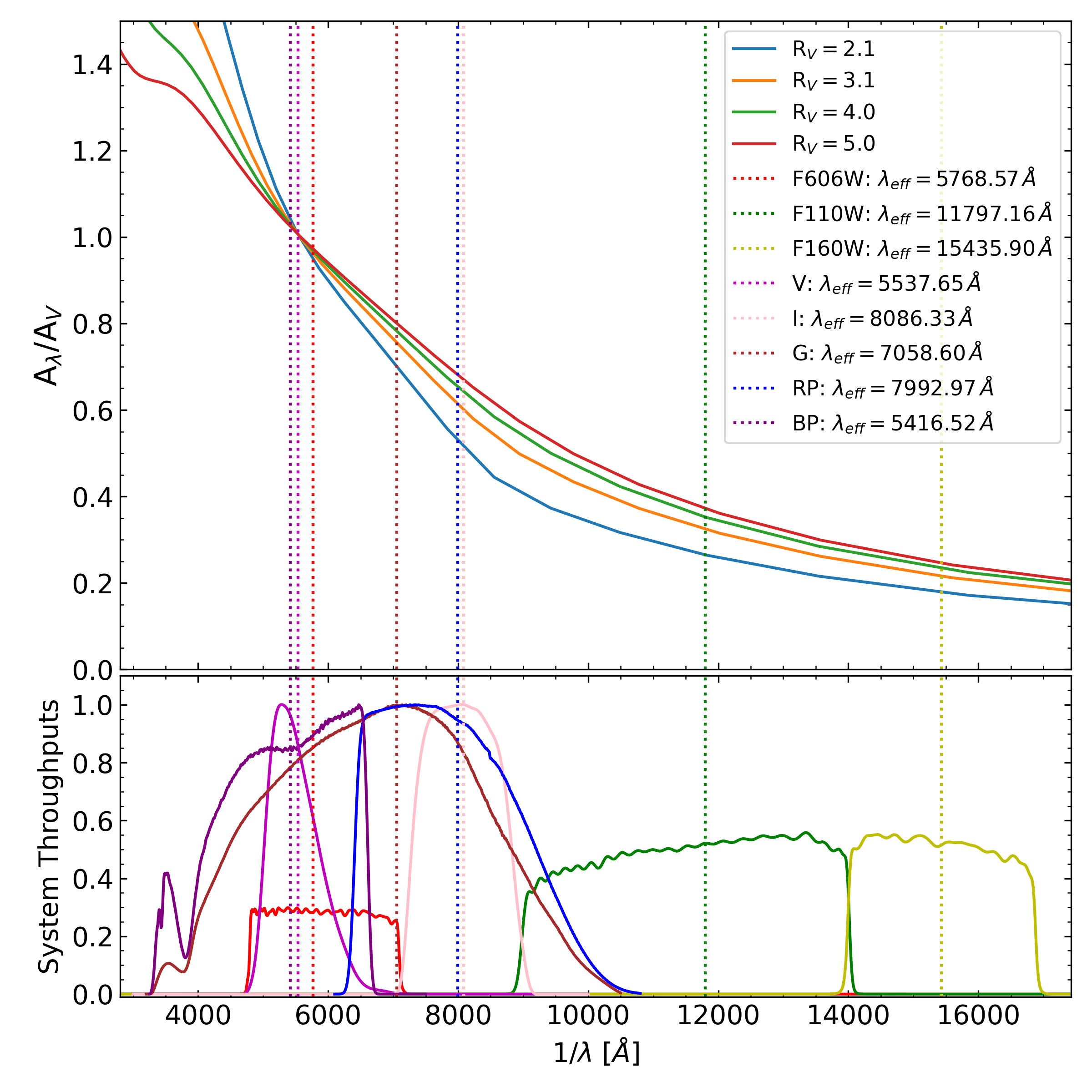}
    \caption{Extinction law curves' derivation. The effective wavelengths are computed from $\lambda_{\rm eff} = \int \lambda^2 T_\lambda d\lambda / \int \lambda T_\lambda d\lambda$. }
    \label{fig:ECs}
\end{figure}

\begin{figure*}
    \centering
    \includegraphics[width=\textwidth]{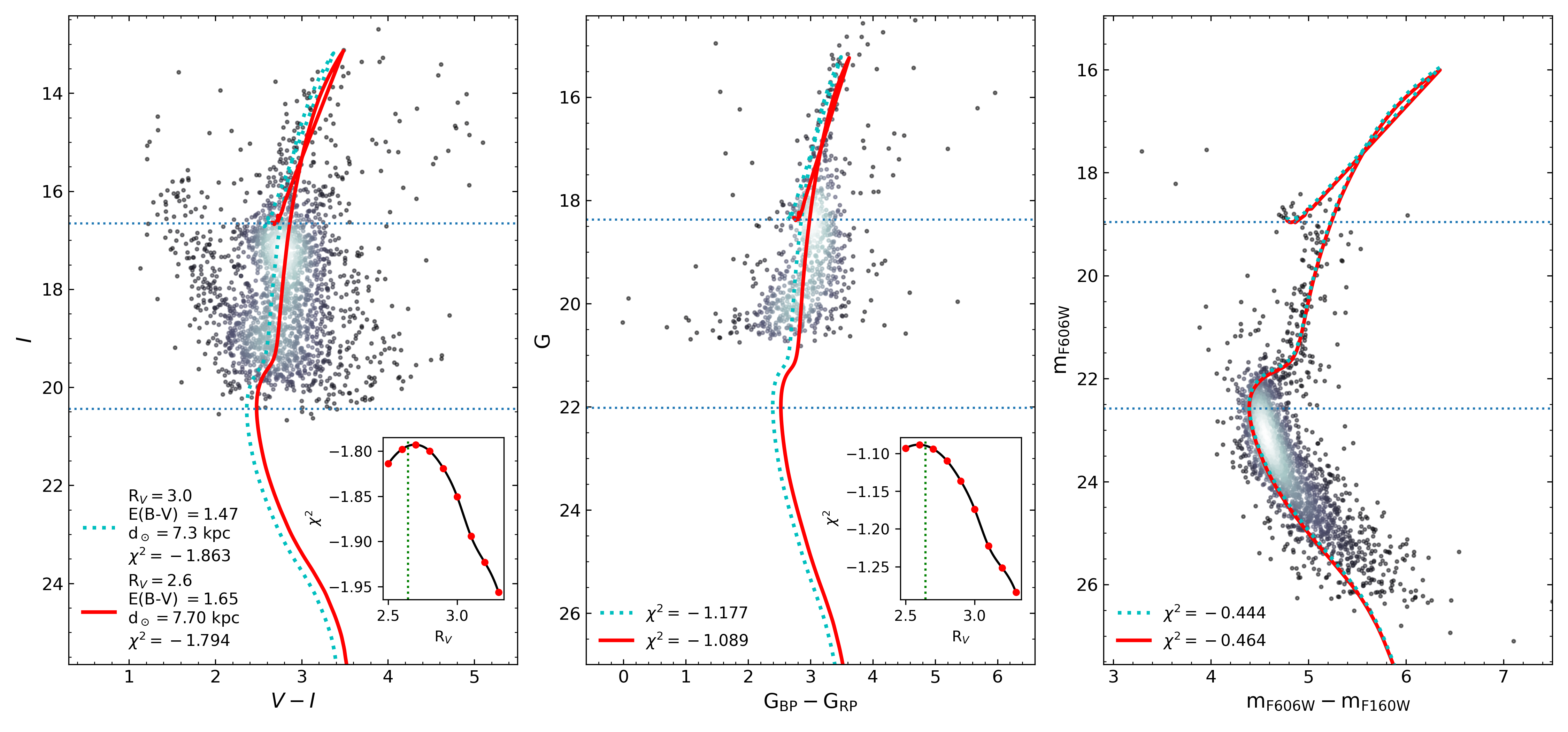}
    \caption{Posterior fitting to obtain the best value of $R_V$. First panel: Optical CMD $V-I$ versus $I$ from \cite{ortolani95}. Second panel: Gaia EDR3 $G_{BP}-G_{RP}$ versus $G$ CMD. Third panel: Corrected NIR HST CMD. The cyan dotted lines are the isochrones considering the best fit from the standard isochrone fitting and standard extinction coefficient ($R_V=3.0$). The embedded plots show the $\chi^2$ function to the variation of $R_V$. The solid red lines are the isochrones with the best $R_V$ value. Finally, the blue horizontal lines denote the isochrone HB and turn-off mean locus. }
    \label{fig:cmdchangerv}
\end{figure*}

\begin{figure}
    \centering
    \includegraphics[width=0.9\columnwidth]{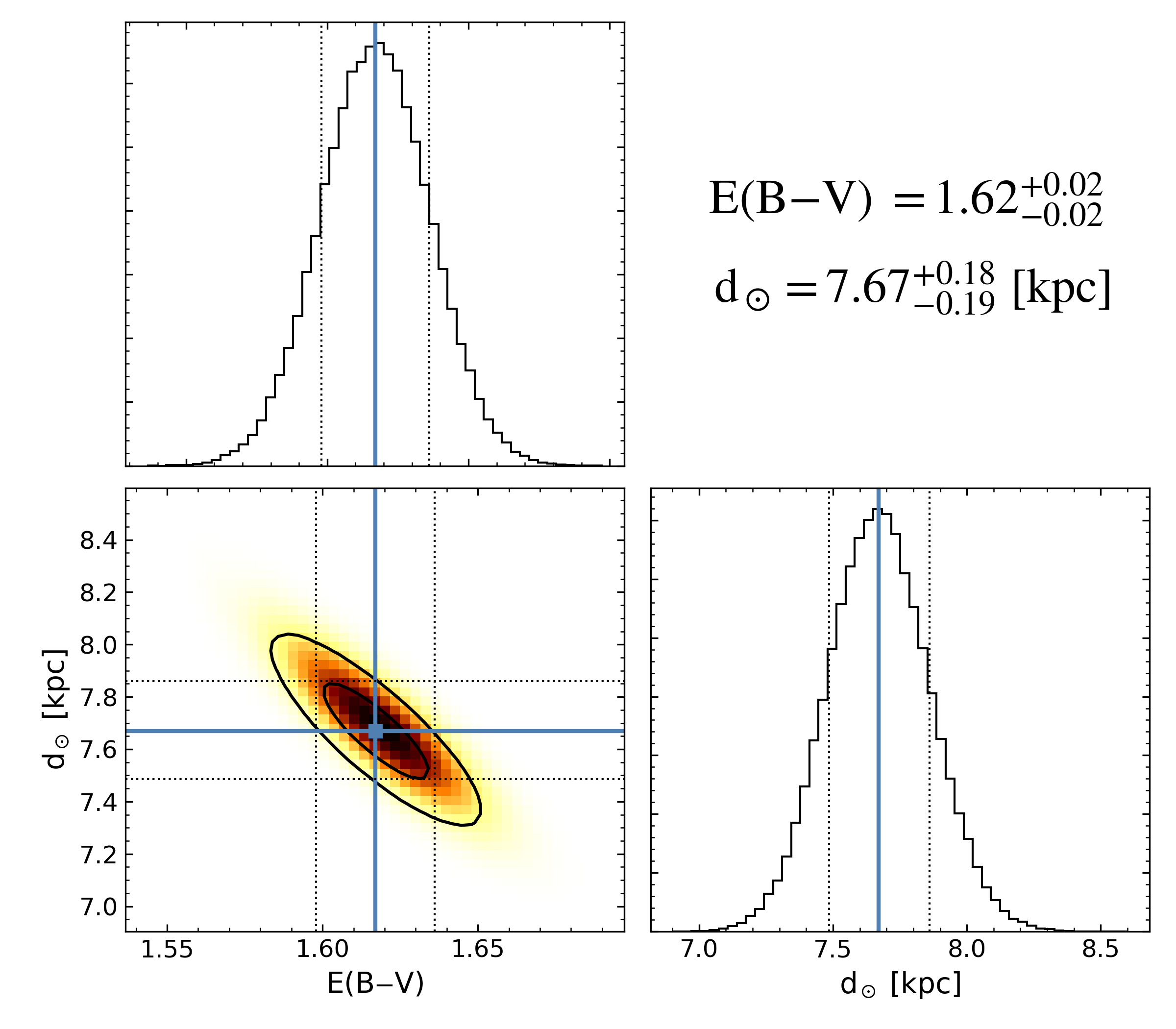}
    \caption{New distance and reddening determinations using the best extinction law. }
    \label{fig:newdistance}
\end{figure}

To determine the value of $R_V$ for Pal~6, we compare the optical \citep[$VI$;][]{ortolani95}, Gaia $G_{BP}-G_{RP}$ versus $G$, and NIR HST CMDs using the best-fit parameters of Figure \ref{fig:corner}. Since we varied the $R_V$, we re-derived the extinction coefficients in the adopted bands (A$_{F606W}$/A$_V$, A$_{F110W}$/A$_V$, A$_{F160W}$/A$_V$, A$_{G_{BP}}$/A$_V$, A$_{G_{RP}}$/A$_V$, A$_{G}$/A$_V$, and A$_{I}$/A$_V$) using the extinction laws from \cite{cardelli89}. The corresponding extinction law to a given $R_V$ value has been done by interpolating the curves in a grid with the values $R_V = {2.1,3.1,4,5}$ (Figure \ref{fig:ECs}). We derived R$_V=2.6$ by maximising the negative $\chi^2$ for the optical and Gaia CMDs (first and second panels of Figure \ref{fig:cmdchangerv}). Finally, we determined an extinction of $A_V=4.21\pm0.05$ ($E(B-V)=1.62\pm0.02$) and a distance of $d_\odot = 7.67\pm0.19$ kpc (Figure \ref{fig:newdistance}), a result within the range between $5.8$ kpc \citep[][2010]{harris96} and  $8.9$ kpc \citep{ortolani95} and very close to  the \cite{lee02}  value of $7.2$ kpc. We stress that the latter distance determination derived from near-IR JHK photometry is independent of the $R_V$ optical value.

To confirm our determination, we performed the distance calculation using two other methods. From the relation Mv$-$[Fe/H] derived by \cite{oliveira21s} for RR Lyrae stars, we can obtain the HB absolute magnitude of $0.758\pm0.086$ in the V band. Assuming the apparent V-magnitude value of $19.70\pm0.15$ calculated for the HB of Pal~6 by \cite{ortolani95}, we obtain the distance modulus of (m$-$M)$_V = 18.94\pm0.18$. Finally, with the extinction value found in the present work ($A_V = 4.21\pm0.05$, compatible with the average calculated with dust map of the Galaxy using the DUST web tool\footnote{\url{https://irsa.ipac.caltech.edu/applications/DUST/}}), we have a distance of $d_\odot ^{\rm \left(HB\right)} = 8.73\pm0.75$ kpc.

Using the Gaia EDR3 membership analysis (Section \ref{sec:memb}), we identified five stars with distances derived by the StarHorse calculations \citep[with Gaia EDR3 and APOGEE DR16;][Queiroz et al. in prep]{queiroz20a}. Due to the low statistics, we expanded the sample with a bootstrapping method taking into account the uncertainties. The mean distance derived from the expanded sample is $d_\odot ^{\rm \left(SH\right)} = 7.3\pm0.8$ kpc.

The individual distance determinations through the HB and StarHorse are already compatible within $1.5\sigma$ with our determination from the isochrone fitting of $7.67\pm0.19$ kpc. In addition, the average of these determinations results in a distance of $<d_\odot>_{\rm \left( HB, SH \right)} = 8.0\pm1.1$ kpc, compatible with the distance of the present work. Finally, we added the average of the literature of $7.05\pm0.46$ kpc given by \cite{baumgardt21}\footnote{They considered the distances from \cite{ortolani95}, \cite{barbuy98}, \cite{lee02}, and \cite{lee04}. Because of the expected distance for Pal~6, they did not consider the inverse of the parallax given by \cite{vasiliev21}.}, resulting in $<d_\odot>_{\rm \left( HB, SH, B \& V21 \right)} = 7.5\pm1.0$kpc, which is in good agreement with our determination. Therefore, these results reinforce the one we found through the isochrone fitting of $d_{\odot}^{\rm \left(Pal~6\right)} = 7.67\pm0.19$ kpc calculated with the derived extinction law ($R_V=2.6$). 

\section{The origin of Pal~6}\label{sec:discussion}
    With the chemical information, age, and distance obtained in the previous sections, it is possible to infer a plausible origin of Pal~6.
Using the distance of Harris  \citep[][]{harris96}, which was adopted by \cite{Baumgardt19}, \cite{perezvillegas2020} classified Pal~6 as belonging to the Galaxy thick disc with a probability of $98$\%. Caution was recommended given other distance estimations in the literature
\citep[e.g.][]{ortolani95}.
    
    Given the much more reliable distance now derived in the present paper, we carried out the calculations of orbits for the cluster. We employed the same Galactic model of \cite{perezvillegas2018,perezvillegas2020} that includes a triaxial Ferrers bar of 3.5 kpc (major axis). The total mass of the bar is $1.2 \times 10^{10}$ M$_{\odot,}$ with an angle of $25^{\circ}$ with the Sun-major axis. We also assume three pattern speeds of the bar: $\Omega_b= 40$, 45, and 50 km s$^{-1}$ kpc$^{-1}$. 

    We generated a set of 1000 initial conditions employing a Monte Carlo approach. In order to do that, we considered the observational uncertainties of distance, heliocentric radial velocity, and absolute proper motion components, with the purpose of evaluating the errors in those observational parameters.  We integrated the orbits forward for 10 Gyr using the NIGO tool \citep{rossi15}. In Table \ref{tab:orbits}, we give the new orbital parameters as the median values of the perigalactic distance $<r_{min}>$, apogalactic distance $<r_{max}>$, mean eccentricity $<e>$ (where the eccentricity is defined as $e = (r_{\rm max} - r_{\rm min} )/(r_{\rm max} + r_{\rm min})$ ), and maximum vertical excursion from the Galactic plane $<|z|_{max}>$. The error of each orbital parameter is given as the standard deviation of the distribution.

\begin{figure}
    \centering
    \includegraphics[width=0.92\columnwidth]{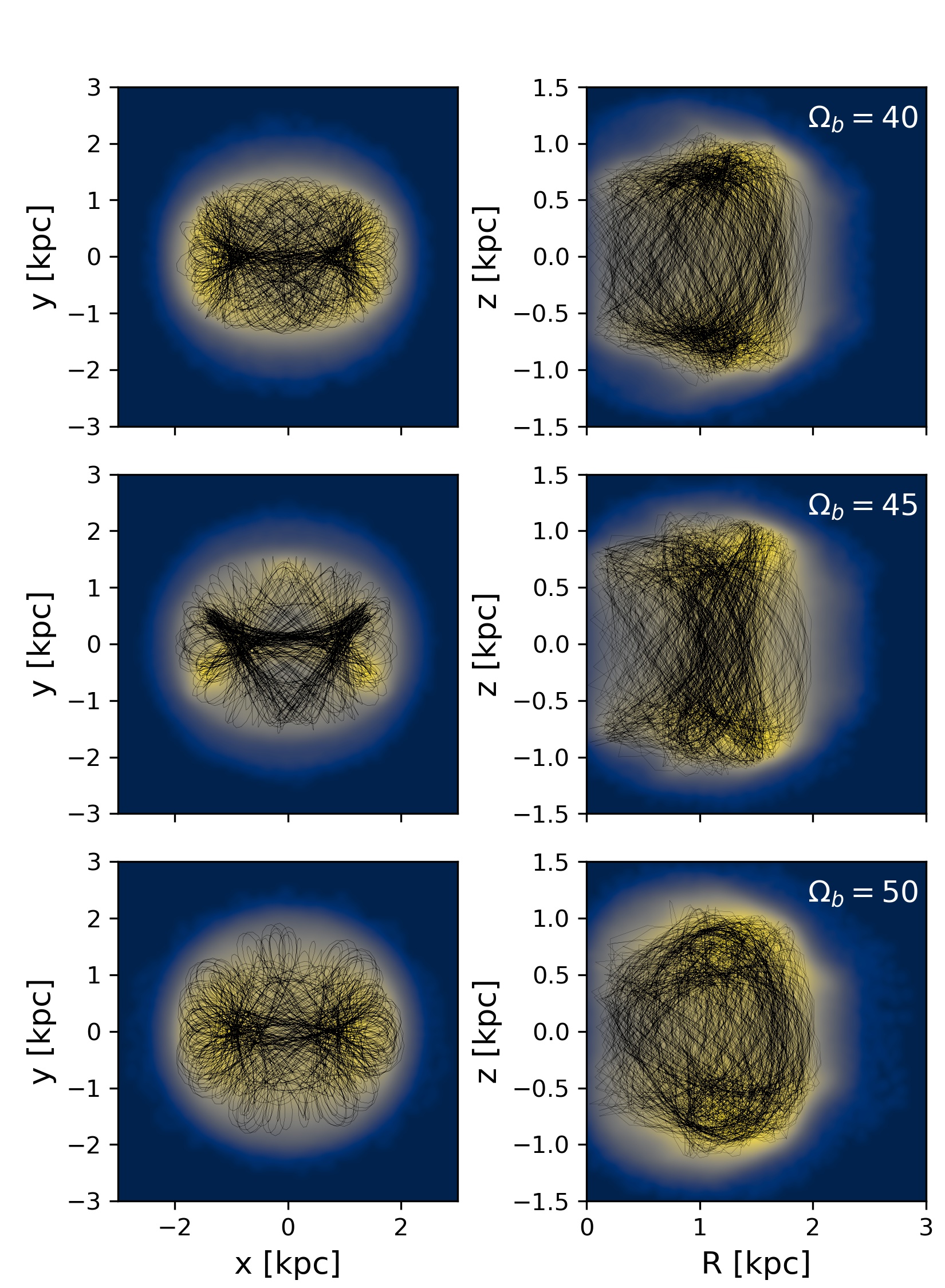}
    \caption{Probability density map for the $x-y$ and $R-z$ projections of the set of orbits for Pal~6 using three different values of $\Omega_b= 40, 45,$ and $50$ km s$^{-1}$ kpc$^{-1}$. The orbits are co-rotating with the bar frame. The gold colour corresponds to the higher probabilities, while the black lines show the orbits using the central observational parameters.}
    \label{fig:orbits}
    \end{figure}

    In Figure \ref{fig:orbits}, we show the probability density map of the orbits of Pal~6 in the $x-y$ and $R-z$ projections co-rotating with the bar. The gold colour displays the space region that the orbits of Pal~6 cross more frequently, while the black curves are the orbits considering the central values of the observational parameters. We can observe that Pal~6 is mostly confined within $\sim 2.1$ kpc, and therefore has a high probability of belonging to the bulge component ($> 99 \%$), when we adopt the distance of $7.67$ kpc estimated in this work. Our new distance determination points out that Pal~6 is also a very inner cluster due to its maximum height of $|z|<1.3$ kpc and a high eccentric orbit. Those characteristics were also found for the GCs NGC~6522, NGC~6558, and HP~1, which are very old and moderately metal-poor GCs of the Galactic bulge.

    Based on our analysis, we are confident that Pal~6 is confined within the Galactic bulge. However, it remains unclear if this cluster is originated from the Galaxy or due to some merger process that occurred in the early stages of the Milky Way. To elucidate whether Pal~6 was formed in situ or accreted, we followed the method described in \cite{massari19} to determine the probable progenitor of Pal~6. The classification is based on the integrals of motion space $L_{\rm z}$ and E, and the age-metallicity relation (AMR). It is important to mention that these integrals of motion are only conserved by Galactic potential in an axisymmetric model. Because of that, we employed the axisymmetric potential of \cite{mcmillan17} and recalculated the orbital parameters of Pal~6 using the python-package \texttt{galpy} \citep{bovy15}. For this case, we also integrated forward for 10 Gyr and employed a set of 1000 initial conditions. The reason for adopting the \cite{mcmillan17} Galactic potential is to compare our results with \cite{massari19} and to relate Pal~6 with its plausible progenitor. The orbital parameters with the axisymmetric potential are listed in last column of Table \ref{tab:orbits}.

\begin{figure}
    \centering
    \includegraphics[width=0.8\columnwidth]{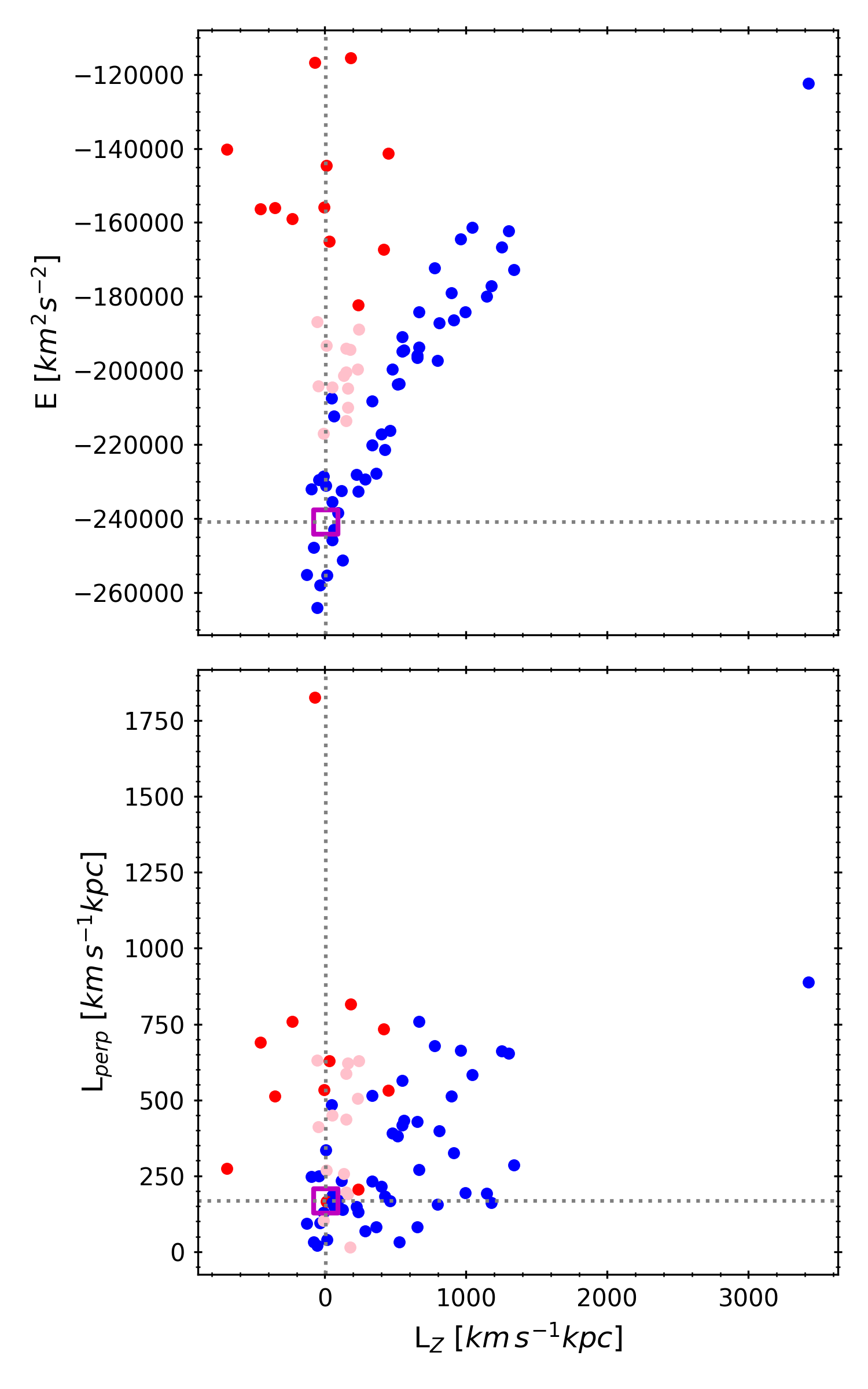}
    \caption{Integrals of motion space of 82 Galactic GCs. The colour-code relates to the association with their probable progenitor \citep{massari19}; Gaia-Enceladus is marked in red, low-energy in pink, and the main progenitor in blue. The magenta square represents Pal~6 with the values of the present work.}
    \label{fig:iom}
\end{figure}

    We found L$_{\rm z} = 7.88\pm13.22$ km\,s$^{-1}$\,kpc and E$=\left(-2.40\pm0.05\right)\times10^5$ km$^2$\,s$^{-2}$. Pal~6 is compatible with three progenitors due to its low values of E, L$_{z}$, and z-perpendicular angular momentum L$_{perp}$ (see Figure \ref{fig:iom}): the main progenitor (in situ), a low-energy progenitor (low-energy), and Gaia-Enceladus. \cite{massari19} classified Pal~6 as having been formed by the low-energy progenitor due to the previous values of distance employed. Since the classification of a main-bulge progenitor is related to the apogalactic distance $r_{\rm max}< 3.6$ kpc (maximum 3D radius of the orbit), Pal~6 is clearly compatible with this definition.  

\begin{figure}
    \centering
    \includegraphics[width=0.8\columnwidth]{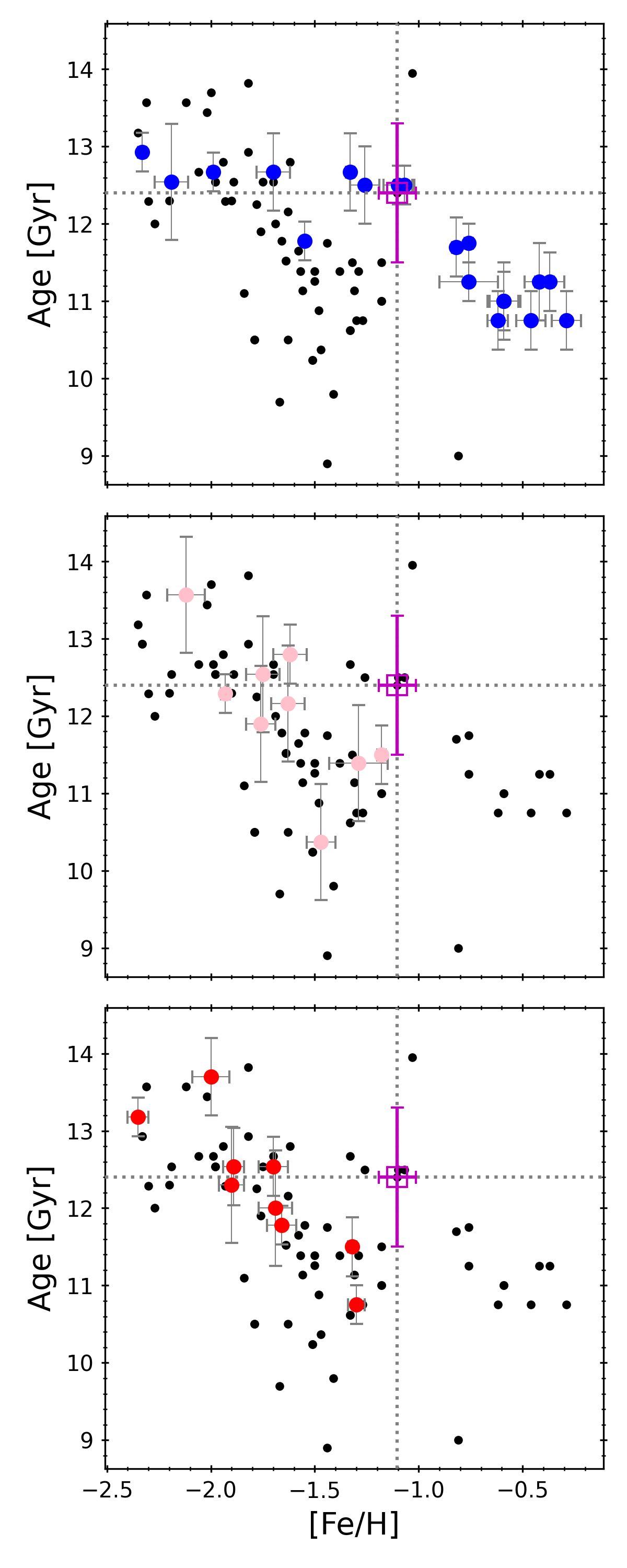}
    \caption{Age-metallicity relation (AMR) for 82 GCs (black dots) analysed by \cite{massari19} (plot based on their Figure 4). The ages are taken from \cite{vandenberg13} and the metallicities from \cite{carretta09b}. In each panel, one progenitor is highlighted. From top to bottom: First panel, main-progenitor (disc and bulge); second panel, low energy; third panel, Gaia-Enceladus.  The magenta square represents Pal~6 with metallicity and age of the present work.}
    \label{fig:amr}
\end{figure}

    On the other hand, with our new determination of [Fe/H] from high-resolution spectroscopy and the age derivation of Pal~6, we can observe the location of the cluster in the AMRs. Figure \ref{fig:amr} shows the AMRs for the clusters according to its associated progenitors for the ones more compatible with Pal~6. We found that Pal~6 is located in a possible ridge line of the main-progenitor distribution. Therefore, according to the classification of the present work, in combination with the classifications presented in \cite{massari19}, we conclude that Pal~6 is a cluster of the Galactic bulge having been formed in situ.

    This result could also give us an explanation about the possible formation scenario of the MPs in Pal~6. Since the cluster was formed in situ, the most compatible formation scenarios of MPs are those predicted by internal pollution of the cluster. This hypothesis is in agreement with the results described in the chemical abundances of heavy elements found in this work. A more detailed analysis of the stellar ages is required to provide a constraint on the most probable formation scenario \citep{nardiello15,souza20,oliveira20,lucertini21}.

\begin{table}
\caption{Orbital parameters of Pal~6 for the \cite{mcmillan17} potential and the potential employed in \cite{perezvillegas2020} assuming three different bar pattern speed values. The energy units are [$E$]$ = $km$^2$ s$^{-2}$ and angular momentum [$L$]$=$km s$^{-1}$ kpc.}     
\label{tab:orbits}      
\small
\scalefont{0.9}
\tabcolsep 0.15cm
\begin{tabular}{lcccc}
\noalign{\smallskip}
\hline
\noalign{\smallskip}
\hline
\noalign{\smallskip}
   Parameter                    & PV\,$\Omega_b=40$ & PV\,$\Omega_b=45$ & PV\,$\Omega_b= 50$ &     McMillan17   \\ 
\noalign{\smallskip}
\hline
\noalign{\smallskip}
<r$_{min}$>   [kpc]               &  $0.08\pm0.03$  &  $0.09\pm0.04$  &  $0.10\pm0.04$  & $ 0.07\pm0.04$   \\ 
<r$_{max}$>   [kpc]               &  $2.06\pm0.07$  &  $2.06\pm0.07$  &  $2.10\pm0.09$  & $ 2.14\pm0.19$   \\ 
<|z|$_{max}$> [kpc]               &  $1.07\pm0.10$  &  $1.14\pm0.05$  &  $1.16\pm0.07$  & $ 1.29\pm0.04$   \\ 
<e>                               &  $0.92\pm0.03$  &  $0.91\pm0.03$  &  $0.90\pm0.03$  & $ 0.96\pm0.05$   \\ 
<E>         [$10^5E$]             &       ---       &       ---       &       ---       & $-2.40\pm0.05$   \\ 
<L$_z$>       [$L$]               &       ---       &       ---       &       ---       & $ 7.88\pm13.2$   \\ 
<L$_{perp}^{\dagger}$>  [$10^2L$]           &       ---       &       ---       &       ---       & $ 1.67\pm0.22$   \\ 
<L$_{z_{c}}$>  [$10^2L$]       &       ---       &       ---       &       ---       & $ 3.07\pm0.19$   \\ 
$\mathcal{P}_{\rm Bulge}$ [\%]  &       $99.4$    &       $99.4$    &       $99.3$    &      $99.1$      \\ 
$\mathcal{P}_{\rm Disc} $ [\%]  &       $0.6$     &       $0.6$     &       $0.7$     &      $ 0.9$      \\ 

\noalign{\vskip 0.2cm}
\noalign{\hrule}
\noalign{\smallskip} \hline
\end{tabular}
\\
$\dagger$ The L$_{perp}$ is not conserved for axisymmetric potentials; however, it is a good parameter to describe the origin of a group of stars \citep{helmi00,massari19}.
\end{table}

\section{Conclusions}\label{sec:conclusions}

We present a complete and detailed analysis of the GC Pal~6, through the analysis of high-resolution spectra of the UVES spectrograph, HST photometry, and a dynamical analysis. Based on Gaia EDR3, we determined that four of our six sample stars are members of Pal~6 and give a heliocentric radial velocity consistent with values from the literature.

\begin{figure*}
    \centering
    \includegraphics[width=0.95\textwidth]{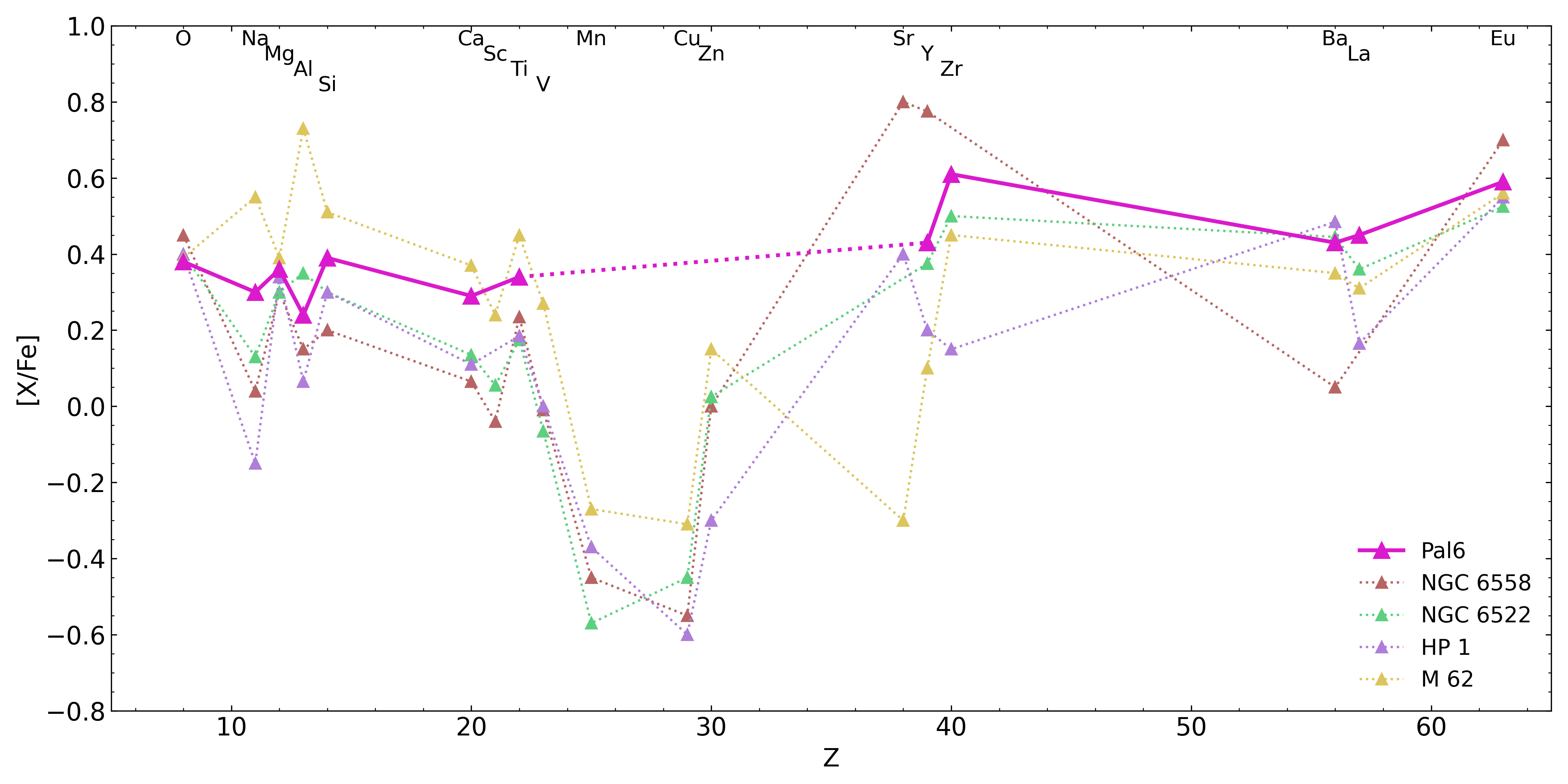}
    \caption{Abundance pattern [X/Fe] vs. atomic number (Z) of the references moderately metal-poor bulge GCs NGC 6558, NGC 6522, M62, and HP~1. The colours are the same as Figure \ref{fig:na_al}. }
    \label{fig:abundances}
\end{figure*}

With the UVES spectroscopic data, we determined the final stellar parameters and abundances for the six sample stars. The metallicity of [Fe/H]$=-1.10\pm0.09$ and $\alpha$-element enhancement of [$\alpha$/Fe]$=+0.35\pm0.06$
were derived. One of the member stars is N-enhanced, indicating a presence of second-generation stars, confirmed from a separation into two populations based on a [Al/Fe]$=+0.30$ threshold. We can also observe that the abundance pattern of Pal~6 is very similar in many aspects to the {GCs typical of the bulge population such as} NGC~6266 \citep[M~62;][]{yong14},  HP~1 \citep{barbuy16}, NGC~6558 \citep{barbuy18a}, and NGC~6522 \citep[][Barbuy et al. 2021]{barbuy14}, as illustrated in Figure \ref{fig:abundances}. The Si, Ca, and Ti abundances are also low enhanced. The abundances of the first-peak of heavy elements are relatively high, while the second-peak of heavy elements is moderately high. Finally, the r-element Eu is enhanced. It is interesting to note that the four reference bulge GCs are representatives of moderately metal-poor GCs with a BHB, whereas Pal~6 has an RHB. 

A photometric analysis combined with dynamics was performed in order to determine the probable progenitor of the cluster. For this, we derived an age of $12.4\pm0.9$ Gyr and  a distance of $7.67\pm0.19$ kpc. Due to the new and more reliable distance value, the orbital analysis indicates that Pal~6 is confined within the Galactic bulge. The dynamical analysis and the values of the age and metallicity of Pal~6 show that the cluster was most probably formed in the main-bulge progenitor of the Galaxy (in situ). Finally, considering that Pal~6 was formed in the Galaxy, it is probable that the second generation of stars in the cluster could be formed from internal pollution of the cluster, which is compatible with both the AGB star \citep{renzini15,calura19} and fast-rotating massive star \citep{decressin07,Chiappini11,frischknecht16} pollution scenarios. 

The study of GCs is of great importance to understand the formation and evolution of the Galaxy. Our analysis shows that Pal~6 is a GC formed in the Galactic bulge progenitor present in the early stages of the Milky Way, and it shares chemical properties with other well-known old-bulge GCs. 

\begin{acknowledgements}
      We acknowledge our referee, Dr. Thomas Masseron, for the detailed review and for the helpful suggestions, which allowed us to improve the manuscript.
      We are grateful to Anna B. A. Queiroz for providing the StarHorse distances with APOGEE DR16 and Gaia EDR3. SOS acknowledges the FAPESP PhD fellowship 2018/22044-3. 
      SOS and APV acknowledge the DGAPA-PAPIIT grant IG100319.
      BB and EB acknowledge grants from FAPESP, CNPq and CAPES - Financial code 001.
      S.O. acknowledges the partial support of the research program DOR1901029, 2019, and the project BIRD191235, 2019 of the University of Padova. MV acknowledges the support of the Deutsche Forschungsgemeinschaft (DFG, project number: 428473034). Part of this work was supported by the German \emph{Deut\-sche For\-schungs\-ge\-mein\-schaft, DFG\/} project number Ts~17/2--1. DN acknowledges the support from the French Centre National d'Etudes Spatiales (CNES).
\end{acknowledgements}

%
%

\begin{appendix}

\section{Fundamental parameter determination}
\begin{figure}
    \centering
    \includegraphics[width=0.9\columnwidth]{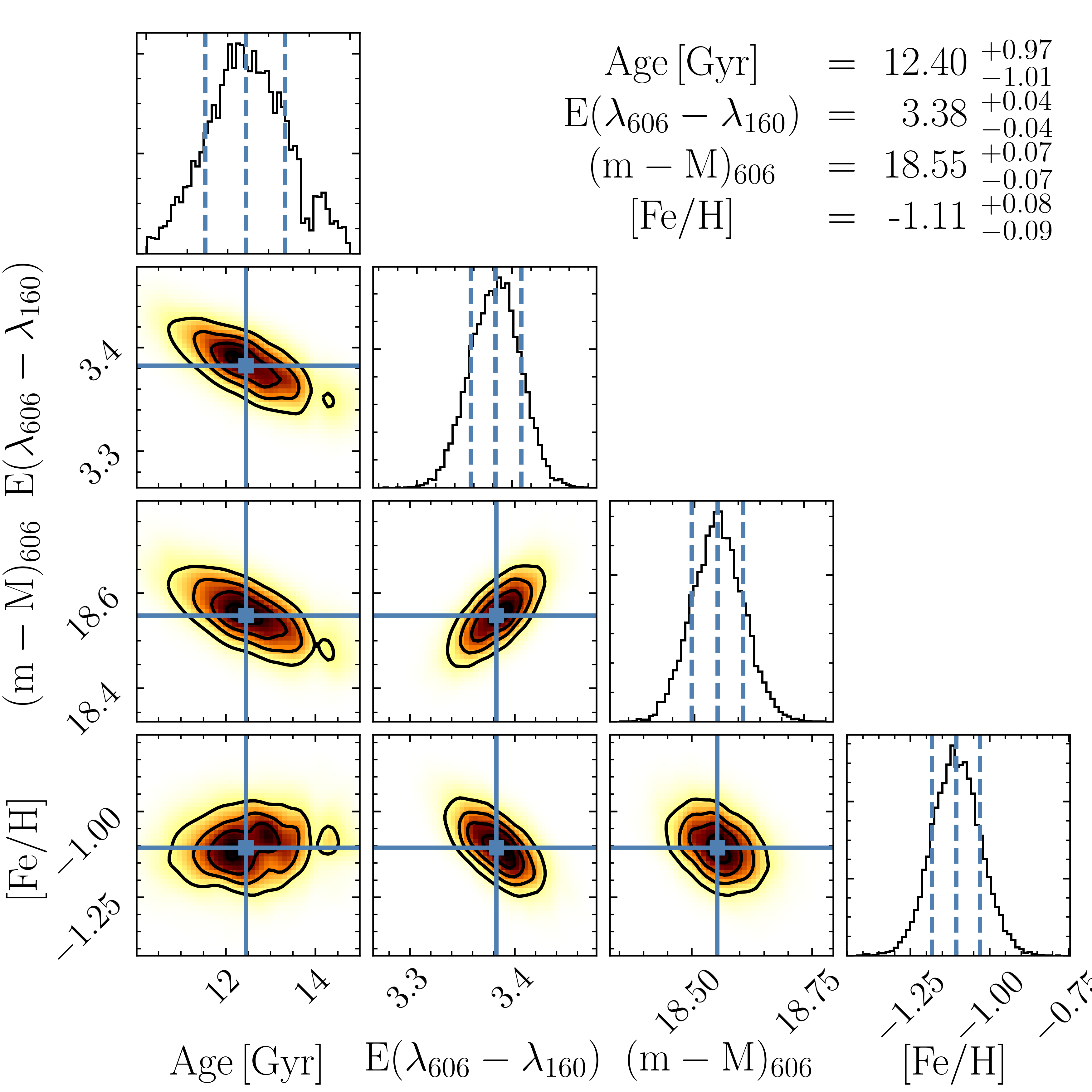}
    \caption{Corner plots with the parameter correlations. }
    \label{fig:corner}
\end{figure}

Figure \ref{fig:corner} shows the corner plots, which represent the 4D parameter space of the isochrone fitting (age, reddening, distance modulus, and metallicity) represented in 2D density distributions. The cumulative best solutions (posterior distributions) of each parameter are represented by the histograms, while the 2D density maps show the correlations between the parameters. To represent the distributions of each parameter, we assumed the region of highest density as the representative value, and the uncertainties were calculated from the 16th and 84th percentiles.

\clearpage
\onecolumn

\section{Line list}

\begin{longtable}{lcccccccccc}
\caption{Equivalent widths for \ion{Fe}{I} and \ion{Fe}{II} lines.}            
\label{tab:fe}\\      
\hline
\multirow{2}{*}{Ion} & $\lambda$ & $\chi_{ex}$  & \multirow{2}{*}{log$gf$} &    star 730   &    star 243   &    star 030   &    star 785   &     star 145  &  star 401 \\ 
                     &  [$\AA$]  &  [eV]        &                          & \multicolumn{6}{c}{[m$\AA$]}\\ 
\hline\hline \endfirsthead
\multicolumn{10}{c}%
{{\bfseries \tablename\ \thetable{} -- continued}} \\ \hline
Ion & $\lambda$ & $\chi_{ex}$  & log$gf$ &    star 730   &    star 243   &    star 030   &    star 785   &     star 145  &   star 401 \\ 
\hline\hline \endhead

\ion{Fe}{II} & $5991.38$ & $3.15$ & $-3.65$ &    $32.60$  & $39.20$  & $25.5$  &   ---   & $32.3$  &   ---    \\
\ion{Fe}{II} & $6084.11$ & $3.20$ & $-3.97$ &    $35.20$  & $26.20$  & $25.6$  & $28.7$  & $14.3$  & $18.70$  \\
\ion{Fe}{II} & $6149.25$ & $3.89$ & $-2.69$ &    $63.80$  & $43.50$  & $22.3$  & $31.1$  & $24.0$  &   ---    \\
\ion{Fe}{II} & $6247.56$ & $3.89$ & $-2.52$ &      ---    &   ---    & $27.3$  & $25.9$  & $49.4$  & $35.20$  \\
\ion{Fe}{II} & $6416.93$ & $3.89$ & $-2.64$ &    $68.90$  & $36.90$  & $27.1$  &   ---   & $35.4$  & $29.10$  \\
\ion{Fe}{II} & $6432.68$ & $2.89$ & $-3.57$ &    $22.70$  & $31.90$  & $10.2$  & $34.7$  & $28.3$  & $30.10$  \\
\ion{Fe}{II} & $6456.39$ & $3.90$ & $-2.31$ &      ---    & $37.30$  & $29.6$  & $23.3$  & $47.4$  & $54.00$  \\
\ion{Fe}{II} & $6516.08$ & $2.89$ & $-3.31$ &    $62.20$  & $59.2$   & $30.8$  & $43.1$  & $26.7$  & $47.80$  \\
\hline                                  
\ion{Fe}{I} & $5902.48$ & $4.59$ & $-1.81$ &      ---    & $5.10$   & $2.6$   &   ---   & $8.7$   &   ---    \\
\ion{Fe}{I} & $5905.67$ & $4.65$ & $-0.73$ &    $40.90$  &   ---    & $28.5$  &   ---   & $31.9$  &   ---    \\
\ion{Fe}{I} & $5905.69$ & $4.65$ & $-0.73$ &      ---    & $40.60$  &   ---   & $46.7$  &   ---   & $43.90$  \\
\ion{Fe}{I} & $5909.97$ & $3.21$ & $-2.78$ &      ---    & $84.90$  &   ---   & $32.7$  &   ---   & $59.80$  \\
\ion{Fe}{I} & $5916.25$ & $2.45$ & $-2.97$ &    $59.50$  &   ---    &   ---   &   ---   &   ---   &   ---    \\
\ion{Fe}{I} & $5916.26$ & $2.45$ & $-2.99$ &      ---    & $91.10$  & $30.9$  & $25.1$  & $32.0$  & $72.10$  \\
\ion{Fe}{I} & $5927.79$ & $4.65$ & $-1.09$ &    $38.50$  & $37.70$  & $44.4$  & $46.7$  &   ---   &   ---    \\
\ion{Fe}{I} & $5929.67$ & $4.55$ & $-1.41$ &    $45.50$  & $44.90$  & $31.3$  &   ---   &   ---   & $30.70$  \\
\ion{Fe}{I} & $5930.18$ & $4.65$ & $-0.23$ &    $51.70$  &   ---    &   ---   &   ---   &   ---   &   ---    \\
\ion{Fe}{I} & $5930.19$ & $4.65$ & $-0.23$ &      ---    & $78.70$  &   ---   & $68.3$  &   ---   & $68.70$  \\
\ion{Fe}{I} & $5934.65$ & $3.93$ & $-1.17$ &    $34.30$  &   ---    & $54.0$  &   ---   & $70.4$  &   ---    \\
\ion{Fe}{I} & $5934.67$ & $3.93$ & $-1.17$ &      ---    & $76.40$  &   ---   & $28.8$  &   ---   & $62.30$  \\
\ion{Fe}{I} & $5952.73$ & $3.98$ & $-1.44$ &    $48.50$  & $50.40$  &   ---   & $30.8$  & $52.7$  & $40.10$  \\
\ion{Fe}{I} & $5956.69$ & $0.86$ & $-4.60$ &    $42.10$  &   ---    & $37.5$  &   ---   &   ---   &   ---    \\
\ion{Fe}{I} & $5956.71$ & $0.86$ & $-4.61$ &      ---    & $79.10$  &   ---   &   ---   &   ---   & $82.50$  \\
\ion{Fe}{I} & $5975.35$ & $4.84$ & $-0.69$ &    $36.80$  &   ---    & $15.6$  &   ---   & $42.3$  &   ---    \\
\ion{Fe}{I} & $5983.69$ & $4.55$ & $-1.47$ &    $30.80$  & $52.80$  & $17.2$  &   ---   & $20.1$  & $43.90$  \\
\ion{Fe}{I} & $5987.06$ & $4.80$ & $-0.43$ &    $43.30$  &   ---    & $27.0$  &   ---   &   ---   &   ---    \\
\ion{Fe}{I} & $5987.07$ & $4.79$ & $-0.15$ &      ---    & $62.90$  &   ---   &   ---   &   ---   & $59.60$  \\
\ion{Fe}{I} & $6003.01$ & $3.88$ & $-1.12$ &    $44.80$  &   ---    &   ---   &   ---   &   ---   &   ---    \\
\ion{Fe}{I} & $6003.02$ & $3.88$ & $-1.12$ &      ---    & $90.40$  & $25.4$  & $44.3$  & $63.7$  & $76.00$  \\
\ion{Fe}{I} & $6005.54$ & $2.59$ & $-3.61$ &    $40.50$  &   ---    & $34.4$  &   ---   & $43.6$  &   ---    \\
\ion{Fe}{I} & $6008.56$ & $3.88$ & $-0.99$ &    $31.50$  &   ---    & $25.3$  &   ---   & $57.2$  &   ---    \\
\ion{Fe}{I} & $6020.17$ & $4.61$ & $-0.27$ &    $38.10$  &   ---    & $29.2$  &   ---   & $64.5$  &   ---    \\
\ion{Fe}{I} & $6024.05$ & $4.55$ & $-0.12$ &    $31.20$  & $90.40$  & $2.00$  & $73.9$  & $45.0$  & $69.90$  \\
\ion{Fe}{I} & $6027.06$ & $4.08$ & $-1.09$ &    $80.00$  & $79.70$  & $29.7$  & $43.2$  & $32.5$  & $59.10$  \\
\ion{Fe}{I} & $6054.08$ & $4.37$ & $-2.31$ &    $46.50$  & $16.50$  &   ---   &   ---   &   ---   &   ---    \\
\ion{Fe}{I} & $6056.01$ & $4.73$ & $-0.46$ &      ---    & $62.30$  & $35.7$  & $41.9$  & $23.8$  & $41.70$  \\
\ion{Fe}{I} & $6065.48$ & $2.61$ & $-1.53$ &    $66.00$  &   ---    &   ---   &   ---   &   ---   &   ---    \\
\ion{Fe}{I} & $6065.49$ & $2.61$ & $-1.53$ &      ---    & $132.60$ &   ---   &   ---   &   ---   & $101.5$  \\
\ion{Fe}{I} & $6078.49$ & $4.80$ & $-0.32$ &    $57.70$  &   ---    &   ---   &   ---   &   ---   &   ---    \\
\ion{Fe}{I} & $6078.50$ & $4.79$ & $-0.40$ &      ---    & $65.90$  & $18.0$  & $49.5$  & $41.5$  & $65.40$  \\
\ion{Fe}{I} & $6079.00$ & $4.65$ & $-1.13$ &      ---    &   ---    &   ---   &   ---   & $44.2$  & $36.90$  \\
\ion{Fe}{I} & $6079.01$ & $4.65$ & $-1.12$ &    $57.10$  &   ---    &   ---   &   ---   &   ---   &   ---    \\
\ion{Fe}{I} & $6082.71$ & $2.22$ & $-3.57$ &    $31.20$  &   ---    &   ---   &   ---   &   ---   &   ---    \\
\ion{Fe}{I} & $6082.72$ & $2.22$ & $-3.57$ &      ---    & $69.20$  &   ---   & $72.2$  &   ---   & $65.10$  \\
\ion{Fe}{I} & $6093.64$ & $4.61$ & $-1.50$ &    $25.10$  &   ---    & $11.3$  &   ---   & $41.9$  &   ---    \\
\ion{Fe}{I} & $6093.67$ & $4.60$ & $-1.51$ &      ---    & $15.50$  &   ---   &   ---   &   ---   & $25.70$  \\
\ion{Fe}{I} & $6094.36$ & $4.65$ & $-1.94$ &    $46.10$  & $18.80$  &   ---   &   ---   &   ---   &   ---    \\
\ion{Fe}{I} & $6096.66$ & $3.98$ & $-1.93$ &      ---    & $43.00$  &   ---   & $39.1$  &   ---   &   ---    \\
\ion{Fe}{I} & $6105.15$ & $4.54$ & $-2.07$ &      ---    & $16.50$  &   ---   &   ---   &   ---   & $10.40$  \\
\ion{Fe}{I} & $6137.70$ & $2.59$ & $-1.40$ &    $40.00$  & $142.10$ &   ---   &   ---   &   ---   & $146.5$  \\
\ion{Fe}{I} & $6151.62$ & $2.18$ & $-3.30$ &    $59.50$  & $114.00$ & $56.4$  & $48.4$  & $37.3$  & $110.8$  \\
\ion{Fe}{I} & $6157.73$ & $4.08$ & $-1.25$ &      ---    & $67.10$  & $27.9$  & $37.8$  & $35.1$  & $58.30$  \\
\ion{Fe}{I} & $6159.38$ & $4.61$ & $-1.97$ &    $26.40$  &   ---    &   ---   &   ---   &   ---   &   ---    \\
\ion{Fe}{I} & $6165.36$ & $4.14$ & $-1.47$ &    $43.50$  & $64.90$  & $43.1$  &   ---   & $17.7$  &   ---    \\
\ion{Fe}{I} & $6173.34$ & $2.22$ & $-2.88$ &      ---    & $122.80$ &   ---   & $46.9$  &   ---   & $96.80$  \\
\ion{Fe}{I} & $6180.21$ & $2.73$ & $-2.59$ &    $56.30$  & $89.50$  & $36.3$  &   ---   & $58.5$  & $63.90$  \\
\ion{Fe}{I} & $6187.99$ & $3.94$ & $-1.72$ &    $63.80$  & $60.10$  & $34.2$  & $48.4$  &   ---   & $62.90$  \\
\ion{Fe}{I} & $6200.32$ & $2.61$ & $-2.44$ &      ---    & $88.40$  &   ---   &   ---   &   ---   & $80.40$  \\
\ion{Fe}{I} & $6213.44$ & $2.22$ & $-2.48$ &    $88.90$  & $129.90$ & $39.5$  &   ---   &   ---   & $105.50$ \\
\ion{Fe}{I} & $6219.29$ & $2.20$ & $-2.43$ &    $96.80$  & $134.10$ &   ---   &   ---   &   ---   & $116.50$ \\
\ion{Fe}{I} & $6220.78$ & $3.88$ & $-2.46$ &    $65.00$  &   ---    &   ---   &   ---   &   ---   & $24.60$  \\
\ion{Fe}{I} & $6226.73$ & $3.88$ & $-2.22$ &    $47.40$  & $32.00$  &   ---   & $18.4$  &   ---   &   ---    \\
\ion{Fe}{I} & $6229.23$ & $2.84$ & $-2.97$ &      ---    & $83.70$  & $11.7$  & $58.5$  & $31.3$  & $52.30$  \\
\ion{Fe}{I} & $6240.65$ & $2.22$ & $-3.21$ &      ---    & $97.30$  & $17.9$  & $44.6$  & $40.6$  & $77.20$  \\
\ion{Fe}{I} & $6246.33$ & $3.60$ & $-0.88$ &      ---    & $34.30$  &   ---   & $89.4$  &   ---   & $87.30$  \\
\ion{Fe}{I} & $6252.57$ & $2.40$ & $-1.69$ &    $54.20$  & $162.20$ &   ---   & $137.8$ &   ---   & $110.40$ \\
\ion{Fe}{I} & $6254.25$ & $2.28$ & $-2.44$ &    $24.30$  &   ---    &   ---   &   ---   &   ---   &   ---    \\
\ion{Fe}{I} & $6265.14$ & $2.18$ & $-2.55$ &    $61.90$  & $119.90$ &   ---   & $121.9$ &   ---   & $105.40$ \\
\ion{Fe}{I} & $6270.23$ & $2.86$ & $-2.46$ &    $36.70$  & $61.10$  & $21.2$  & $32.4$  & $90.8$  & $61.50$  \\
\ion{Fe}{I} & $6271.28$ & $3.32$ & $-2.70$ &    $23.20$  & $40.70$  & $19.0$  & $36.4$  & $24.6$  & $39.70$  \\
\ion{Fe}{I} & $6297.80$ & $2.22$ & $-2.74$ &      ---    & $130.10$ &   ---   & $58.6$  & $84.8$  & $101.60$ \\
\ion{Fe}{I} & $6301.51$ & $3.65$ & $-0.72$ &    $138.40$ & $118.20$ &   ---   & $127.7$ &   ---   &   ---    \\
\ion{Fe}{I} & $6302.50$ & $3.69$ & $-0.91$ &      ---    & $94.40$  &   ---   & $148.5$ &   ---   &   ---    \\
\ion{Fe}{I} & $6311.50$ & $2.83$ & $-3.14$ &    $85.60$  & $81.20$  &   ---   & $26.2$  & $63.2$  & $36.50$  \\
\ion{Fe}{I} & $6315.31$ & $4.14$ & $-1.23$ &    $51.70$  & $79.10$  & $5.7$   & $42.9$  & $25.1$  & $53.60$  \\
\ion{Fe}{I} & $6315.81$ & $4.08$ & $-1.71$ &    $35.30$  & $48.60$  & $22.4$  & $58.1$  & $37.2$  & $50.10$  \\
\ion{Fe}{I} & $6322.69$ & $2.59$ & $-2.43$ &      ---    & $80.80$  &   ---   & $36.9$  &   ---   & $92.80$  \\
\ion{Fe}{I} & $6330.84$ & $4.73$ & $-1.74$ &      ---    & $44.60$  &   ---   &   ---   &   ---   &   ---    \\
\ion{Fe}{I} & $6335.34$ & $2.20$ & $-2.18$ &    $43.4$   & $143.30$ & $1.5$   & $88.4$  &   ---   & $119.30$ \\
\ion{Fe}{I} & $6336.83$ & $3.69$ & $-1.05$ &      ---    & $105.70$ & $38.7$  & $84.5$  & $60.5$  & $102.80$ \\
\ion{Fe}{I} & $6344.16$ & $2.43$ & $-2.92$ &    $87.20$  & $127.10$ & $22.8$  & $75.8$  & $82.3$  & $90.70$  \\
\ion{Fe}{I} & $6355.04$ & $2.84$ & $-2.29$ &      ---    & $128.30$ &   ---   &   ---   &   ---   & $92.70$  \\
\ion{Fe}{I} & $6358.69$ & $0.86$ & $-4.47$ &      ---    & $160.90$ &   ---   &   ---   &   ---   & $149.70$ \\
\ion{Fe}{I} & $6380.75$ & $4.19$ & $-1.38$ &    $50.80$  & $42.00$  & $15.9$  & $23.8$  & $36.2$  & $50.30$  \\
\ion{Fe}{I} & $6392.54$ & $2.28$ & $-4.03$ &    $52.50$  & $63.10$  & $11.7$  &   ---   &   ---   &   ---    \\
\ion{Fe}{I} & $6393.61$ & $2.43$ & $-1.43$ &    $46.5$   & $158.70$ &   ---   &   ---   &   ---   &   ---    \\
\ion{Fe}{I} & $6408.03$ & $3.69$ & $-1.00$ &      ---    & $101.50$ &   ---   &   ---   & $74.9$  & $95.80$  \\
\ion{Fe}{I} & $6411.11$ & $4.73$ & $-1.92$ &    $40.10$  & $13.70$  &   ---   &   ---   &   ---   & $5.0$    \\
\ion{Fe}{I} & $6411.66$ & $3.65$ & $-0.60$ &    $48.50$  & $121.10$ &   ---   &   ---   & $105.$  & $98.70$  \\
\ion{Fe}{I} & $6419.94$ & $4.73$ & $-0.24$ &    $44.40$  & $78.70$  & $15.6$  & $75.9$  & $44.2$  & $80.60$  \\
\ion{Fe}{I} & $6421.35$ & $2.28$ & $-2.03$ &    $71.10$  & $158.20$ &   ---   &   ---   & $107.$  & $113.50$ \\
\ion{Fe}{I} & $6430.86$ & $2.18$ & $-2.01$ &    $51.30$  & $174.20$ &   ---   &   ---   &   ---   & $123.40$ \\
\ion{Fe}{I} & $6469.21$ & $4.83$ & $-0.77$ &    $73.50$  & $84.00$  &   ---   & $56.5$  & $27.7$  &   ---    \\
\ion{Fe}{I} & $6475.63$ & $2.56$ & $-2.94$ &    $63.00$  & $127.30$ & $18.3$  & $70.5$  & $31.2$  &   ---    \\
\ion{Fe}{I} & $6481.88$ & $2.28$ & $-2.98$ &      ---    & $140.40$ & $16.6$  & $55.1$  & $70.5$  & $97.50$  \\
\ion{Fe}{I} & $6494.99$ & $2.40$ & $-1.27$ &    $143.20$ &   ---    & $57.1$  &   ---   &   ---   &   ---    \\
\ion{Fe}{I} & $6498.95$ & $0.96$ & $-4.70$ &      ---    & $123.10$ &   ---   & $33.8$  &   ---   & $109.60$ \\
\ion{Fe}{I} & $6518.37$ & $2.83$ & $-2.30$ &      ---    & $83.60$  &   ---   & $51.1$  &   ---   & $81.30$  \\
\ion{Fe}{I} & $6533.93$ & $4.56$ & $-1.45$ &      ---    & $31.30$  & $2.00$  &   ---   & $29.5$  &   ---    \\
\ion{Fe}{I} & $6546.25$ & $2.75$ & $-1.54$ &    $106.10$ & $159.90$ & $63.3$  &   ---   &   ---   & $141.80$ \\
\ion{Fe}{I} & $6556.81$ & $4.79$ & $-1.72$ &      ---    & $19.00$  &   ---   & $28.1$  &   ---   &   ---    \\
\ion{Fe}{I} & $6569.22$ & $4.73$ & $-0.42$ &    $99.40$  & $77.60$  & $31.6$  &   ---   & $55.9$  & $69.30$  \\
\ion{Fe}{I} & $6574.25$ & $0.99$ & $-5.02$ &      ---    & $116.30$ &   ---   & $26.7$  &   ---   & $93.60$  \\
\ion{Fe}{I} & $6575.04$ & $2.59$ & $-2.71$ &      ---    & $121.80$ &   ---   &   ---   &   ---   & $80.90$  \\
\ion{Fe}{I} & $6581.21$ & $1.48$ & $-4.68$ &    $85.20$  & $97.20$  &   ---   & $32.0$  & $23.1$  & $65.70$  \\
\ion{Fe}{I} & $6591.31$ & $4.59$ & $-2.07$ &    $28.10$  & $10.70$  & $1.00$  &   ---   &   ---   & $11.50$  \\
\ion{Fe}{I} & $6593.87$ & $2.43$ & $-2.42$ &    $39.60$  & $136.20$ & $69.4$  &   ---   &   ---   & $103.70$ \\
\ion{Fe}{I} & $6597.56$ & $4.80$ & $-1.07$ &    $45.70$  & $36.50$  & $26.7$  &   ---   &   ---   & $36.00$  \\
\ion{Fe}{I} & $6608.04$ & $2.28$ & $-4.03$ &    $74.30$  & $62.60$  & $10.0$  &   ---   & $49.4$  & $45.40$  \\
\ion{Fe}{I} & $6609.12$ & $2.56$ & $-2.69$ &    $73.40$  & $132.60$ & $24.5$  & $73.0$  &   ---   & $131.60$ \\
\ion{Fe}{I} & $6627.54$ & $4.55$ & $-1.68$ &    $36.80$  & $42.40$  & $5.1$   &   ---   & $9.9$   &   ---    \\
\ion{Fe}{I} & $6678.00$ & $2.69$ & $-1.42$ &    $45.00$  & $157.60$ &   ---   &   ---   &   ---   & $118.10$ \\
\ion{Fe}{I} & $6699.14$ & $4.59$ & $-2.10$ &    $24.40$  & $24.20$  & $2.70$  &   ---   &   ---   &   ---    \\
\ion{Fe}{I} & $6705.11$ & $4.61$ & $-1.06$ &      ---    &   ---    & $16.3$  &   ---   & $30.8$  &   ---    \\
\ion{Fe}{I} & $6726.67$ & $4.59$ & $-1.09$ &      ---    &   ---    & $28.0$  &   ---   & $16.0$  &   ---    \\
 \\ \hline \hline \\            
\end{longtable} 

\begin{longtable}{lccccccccc}
\caption{Line-by-line abundance ratios in the six UVES sample stars for the CNO, odd-Z (Na and Al), alpha- (Mg, Si, Ca, and Ti), and heavy elements (Y, Zr, Ba, La, and Eu).}            
\label{tab:lines_ratios}\\   
\hline
\noalign{\smallskip}
\multirow{2}{*}{Species} & $\lambda$    & \phantom{-}\phantom{-}{ $\chi_{ex}$ }&  \multirow{2}{*}{log$gf$} &    star 730   &    star 243   &    star 030   &    star 785   &     star 145  &   star 401 \\
                     &  [$\AA$]  &  [eV]        &                          & \multicolumn{6}{c}{[X/Fe]}\\ 
\hline\hline \endfirsthead
\multicolumn{10}{c}%
{{\bfseries \tablename\ \thetable{} -- continued}} \\ \hline
Species & $\lambda$    & \phantom{-}\phantom{-}{ $\chi_{ex}$ }&  log$gf$ &    star 730   &    star 243   &    star 030   &    star 785   &     star 145  &   star 401 \\ 
\hline\hline \endhead

\ion{Na}{I}   &  $5682.633$  &  $2.10$  &  $-0.71$  &  $+0.07$  &  $+0.28$  &  $-0.09$  &  $+0.27$  &  ---      &  ---      \\ 
\ion{Na}{I}   &  $6154.230$  &  $2.10$  &  $-1.56$  &  ---      &  $+0.48$  &  $+0.43$  &  $+0.42$  &  $+0.42$  &  ---      \\
\ion{Na}{I}   &  $6160.753$  &  $2.10$  &  $-1.26$  &  $+0.51$  &  $+0.51$  &  ---      &  $+0.13$  &  $+0.37$  &  $+0.10$  \\
\ion{Al}{I}   &  $6696.185$  &  $4.02$  &  $-1.58$  &  $+0.47$  &  $+0.37$  &  $+0.59$  &  $+0.09$  &  $+0.08$  &  $+0.33$  \\
\ion{Al}{I}   &  $6698.673$  &  $3.14$  &  $-1.65$  &  $+0.42$  &  $+0.14$  &  $+0.39$  &  $+0.29$  &  $+0.15$  &  $+0.47$  \\
\hline
\ion{Mg}{I}   &  $5528.405$  &  $5.11$  &  $-2.10$  &  ---      &  ---      &  ---      &  $+0.14$  &  ---      &  ---      \\ 
\ion{Mg}{I}   &  $6318.720$  &  $5.11$  &  $-2.36$  &  $+0.41$  &  $+0.48$  &  $+0.58$  &  $+0.34$  &  $+0.49$  &  $+0.24$  \\
\ion{Mg}{I}   &  $6319.242$  &  $5.11$  &  $-2.80$  &  $+0.45$  &  $+0.45$  &  $+0.53$  &  ---      &  $+0.40$  &  $+0.40$  \\
\ion{Mg}{I}   &  $6765.450$  &  $5.75$  &  $-1.94$  &  ---      &  $+0.27$  &  $+0.47$  &  $+0.27$  &  $+0.55$  &  $+0.25$  \\
\ion{Si}{I}   &  $5665.555$  &  $4.92$  &  $-2.04$  &  $+0.48$  &  $+0.19$  &  $+0.38$  &  $+0.60$  &  $+0.51$  &  $+0.61$  \\
\ion{Si}{I}   &  $5666.690$  &  $5.62$  &  $-1.74$  &  $+0.41$  &  $+0.59$  &  $+0.50$  &  $+0.20$  &  ---      &  $+0.44$  \\
\ion{Si}{I}   &  $5690.425$  &  $4.93$  &  $-1.87$  &  $+0.07$  &  $+0.58$  &  $+0.12$  &  ---      &  $+0.43$  &  $+0.60$  \\  
\ion{Si}{I}   &  $5948.545$  &  $5.08$  &  $-1.30$  &  $+0.21$  &  $+0.18$  &  $+0.13$  &  $+0.15$  &  $+0.54$  &  $+0.39$  \\
\ion{Si}{I}   &  $6142.494$  &  $5.62$  &  $-1.50$  &  $+0.24$  &  $+0.22$  &  ---      &  $+0.50$  &  $+0.02$  &  $+0.29$  \\
\ion{Si}{I}   &  $6145.020$  &  $5.61$  &  $-1.45$  &  $+0.49$  &  $+0.19$  &  ---      &  ---      &  $+0.23$  &  $+0.42$  \\
\ion{Si}{I}   &  $6155.142$  &  $5.62$  &  $-0.85$  &  $+0.05$  &  $+0.40$  &  $+0.20$  &  $+0.30$  &  $+0.09$  &  $+0.25$  \\
\ion{Si}{I}   &  $6237.328$  &  $5.61$  &  $-1.01$  &  $+0.43$  &  $+0.48$  &  $+0.31$  &  $+0.29$  &  $+0.43$  &  $+0.48$  \\
\ion{Si}{I}   &  $6243.823$  &  $5.61$  &  $-1.30$  &  $+0.53$  &  $+0.46$  &  ---      &  $+0.42$  &  $+0.46$  &  $+0.06$  \\
\ion{Si}{I}   &  $6414.987$  &  $5.87$  &  $-1.13$  &  $+0.12$  &  $+0.45$  &  $+0.42$  &  $+0.51$  &  $+0.67$  &  $+0.64$  \\
\ion{Si}{I}   &  $6721.844$  &  $5.86$  &  $-1.17$  &  $+0.60$  &  $+0.43$  &  $+0.52$  &  $+0.49$  &  $+0.40$  &  $+0.31$  \\
\ion{Ca}{I}   &  $5601.277$  &  $2.53$  &  $-0.52$  &  $-0.33$  &  $+0.07$  &  ---      &  ---      &  ---      &  $+0.22$  \\   
\ion{Ca}{I}   &  $5867.562$  &  $2.93$  &  $-1.55$  &  $+0.50$  &  $+0.02$  &  $+0.49$  &  $+0.40$  &  $+0.41$  &  $+0.40$  \\ 
\ion{Ca}{I}   &  $6156.030$  &  $2.52$  &  $-2.39$  &  ---      &  ---      &  ---      &  $+0.40$  &  $+0.56$  &  $+0.44$  \\ 
\ion{Ca}{I}   &  $6161.295$  &  $2.51$  &  $-1.02$  &  $+0.26$  &  ---      &  $+0.50$  &  $-0.10$  &  $-0.09$  &  $+0.48$  \\ 
\ion{Ca}{I}   &  $6166.440$  &  $2.52$  &  $-0.90$  &  $+0.37$  &  $+0.53$  &  $-0.39$  &  $+0.30$  &  $+0.27$  &  $+0.11$  \\
\ion{Ca}{I}   &  $6169.044$  &  $2.52$  &  $-0.54$  &  $+0.45$  &  $+0.43$  &  ---      &  $+0.00$  &  $+0.11$  &  $+0.28$  \\ 
\ion{Ca}{I}   &  $6169.564$  &  $2.52$  &  $-0.27$  &  $+0.55$  &  $+0.44$  &  ---      &  ---      &  $+0.15$  &  $+0.42$  \\ 
\ion{Ca}{I}   &  $6439.080$  &  $2.52$  &  $+0.30$  &  $+0.24$  &  $+0.55$  &  ---      &  ---      &  ---      &  $+0.50$  \\   
\ion{Ca}{I}   &  $6455.605$  &  $2.52$  &  $-1.35$  &  $+0.47$  &  $+0.48$  &  $+0.43$  &  $+0.10$  &  $+0.40$  &  $+0.47$  \\ 
\ion{Ca}{I}   &  $6464.679$  &  $2.52$  &  $-2.10$  &  ---      &  ---      &  ---      &  $+0.49$  &  $+0.37$  &  $+0.55$  \\ 
\ion{Ca}{I}   &  $6493.788$  &  $2.52$  &  $-2.44$  &  $-0.07$  &  $+0.14$  &  ---      &  ---      &  ---      &  $-0.10$  \\   
\ion{Ca}{I}   &  $6499.654$  &  $2.52$  &  $-0.85$  &  $+0.36$  &  $+0.58$  &  $+0.15$  &  ---      &  $+0.20$  &  $+0.43$  \\ 
\ion{Ca}{I}   &  $6572.779$  &  $0.00$  &  $-4.32$  &  ---      &  ---      &  $-0.16$  &  $+0.20$  &  $+0.32$  &  $+0.08$  \\ 
\ion{Ca}{I}   &  $6717.687$  &  $2.71$  &  $-0.61$  &  $+0.46$  &  ---      &  $-0.23$  &  $-0.10$  &  $+0.33$  &  $+0.43$  \\ 
\ion{Ti}{I}   &  $5689.459$  &  $2.29$  &  $-0.44$  &  $+0.39$  &  $+0.39$  &  $+0.42$  &  $+0.49$  &  $+0.44$  &  $+0.58$  \\
\ion{Ti}{I}   &  $5866.449$  &  $1.07$  &  $-0.84$  &  $+0.36$  &  $+0.51$  &  $+0.12$  &  $+0.09$  &  $+0.48$  &  $+0.29$  \\
\ion{Ti}{I}   &  $5922.108$  &  $1.05$  &  $-1.46$  &  ---      &  $+0.43$  &  $+0.32$  &  $+0.40$  &  $+0.27$  &  $+0.35$  \\
\ion{Ti}{I}   &  $5941.750$  &  $1.05$  &  $-1.50$  &  $+0.49$  &  $+0.28$  &  $+0.50$  &  $+0.30$  &  $+0.23$  &  $+0.35$  \\
\ion{Ti}{I}   &  $5965.825$  &  $1.88$  &  $-0.42$  &  $+0.54$  &  $+0.53$  &  $+0.47$  &  $+0.20$  &  $+0.30$  &  $+0.33$  \\
\ion{Ti}{I}   &  $5978.539$  &  $1.87$  &  $-0.53$  &  $+0.02$  &  $+0.36$  &  $+0.46$  &  $+0.60$  &  $+0.56$  &  $+0.13$  \\
\ion{Ti}{I}   &  $6064.623$  &  $1.05$  &  $-1.94$  &  ---      &  $+0.58$  &  $+0.48$  &  ---      &  $+0.44$  &  $+0.35$  \\
\ion{Ti}{I}   &  $6091.169$  &  $2.27$  &  $-0.42$  &  $+0.36$  &  $+0.43$  &  $+0.49$  &  $+0.40$  &  $+0.49$  &  $+0.33$  \\
\ion{Ti}{I}   &  $6126.214$  &  $1.07$  &  $-1.43$  &  $+0.46$  &  $+0.57$  &  $+0.36$  &  $+0.00$  &  $+0.33$  &  $+0.22$  \\
\ion{Ti}{I}   &  $6258.110$  &  $1.44$  &  $-0.36$  &  $+0.11$  &  $+0.31$  &  $+0.22$  &  $+0.00$  &  $-0.08$  &  $-0.22$  \\  
\ion{Ti}{I}   &  $6261.106$  &  $1.43$  &  $-0.48$  &  $+0.57$  &  $+0.45$  &  $+0.13$  &  $+0.00$  &  $-0.04$  &  $+0.06$  \\  
\ion{Ti}{I}   &  $6303.767$  &  $1.44$  &  $-1.57$  &  $+0.56$  &  $+0.56$  &  ---      &  $+0.20$  &  $+0.57$  &  $+0.39$  \\
\ion{Ti}{I}   &  $6312.240$  &  $1.46$  &  $-1.60$  &  ---      &  ---      &  $+0.09$  &  $+0.50$  &  ---      &  $+0.26$  \\
\ion{Ti}{I}   &  $6336.113$  &  $1.44$  &  $-1.74$  &  ---      &  $+0.25$  &  ---      &  ---      &  ---      &  $+0.56$  \\
\ion{Ti}{I}   &  $6508.150$  &  $1.43$  &  $-2.05$  &  ---      &  $+0.60$  &  ---      &  ---      &  ---      &  $+0.28$  \\
\ion{Ti}{I}   &  $6554.238$  &  $1.44$  &  $-1.22$  &  $-0.10$  &  $+0.47$  &  $+0.33$  &  $+0.20$  &  $+0.37$  &  $+0.38$  \\
\ion{Ti}{I}   &  $6556.077$  &  $1.46$  &  $-1.07$  &  $+0.25$  &  ---      &  $+0.42$  &  $+0.40$  &  $+0.46$  &  $+0.59$  \\
\ion{Ti}{I}   &  $6599.113$  &  $0.90$  &  $-2.09$  &  $+0.52$  &  $+0.50$  &  $+0.46$  &  $+0.60$  &  $+0.68$  &  $+0.40$  \\
\ion{Ti}{I}   &  $6743.127$  &  $0.90$  &  $-1.73$  &  $+0.39$  &  $+0.47$  &  $+0.10$  &  $+0.40$  &  $+0.49$  &  $+0.52$  \\
\ion{Ti}{II}  &  $5418.751$  &  $1.58$  &  $-2.13$  &  ---      &  ---      &  $-0.23$  &  ---      &  $+0.20$  &  $+0.27$  \\
\ion{Ti}{II}  &  $6491.580$  &  $2.06$  &  $-2.10$  &  $+0.18$  &  $+0.38$  &  $+0.23$  &  $-0.10$  &  $+0.00$  &  $+0.41$  \\
\ion{Ti}{II}  &  $6559.576$  &  $2.05$  &  $-2.35$  &  $+0.14$  &  $+0.26$  &  $+0.48$  &  $+0.10$  &  $+0.26$  &  ---      \\
\ion{Ti}{II}  &  $6606.970$  &  $2.06$  &  $-2.85$  &  $+0.26$  &  $+0.39$  &  $+0.02$  &  $+0.40$  &  $+0.36$  &  $+0.23$  \\  
\hline
\ion{Y}{I}    &  $6435.004$  &  $0.07$  &  $-0.82$  &  $+0.02$  &  $+0.44$  &  $+1.12$  &  $+0.78$  &  $+0.89$  &  $-0.01$ \\
\ion{Y}{II}   &  $6795.414$  &  $1.74$  &  $-1.19$  &  $+0.33$  &  $+0.23$  &  $+0.48$  &  $+0.84$  &  $+0.57$  &  $+0.09$ \\
\ion{Zr}{I}   &  $6127.475$  &  $0.15$  &  $-1.06$  &  $+0.71$  &  $+0.74$  &  $+0.58$  &  $+0.60$  &  $+0.85$  &  $+0.29$ \\
\ion{Zr}{I}   &  $6134.585$  &  $0.00$  &  $-1.42$  &  $+0.86$  &  $+0.37$  &  $+0.84$  &   ---     &  $+0.33$  &  $+0.48$ \\
\ion{Zr}{I}   &  $6140.535$  &  $0.52$  &  $-1.60$  &   ---     &  $+0.86$  &   ---     &   ---     &  $+0.95$  &  $+0.92$ \\
\ion{Zr}{I}   &  $6143.252$  &  $0.07$  &  $-1.10$  &  $+0.71$  &  $+0.75$  &  $+0.54$  &  $+0.63$  &  $+0.84$  &  $-0.04$ \\
\ion{Ba}{II}  &  $5853.675$  &  $0.60$  &  $-1.10$  &   ---     &  $+0.52$  &   ---     &   ---     &   ---     &  $+0.36$ \\
\ion{Ba}{II}  &  $6496.897$  &  $0.60$  &  $-0.32$  &   ---     &  $+0.65$  &   ---     &  $+0.23$  &   ---     &  $+0.62$ \\
\ion{La}{II}  &  $6172.721$  &  $0.13$  &  $-2.25$  &  $+0.72$  &  $+0.03$  &  $+0.86$  &  $+0.78$  &  $+0.81$  &  $+0.32$ \\
\ion{La}{II}  &  $6262.287$  &  $0.40$  &  $-1.60$  &  $+0.55$  &  $+0.30$  &   ---     &   ---     &  $+0.37$  &  $+0.16$ \\
\ion{La}{II}  &  $6296.079$  &  $1.25$  &  $-0.84$  &  $+0.37$  &   ---     &   ---     &   ---     &   ---     &  $+0.19$ \\
\ion{La}{II}  &  $6320.376$  &  $0.17$  &  $-1.56$  &  $+0.37$  &  $+0.29$  &  $+0.19$  &  $+0.52$  &  $+0.79$  &  $+0.21$ \\
\ion{La}{II}  &  $6390.477$  &  $0.32$  &  $-1.41$  &  $+0.31$  &  $+0.10$  &  $+0.67$  &  $+0.77$  &  $+0.75$  &  $+0.30$ \\
\ion{Eu}{II}  &  $6437.640$  &  $1.32$  &  $-0.32$  &  $+0.55$  &  $+0.50$  &  $+0.62$  &  $+0.65$  &  $+0.77$  &  $+0.57$ \\
\ion{Eu}{II}  &  $6645.064$  &  $1.38$  &  $+0.12$  &  $+0.31$  &  $+0.12$  &  $+0.37$  &  $+0.82$  &  $+0.72$  &  $+0.59$ \\ 
\\ \hline \hline \\            
\end{longtable} 

\end{appendix}

\end{document}